\documentclass{jfm}
\pdfoutput=1
\usepackage{graphicx}
\usepackage{epstopdf, epsfig}
\usepackage{float,bm,amsmath,amssymb}
\usepackage{url}

\shorttitle{VSHF Formation in 2D Stratified Turbulence}
\shortauthor{J. G. Fitzgerald and B. F. Farrell}

\title{Statistical State Dynamics of Vertically Sheared Horizontal Flows in Two-Dimensional Stratified Turbulence}

\author{Joseph G. Fitzgerald\aff{1}
  \corresp{\email{jfitzgerald@fas.harvard.edu}},
 \and Brian F. Farrell\aff{1}}

\affiliation{\aff{1}Department of Earth and Planetary Sciences, Harvard University,
Cambridge, MA 02138, USA}

\begin{document}

\maketitle

\begin{abstract}
Simulations of strongly stratified turbulence often exhibit coherent large-scale structures called vertically sheared horizontal flows (VSHFs). 
VSHFs emerge in both two-dimensional (2D) and three-dimensional (3D) stratified turbulence with similar vertical structure. 
The mechanism responsible for VSHF formation is not fully understood. 
In this work, the formation and equilibration of VSHFs in a 2D Boussinesq model of stratified turbulence is studied using statistical state dynamics (SSD). 
In SSD, equations of motion are expressed directly in the statistical variables of the turbulent state. 
Restriction to 2D turbulence makes available an analytically and computationally attractive implementation of SSD referred to as S3T, in which the SSD is expressed by coupling the equation for the horizontal mean structure with the equation for the ensemble mean perturbation covariance. 
This second order SSD produces accurate statistics, through second order, when compared with fully nonlinear simulations. 
In particular, S3T captures the spontaneous emergence of the VSHF and associated density layers seen in simulations of turbulence maintained by homogeneous large-scale stochastic excitation. 
An advantage of the S3T system is that the VSHF formation mechanism, which is wave-mean flow interaction between the emergent VSHF and the stochastically excited large-scale gravity waves, is analytically understood  in the S3T system. 
Comparison with fully nonlinear simulations verifies that S3T solutions accurately predict the scale selection, dependence on stochastic excitation strength, and nonlinear equilibrium structure of the VSHF.  
These results facilitate relating VSHF theory and geophysical examples of turbulent jets such as the ocean's equatorial deep jets.

\end{abstract}


\section{Introduction}\label{sec:introduction}

Understanding turbulence in stable density stratification is a central problem in atmosphere, ocean, and climate dynamics, as well as in the context of engineering flows \citep{Riley:2000bh}. 
In strongly stratified turbulence a common phenomenon is the emergence of a vertically sheared horizontal flow (VSHF). 
VSHFs have commonly been observed in numerical simulations of strongly stratified Boussinesq turbulence maintained by stochastic excitation \citep{Herring:1989fx,Smith:2001uo,Smith:2002wg,Laval:2003ko,Waite:2004bt,WAITE:2005bj,Brethouwer:2007uk,Marino:2014id,Rorai:2015bw,Herbert:2016cc,Kumar:2017ie}.

The VSHF formation mechanism and the mechanism determining the equilibrium VSHF structure remain incompletely understood. 
%Mechanisms that have been advanced include resonant and near-resonant interactions among gravity waves \citep{Holloway:1986wg,Smith:2001uo,Smith:2002wg} and rapid distortion theory \citep{Galmiche:2002cw}.
Mechanisms that have previously been advanced include rapid distortion theory \citep{Galmiche:2002cw} and resonant interactions among gravity waves \citep{Holloway:1986wg,Smith:2001uo,Smith:2002wg}. 
Although resonant interactions cannot transfer energy directly into the VSHF due to its vanishing frequency, a mechanism has been suggested in which resonant interactions transfer energy toward the VSHF which is subsequently transferred into the VSHF by non-resonant interactions \citep{Smith:2001uo,Smith:2002wg}. 
Studying the VSHF equilibration process has proven difficult because the VSHF development timescale is long compared to the timescale for establishment of equilibrium between the turbulence and the VSHF so that obtaining a statistically steady VSHF requires long simulations \citep{Brethouwer:2007uk,Herbert:2016cc}. 

Computational impediments associated with equilibrating the VSHF in simulation are mitigated by investigating VSHF behaviour in the simplified model of two-dimensional (2D) stratified turbulence. 
This approach is predicated on establishing that the dynamics of VSHF emergence in the 2D system is similar to that in the 3D system. 
A potentially important physical difference between 2D and 3D Boussinesq dynamics is that the 3D system admits modes with vertically oriented vorticity, referred to as vortical modes, while the 2D system does not. 
However, \citet{Remmel:2013ce} recently compared VSHF formation in the full 3D system to that in a reduced 3D system in which these vortical modes were removed from the dynamics and found that similar VSHFs form with or without vortical modes.
This result suggests that VSHF formation results from interactions that can be captured in the 2D system.
Another important physical difference between 2D and 3D stratified turbulence is that 3D turbulence maintains vorticity by vortex stretching and exhibits a direct cascade of energy toward small scales \citep[\emph{e.g.,}][]{LINDBORG:2006eg}, whereas 2D turbulence does not permit vortex stretching and exhibits an inverse cascade of energy toward large scales \citep[\emph{e.g.,}][]{Kumar:2017ie}. 
%However, previous investigations into the dynamics of VSHFs have suggested that a direct and spectrally nonlocal interaction between large-scale gravity waves and the VSHF \citep{Galmiche:2002cw} constitutes the central driver of VSHF formation and maintenance, and that the details of the route to dissipation are not primarily involved in VSHF dynamics. 
However, previous analysis of stratified turbulence using rapid distortion theory has identified a direct and spectrally nonlocal interaction between large-scale gravity waves and large-scale shear flows that is viable as the central driver of VSHF formation and maintenance, suggesting that the details of the route to dissipation are not primarily involved in VSHF dynamics \citep{Galmiche:2002cw}. 
Numerical simulations have also demonstrated that VSHFs form robustly in strongly stratified 2D turbulence and that these structures have similar properties to those seen in the 3D case \citep{Smith:2001uo,Kumar:2017ie}, further indicating that VSHF dynamics can be usefully studied in 2D. 
In view of the great analytic and computational advantage afforded by the availability in 2D of statistical state dynamics to elucidate the mechanisms involved we are motivated to begin our study of VSHF dynamics by exploiting this method, which has proven successful in addressing similar problems of structure formation in related systems \citep[see][and references therein]{FARRELL:2014we}. 

Spontaneous emergence of large-scale shear flows from small-scale turbulence has been extensively studied in the context of geophysical fluid dynamics, where the emergent structures are referred to as turbulent jets. 
The banded winds of Jupiter \citep{Vasavada:2005gs} provide a striking example in which the jet structure takes the form of statistically steady planetary-scale zonal (east-west oriented) winds which oscillate in sign as a function of latitude.  
Layered shear flows are also found in the weakly rotating, strongly stratified environment of the Earth's equatorial oceans. 
The equatorial deep jets (EDJs) are zonal currents observed below approximately $1000$ metres depth and within $1^{\circ}$ of latitude of the equator in all ocean basins that are characterized by a vertically sheared structure in which the zonal flow oscillates in sign as a function of depth \citep{Eden:2008hl}. 
Although the EDJs are reminiscent of the VSHFs that emerge in stratified turbulence simulations, these geophysical jets differ from VSHFs in that they are time dependent and exhibit phase propagation in the vertical direction \citep{Brandt:2011fp}. 
Nonetheless, understanding VSHF emergence in Boussinesq stratified turbulence may provide insight into the EDJs in a manner analogous to the insight provided by barotropic beta-plane turbulence into planetary-scale baroclinic jet formation \citep[\emph{e.g.,}][]{Farrell:2003ud}. 

As in the case of VSHFs, theoretical understanding of the origin and maintenance of geophysical planetary-scale turbulent jets is not yet secure. 
Attempts to theoretically explain the formation of large-scale structure from turbulence date back to \citet{FJoRTOFT:1953km} and \citet{Kraichnan:1967us}, who showed that nonlinear spectral broadening together with energy and vorticity conservation implies that energy is transferred, on average, from small scales to large scales in 2D inviscid unstratified turbulence (a similar inverse cascade may occur in 3D rotating turbulence \citep{Sukoriansky:2016ce}).
The mechanism of 2D inverse cascade is consistent with the observed concentration of energy at large scales on Jupiter \citep{Galperin:2014tw}, as the planetary-scale flow of the weather layer is believed to be both lightly damped and nearly 2D. 
Similar arguments have been made for the EDJs in which the jets are suggested to result from a nonlinear cascade in which baroclinic mode energy is funneled toward the equator \citep{Salmon:1982up}. 
However, the Jovian jets have an intricate and nearly time-invariant structure \citep{Vasavada:2005gs}, and while the vertical structure of the EDJs has not been as well established, they are also observed to be phase coherent over long times and large length scales \citep{Youngs:2015gh}. 
While general arguments based on the direction of spectral energy transfer predict that the large scales will be energized in these systems, they do not predict the form of these coherent structures. 
Other theoretical proposals for the origins of the EDJs have been based on instabilities of finite-amplitude equatorial waves \citep{Hua:2008wd} and on the linear response of the equatorial ocean to periodic wind forcing \citep{Wunsch:1977jl,McCrearyJr:1984bb}. 
Although these mechanisms can produce high-wavenumber baroclinic structure near the equator, how this structure would remain coherent in the presence of turbulence remains an open question.

Improving understanding of the formation and maintenance of shear flows in strongly stratified turbulence is the subject of this paper. 
We focus on a simple example, VSHF emergence in 2D stratified turbulence, which has at least suggestive connection to geophysical systems such as the EDJs. 

Statistical state dynamics (SSD) refers to a class of theoretical approaches to the analysis of chaotic dynamical systems in which the dynamics are expressed directly in terms of the statistical quantities of the system \citep{FARRELL:2014we}. 
A familiar example of SSD is the Fokker-Planck equation for the evolution of the probability distribution function of a system whose realizations evolve according to a stochastic differential equation. 
In this work we apply the simplest nontrivial form of SSD, known as stochastic structural stability theory (S3T) \citep{Farrell:2003ud}, to investigate VSHF formation in the stochastically excited 2D Boussinesq system. 
By comparing the results of analysis of the S3T system to simulations made with the full nonlinear equations (NL), we show that S3T captures the essential features of the full system, including the emergence and structure of the VSHF and associated density layers.
%In S3T, and the related system referred to as CE2 (second-order cumulant expansion) \citep{Marston:2010ew}, nonlinearity due to perturbation-perturbation advection is either set to zero or stochastically parameterized, so that the SSD is closed at second order.   
In S3T, and the related system referred to as CE2 \citep[second-order cumulant expansion,][]{Marston:2010ew}, nonlinearity due to perturbation-perturbation advection is either set to zero or stochastically parameterized, so that the SSD is closed at second order.   
This second-order closure has proven useful in the study of coherent structure emergence in barotropic turbulence \citep{Farrell:2007fq,Marston:2008gx,Srinivasan:2012im,Tobias:2013hk,Bakas:2013ft,Constantinou:2014fh,Parker:2014ui,Bakas:2017uh}, two-layer baroclinic turbulence \citep{Farrell:2008fd,Farrell:2009cq,Marston:2010ew,Marston:2012co,Farrell:2017ed}, turbulence in the shallow-water equations on the equatorial beta-plane \citep{Farrell:2009iu}, drift wave turbulence in plasmas \citep{Farrell:2009dt,Parker:2013hy}, unstratified 2D turbulence \citep{Bakas:2011bt}, rotating magnetohydrodynamics \citep{Tobias:2011cn,Squire:2015kb,Constantinou:2018ut}, 3D wall-bounded shear flow turbulence \citep{Farrell:2012jm,Thomas:2014ek,Thomas:2015dl,Farrell:2017dx,Farrell:2017iz,Farrell:2017br}, and the turbulence of stable ion-temperature-gradient modes in plasmas \citep{StOnge:2017tu}. 
In the present work we place 2D stratified Boussinesq turbulence into the mechanistic and phenomenological context of the mean flow-turbulence interaction mechanism that has been identified in these other turbulent systems.

In formulating the S3T dynamics for the Boussinesq system the perturbation vorticity and buoyancy variables are expressed in terms of ensemble mean two-point covariance functions. 
%When coupled to the dynamics of the mean state this second-order perturbation dynamics contains the statistical wave-mean flow interaction between the turbulent perturbation fluxes and the horizontal mean structure, but greatly simplifies the dynamics by discarding the phase information in the horizontal direction pertaining to the detailed configuration of the turbulent perturbation fields, which we will demonstrate is inessential to the VSHF formation mechanism. %, greatly simplifying the dynamics.  
When coupled to the dynamics of the mean state this second-order perturbation dynamics contains the statistical wave-mean flow interaction between the turbulent perturbation fluxes and the horizontal mean structure. 
The dynamics is greatly simplified by discarding the phase information in the horizontal direction pertaining to the detailed configuration of the turbulent perturbation fields, which we will demonstrate to be inessential to the VSHF formation mechanism. %, resulting in a greatly simplified dynamics. 
Because the S3T dynamics is written in terms of two-point covariance functions the state space of the S3T system is of higher dimension than that of the underlying system, and use of the 2D, rather than 3D, Boussinesq system substantially reduces the resulting computational burden. 

The SSD approach used in S3T permits identification and analysis of cooperative phenomena and mechanisms operating in turbulence that cannot be expressed using analysis based on a single realization. 
For example, we will show that in the S3T system, the initial formation of the VSHF occurs through a bifurcation associated with the onset of a linear instability caused by a statistical wave-mean flow interaction mechanism in which turbulent fluxes are organized by a perturbatively small mean flow in such a way as to reinforce that flow. 
The resulting instability is a statistical phenomenon that lacks analytical expression in the dynamics of a single realization and therefore cannot be fundamentally understood through analysis of single realizations of the turbulent state. 
However, the reflection of this phenomenon is strikingly apparent in single realizations of the system, and we will demonstrate that the VSHF structures predicted to arise via S3T instabilities emerge in NL simulations of realizations. 
The S3T system also reveals subtle details of turbulent equilibrium structures that might not otherwise be detected from observing the NL simulations, including the turbulent modification of the horizontal mean stratification producing density layers. Although these density layers are obscured by fluctuations in snapshots of the flow, time-averaging reveals that they coincide with the structure predicted by the S3T system.

The present work is closely related to our recent work, \citet{Fitzgerald:2018vx}, in which we apply the linearized differential formulation of S3T, originally developed by \citet{Srinivasan:2012im}, to analyze VSHF formation in 2D Boussinesq turbulence. 
In \citet{Fitzgerald:2018vx} we focus on the initial linear formation process of VSHFs and analyze how this process depends on the structure of the underlying turbulence and how individual physical processes contribute to the VSHF formation mechanism. 
In the present work, we instead apply the conventional matrix formulation of S3T and focus on analyzing the structure and maintenance mechanisms of finite amplitude equilibria in 2D Boussinesq turbulence. 
 
The structure of the paper is as follows. In \textsection \ref{sec:NLphenom} we introduce the 2D stochastically excited Boussinesq equations and present NL simulation results demonstrating VSHF formation. In \textsection \ref{sec:testfunction} we use SSD to illustrate the wave-mean flow interaction mechanism underlying VSHF formation and maintenance. In \textsection \ref{sec:QL_and_SSD_formulation} we formulate the deterministic S3T system in its conventional matrix form and also the intermediate quasilinear (QL) system, which provides a stochastic approximation to the second-order closure and bridges the gap between NL simulations and the S3T system.
In \textsection \ref{sec:modelcomparison} we show that the primary phenomena observed in NL simulations are captured by the QL and S3T systems. In \textsection \ref{sec:scaleselection} we carry out a linear stability analysis of the S3T system and relate the results to the scale selection of the initially emergent VSHF in NL simulations. In \textsection \ref{sec:equilibration} we analyze the finite-amplitude equilibration of the VSHF as a function of the strength of the stochastic excitation. In \textsection \ref{sec:multiple_equilibria} we show that multiple simultaneously stable turbulent equilibrium states exist in this system, a phenomenon which is predicted by S3T and verified in the NL simulations. In \textsection \ref{sec:bifurcation_comparison} we compare the NL, QL, and S3T systems as a function of the excitation strength and show that the VSHF-forming bifurcation predicted by S3T is reflected in the NL and QL systems. We conclude and discuss these results in \textsection \ref{sec:conclusions}. Appendix \ref{sec:appendixA} describes a  simplified model system in which the mathematical structure and conceptual utility of S3T is revealed simply. Appendix \ref{sec:appendixB} provides analytical details required for the linear stability analysis.

\section{VSHF Formation in Simulations of 2D Boussinesq Turbulence}\label{sec:NLphenom}

We study VSHF formation using the 2D stochastically excited Boussinesq equations using a unit aspect ratio $(x,z)$ computational domain doubly periodic with length $L$ in the $x$ and $z$ directions. 
Anticipating the development of horizontal mean structure we use a Reynolds decomposition in which the averaging operator is the horizontal mean. 
The resulting equations, in terms of the mean velocity, perturbation vorticity, and mean and perturbation buoyancy are
\begin{align}
\frac{\partial U}{\partial t} &= -\frac{\partial}{\partial z}\overline{u'w'}-r_m U + \nu \frac{\partial^2 U}{\partial z^2}, \label{eq:NL1} \\
\frac{\partial B}{\partial t} &=  -\frac{\partial}{\partial z}\overline{w'b'}-r_m B + \nu \frac{\partial^2 B}{\partial z^2}, \label{eq:NL2} \\
\frac{\partial \Delta \psi'}{\partial t} &= - U\frac{\partial \Delta \psi'}{\partial x} +w'\frac{\partial^2 U}{\partial z^2} + \frac{\partial b'}{\partial x} -\left[J(\psi',\Delta \psi')-\overline{J(\psi',\Delta \psi')}\right]-r \Delta \psi'+\nu \Delta^2 \psi' + \sqrt{\varepsilon}S, \label{eq:NL3} \\
\frac{\partial b'}{\partial t} &= -U \frac{\partial b'}{\partial x}-w'\left(N_0^2+\frac{\partial B}{\partial z}\right) -\left[J(\psi',b')-\overline{J(\psi',b')}\right]-rb'+\nu \Delta b'. \label{eq:NL4}
\end{align}

In these equations an overline indicates a horizontal mean and primes indicate deviations from the horizontal mean so that $f'=f-\overline{f}$. The velocity is $\boldsymbol{u}=(u,w)$ with $u$ and $w$ the horizontal and vertical velocity components, $U=\overline{u}$ is the horizontal mean horizontal velocity, $b$ is the buoyancy with $B=\overline{b}$ the horizontal mean buoyancy, $\psi$ is the streamfunction satisfying $(u,w)=(-\partial_z \psi,\partial_x \psi)$, and the vorticity is $\Delta \psi=\partial_x w - \partial_z u$ where $\Delta = \partial_{xx}^2+\partial_{zz}^2$ is the Laplacian operator. Perturbation-perturbation advection terms are written using the Jacobian $J(f,g)=(\partial_x f )(\partial_z g)- (\partial_x g)(\partial_z f)$. 
$\sqrt{\varepsilon}S$ denotes the stochastic excitation, which has zero horizontal mean and excites the perturbation vorticity only. 
$\varepsilon$ controls the strength of the excitation. 
$N_0$ is the constant background buoyancy frequency. 
Dissipation is provided by Rayleigh drag and diffusion acting on both the buoyancy and vorticity fields. 
Consistent with previous studies of VSHF formation, dissipation coefficients are assumed equal for vorticity and buoyancy, \emph{i.e.,} the Prandtl numbers associated with the Rayleigh drag and with the diffusive dissipation are each set equal to one. 
To approximate the effects of diffusive turbulent dissipation, which damps the large scales less rapidly, the Rayleigh drag on the mean fields (with coefficient $r_m$) is typically taken to be weaker than that on the perturbation fields (with coefficient $r$). 
We refer to equations (\ref{eq:NL1})-(\ref{eq:NL4}) as the NL equations (for fully nonlinear) to distinguish them from the quasilinear (QL) and S3T systems which we formulate in \textsection \ref{sec:QL_and_SSD_formulation}. 

Use of Rayleigh drag in (\ref{eq:NL1})-(\ref{eq:NL4}) departs from the diffusive dissipation commonly used in simulating stratified turbulence \citep[\emph{e.g.},][]{Smith:2001uo}. 
Rayleigh drag provides a simplified parameterization of dissipation that allows the system to reach statistical equilibrium quickly, enabling simulations to obtain the asymptotic state of the VSHF that is difficult to study comprehensively using diffusive dissipation. 
We emphasize that the essential phenomenon of VSHF formation does not depend crucially on the details of the dissipation, which we demonstrate using examples near the end of the present section. 

\begin{figure}
    \centerline{\includegraphics{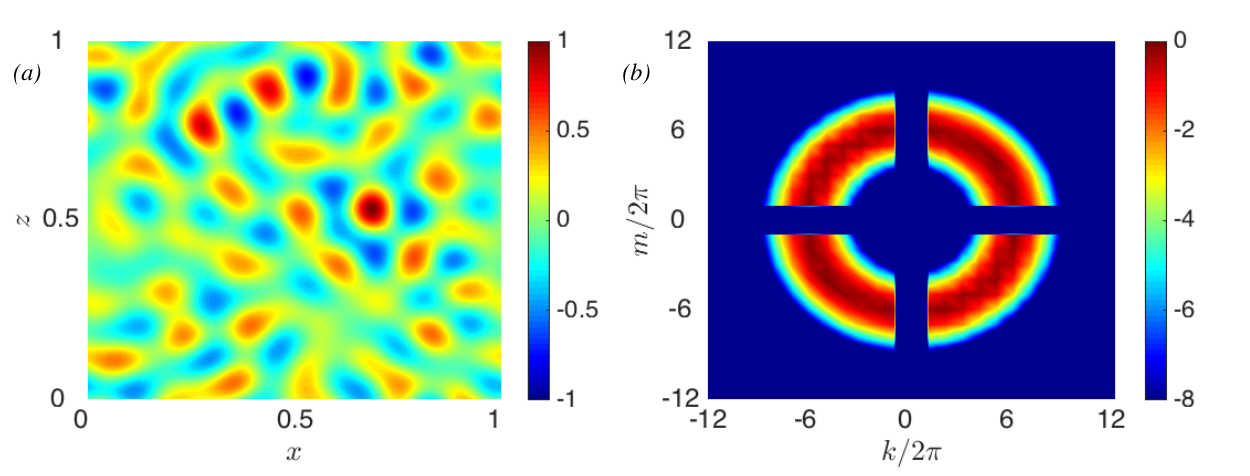}}
    \caption{Spatial structure of the stochastic excitation of the vorticity field, $\sqrt{\varepsilon}S$. (a) A sample realization of the excitation pattern, shown in normalized form as $S(x,z,t)/\text{max}[S(x,z,t)]$. (b) The wavenumber power spectrum of $S$, shown in normalized logarithmic form as $\ln(P(k,m)/\text{max}[P(k,m)])$. 
Here we define $P(k,m)=\langle |\tilde{S}_{k,m}|^2\rangle$ in which $\tilde{S}_{k,m}(t)$ is the Fourier coefficient of the excitation when $S$ is expanded as $S(x,z,t)=\sum_{k,m}\tilde{S}_{k,m}(t)e^{\text{i}(kx+mz)}$. 
Angle brackets indicate the ensemble average over noise realizations. The parameters of the excitation are $k_e/2\upi=6$ and $\delta k /2\upi=1$.}
\label{fig:Z_Forcing_Plot}
\end{figure}

We choose the stochastic excitation, $\sqrt{\varepsilon}S$ in (\ref{eq:NL3}), to have the spatial structure of an isotropic ring in Fourier space and to be delta-correlated in time. 
Figure \ref{fig:Z_Forcing_Plot} shows a snapshot of $\sqrt{\varepsilon}S$ (panel (a)) and its wavenumber power spectrum (panel (b)), in which $\boldsymbol{k}=(k,m)$ is the vector wavenumber with $k$ and $m$ the horizontal and vertical wavenumber components. 
The excitation is homogeneous in space and approximately isotropic, with some anisotropy being introduced by the omission of the horizontal mean ($k=0$) and vertical mean ($m=0$) components of the excitation and also by the finite domain size. 
We set the total wavenumber of the ring, $k_e$, to be global wavenumber six, $k_e/(2\upi L)=6$. 
As the excitation is delta-correlated in time, the rate at which energy is injected into the flow by the vorticity excitation is a control parameter that is independent of the system state. 
Here we define the kinetic energy, $K$, the potential energy, $V$, and the total energy, $E$, of the flow as 
\begin{align}
K=[\frac{1}{2}\boldsymbol{u \cdot u}], && V=[\frac{1}{2}N_0^{-2}b^2], && E=K+V,
\end{align}
in which square brackets indicate the domain average. 
The energy injection rate as a function of wavenumber, denoted $\varepsilon_{k,m}$, follows a Gaussian distribution centred at $k_e$ so that $\varepsilon_{k,m}=\alpha\exp [-(|\boldsymbol{k}|-k_e)^2/\delta k^2]$, where $\delta k=2\upi/L$ sets the ring thickness and $\alpha$ is a normalization factor chosen so that the total energy injection rate summed over all wavenumbers, $\sum_{k,m}\varepsilon_{k,m}$, is equal to the value of the parameter $\varepsilon$ appearing in (\ref{eq:NL3}). 
With this normalization $\varepsilon$ corresponds to the rate at which the vorticity excitation injects energy into the system. Global horizontal wavenumbers $1-8$ have nonzero excitation and all higher horizontal wavenumbers are omitted from the excitation. 

Equations (\ref{eq:NL1})-(\ref{eq:NL4}) are nondimensionalized by choosing the unit of length to be the domain size, $L$, and the unit of time to be the Rayleigh damping time of the perturbations, $1/r$. 
The nondimensional parameters of the problem are $\widehat{k_e}=Lk_e$, $\widehat{\delta k}=L\delta k$, $\widehat{r_m}=r_m/r$, $\widehat{\nu}=\nu/(L^2 r)$, $\widehat{\varepsilon}=\varepsilon/(r^3 L^2)$, and $\widehat{N_0^2}=N_0^2/r^2$. 
We hold fixed the parameters $\widehat{k_e}/2\upi=6$, $\widehat{\delta k }/2\upi=1$, $\widehat{r_m}=0.1$, $\widehat{\nu}=2.4\times 10^{-5}$, and $\widehat{N_0^2}=10^3$ unless otherwise stated. 
The choice of $\widehat{k_e}$ represents a compromise between providing separation between the excitation scale and the domain scale while minimizing the effects of diffusion on perturbations at the excitation scale. 
Modelling scale-selective diffusive dissipation motivates setting $\widehat{r_m}<1$ and our specific choice to set $\widehat{r_m}=0.1$, so that the mean fields are damped ten times less rapidly than the perturbation fields, is made for computational convenience. 
We examine the sensitivity of the system to this choice in figure \ref{fig:newfig1} (a). 
The value of $\widehat{\nu}$ is small and was selected to ensure numerical convergence. 
The rate of energy injection by the excitation, $\widehat{\varepsilon}$, is the primary control parameter which is varied to determine the response of the system to changes in excitation.

We choose $\widehat{N_0^2}=10^3$ to place the system in the strongly stratified regime in which VSHFs have previously been found to form \citep{Smith:2001uo,Smith:2002wg}. 
The strongly stratified regime is also the regime relevant to the EDJs. 
For example, taking the equatorial deep stratification as $N_{\text{deep}} \sim 2\times10^{-3} \text{ s}^{-1}$, a typical gravity wave wavelength of $\lambda_{\text{GW}}\sim 10 \text{ km}$, and a lateral eddy viscosity of $\nu_{\text{eddy}}\sim 100 \text{ m}^2 \text{ s}^{-1}$ gives an effective Rayleigh drag coefficient of $r_{\text{eff}}\sim (2\upi/\lambda_{\text{GW}})^2 \nu_{\text{eddy}}\sim 4\times10^{-5}\text{ s}^{-1}$ and so $\widehat{N_0^2}_{\text{,EDJ}}=N_{\text{deep}}^2/r_{\text{eff}}^2\sim2500$. 
Although we do not attempt in this work to model the EDJs, which have 3D structure and are influenced by rotation and boundaries, this estimate suggests that the presently studied idealized turbulence is in the appropriate parameter regime to allow comparison between our VSHF dynamics and EDJ phenomena. 
For the remainder of the paper we work exclusively in terms of nondimensional parameters and drop hats in our notation. 

We now summarize the behaviour of an NL simulation exhibiting VSHF formation in which the system was integrated from rest over $t\in[0,60]$ with $\varepsilon=0.25$ and the other parameters as described above, which we refer to as the standard parameter case. 
The standard case value of $\varepsilon$ places the system in the parameter regime in which strong VSHF formation occurs; the sensitivity of the system to $\varepsilon$ is examined in \textsection 6, \textsection 7, and \textsection 9. 
In \textsection \ref{sec:modelcomparison} we compare the first- and second-order statistical features of NL simulations with the results of the QL and S3T simulations. 
To perform the numerical integration we use a 2D finite-difference configuration of DIABLO \citep{Taylor:2008um} with 512 gridpoints in both the $x$ and $z$ directions. 

To estimate the canonical scales and nondimensional parameters of the standard case simulation we use the estimates $U_0\sim\sqrt{\varepsilon}$ for the velocity scale and $L_0\sim1/k_e$ for the length scale. 
The velocity scale is estimated based on the approximate energy balance in the absence of a VSHF, $\dot{E}\approx -2E+\varepsilon$, together with the estimate $U_0\sim\sqrt{E}$. 
Using these estimates, the Froude number of the standard parameter case is $Fr\equiv U_0/(L_0N_0)\approx 0.6$, the Ozmidov wavenumber is $k_O/(2\upi)\approx 56$ where $k_O\equiv (N_0^3/\varepsilon)^{1/2}$, and the buoyancy wavenumber is $k_b/(2\upi)\approx 10$ where $k_b\equiv N_0/U_0 \sim N_0/\sqrt{\varepsilon}$. 
The buoyancy Reynolds number is conventionally defined as $Re_b \equiv \varepsilon/(\nu N_0^2)$ and is used to estimate the ratio of the vertical advection term to the viscous damping term in the horizontal momentum equation in 3D stratified turbulence \citep{Brethouwer:2007uk}. 
Using this definition, the value of $Re_b$ in the standard parameter case is $Re_b \approx 10.4$. 
Although our system is 2D and includes Rayleigh drag, this estimate of $Re_b$ is consistent with the time average value in the standard case simulation of the ratio of interest, $(w' \partial_z u')_{RMS}/(-u'+\nu\Delta u')_{RMS}\approx10.7$, where the time average is calculated over the final 15 time units of the simulation and the subscript RMS denotes the root mean square average over space. 

\begin{figure}
           \centerline{\includegraphics{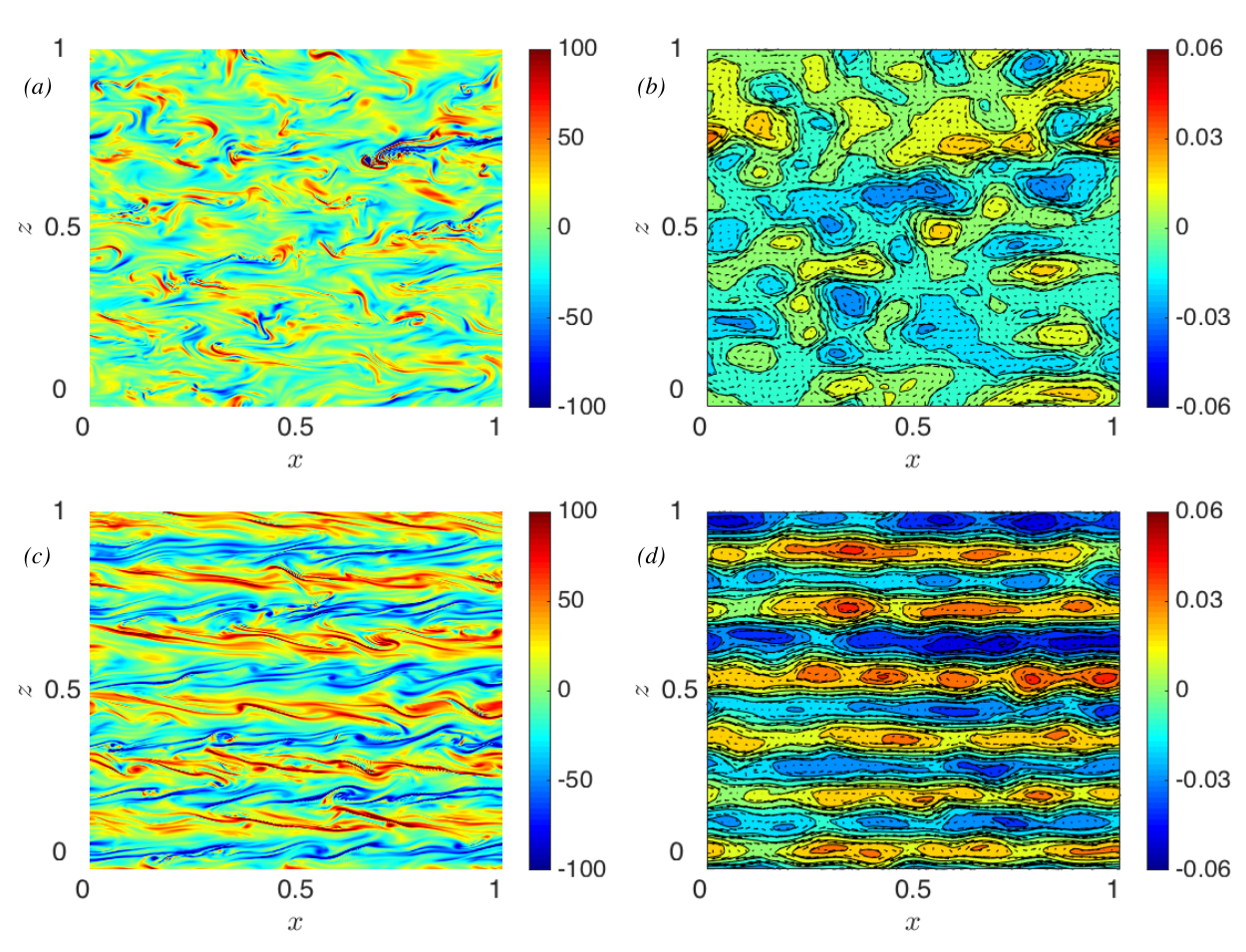}} 
      \caption{Snapshots of the vorticity, streamfunction, and velocity fields for the standard case NL simulation showing the development of the VSHF in turbulence. Just after initialization ($t=2.5$), the vorticity field (a) and the streamfunction and associated velocity field (b) are characterized by perturbations at the scale of the excitation. The system evolves into a statistical equilibrium state by $t=60$ in which the vorticity field (c) is dominated by horizontal stripes with alternating sign indicative of a strong VSHF. The streamfunction and velocity field at $t=60$ (d) show that the VSHF is the dominant feature of the instantaneous flow. Parameters are set to the standard values $r_m=0.1$, $N_0^2=10^3$, $k_e/2\upi=6$, $\delta k/2\upi=1$, $\nu=2.4\times 10^{-5}$, and $\varepsilon=0.25$. The buoyancy Reynolds number is $Re_b=10.4$ and the Froude number is $Fr=0.6$.}
\label{fig:NL_Phenom_FlowSnapshots}
\end{figure}

\begin{figure}
      \centerline{\includegraphics{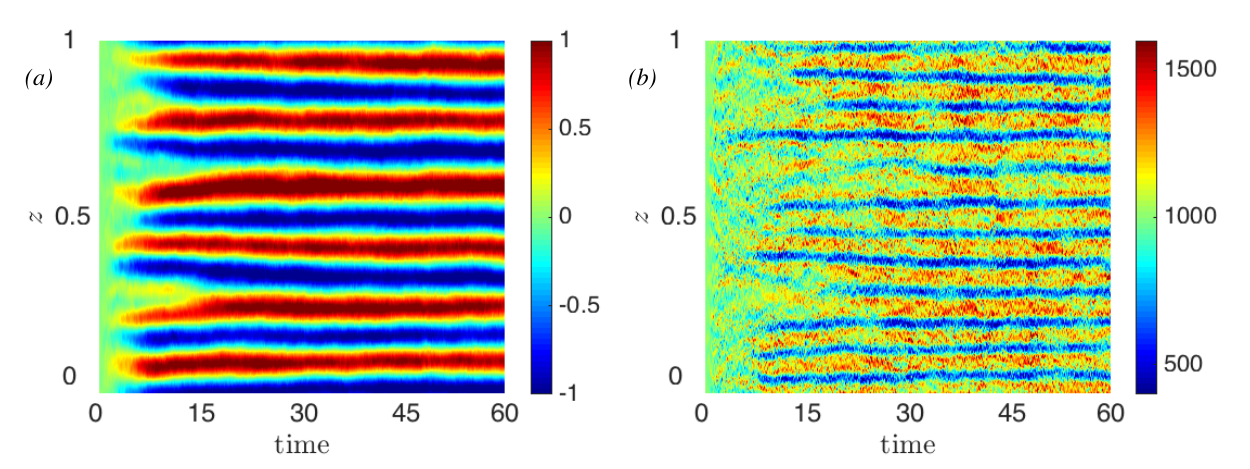}}   
  \caption{Development of the VSHF and associated density layers in the standard case NL simulation. (a) Time evolution of the horizontal mean flow, $U$, which develops from zero at $t=0$ into a persistent VSHF pattern with vertical wavenumber $m_U/2\upi=6$ by $t\approx15$. (b) Time evolution of the horizontal mean stratification, $\overline{N^2}$, which develops into a pattern with vertical wavenumber $m_B/2\upi=12$ that is phase-aligned with $U$ so that regions of weak stratification coincide with the shear regions of the VSHF structure.}
\label{fig:NL_Phenom_Hovmollers}
\end{figure}

\begin{figure}
       \centerline{\includegraphics{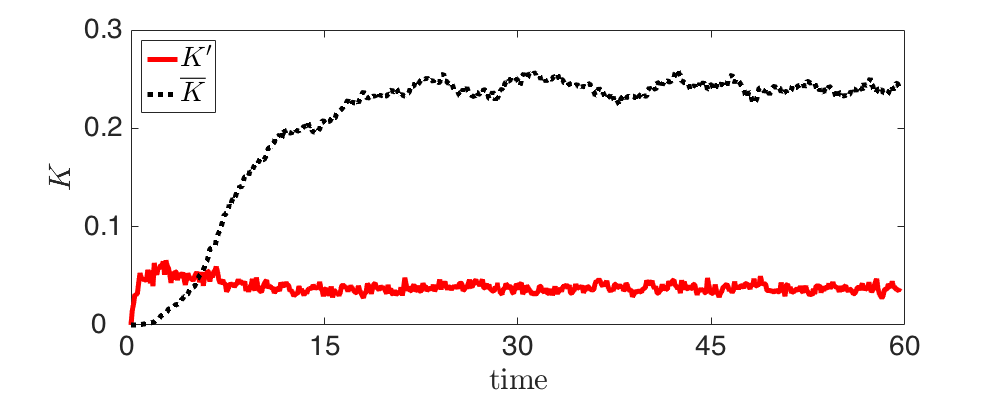}}   
  \caption{Kinetic energy evolution in the standard case NL simulation. In statistically steady state, the kinetic energy of the VSHF (dotted line) is approximately six times that of the perturbations (solid line).}
\label{fig:NL_Phenom_Energy_timeseries}
\end{figure}

\begin{figure}
      \centerline{\includegraphics{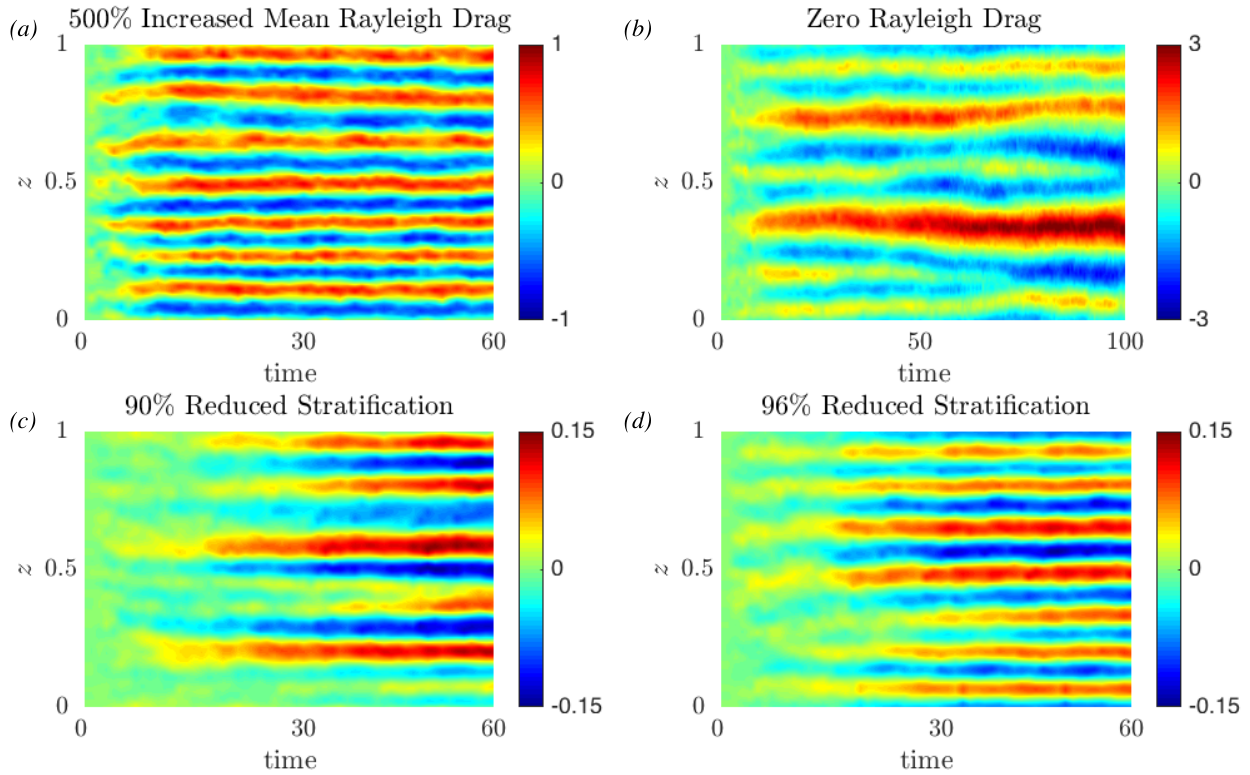}}
  \caption{Time evolution of the VSHF in four additional cases. Unless otherwise stated all parameters are as in figure \ref{fig:NL_Phenom_FlowSnapshots}. (a) An example with enhanced Rayleigh drag on the mean fields, $r_m=0.5$, and excitation strength $\varepsilon=0.5$ ($Re_b = 20.8$, $Fr = 0.8$). (b) An example with zero Rayleigh drag on both the mean and perturbations, $r=r_m=0$ ($Re_b = 10.4$, $Fr = 0.23$). Dissipation is provided solely by diffusion. (c,d) Two examples with reduced stratification and with excitation strength $\varepsilon=1.5\times10^{-2}$: (c) $N_0^2=100$ ($Re_b=6.3$, $Fr=.05$) and (d) $N_0^2=40$, ($Re_b=15.6$, $Fr=.12$). This figure demonstrates that VSHFs form robustly when the dissipation and stratification are varied.}
\label{fig:newfig1}
\end{figure}

Indicative example snapshots and time series of the NL system are shown in figures \ref{fig:NL_Phenom_FlowSnapshots}, \ref{fig:NL_Phenom_Hovmollers}, and \ref{fig:NL_Phenom_Energy_timeseries}. 
Near the start of the integration (figure \ref{fig:NL_Phenom_FlowSnapshots} (a,b)), the structure of the flow reflects the structure of the stochastic excitation and is incoherent with a dominant length scale corresponding to the stochastic excitation scale, $1/k_e$.
By $t=60$ (figure \ref{fig:NL_Phenom_FlowSnapshots} (c,d)) the system has evolved into a state in which the flow is dominated by the VSHF, $U$, which manifests as horizontal `stripes' in both the vorticity and streamfunction fields with vertical wavenumber $m_U/(2\upi)=6$. 
Simulated realizations of the NL system in the standard parameter case are always found to form a VSHF, but the VSHF wavenumber, $m_U$, differs slightly between simulations when the system is initialized from rest. 
We focus, in this section, on an example in which $m_U/(2\upi)=6$ to facilitate comparison with SSD results in \textsection \ref{sec:modelcomparison}. 
However, VSHFs with $m_U/(2\upi)=7$ form somewhat more frequently, which we discuss in \textsection \ref{sec:scaleselection}. % a phenomenon which we discuss in . %, and VSHFs with $m_U/(2\pi)=8$ also emerge in some simulations. 
%The existence of multiple possible VSHF scales is expected from the point of view of SSD and is discussed in \textsection \ref{sec:multiple_equilibria}. 
%
We analyze how the VSHF wavenumber, $m_U$, is related to the parameters in \textsection \ref{sec:scaleselection} and \textsection \ref{sec:equilibration}, but presently note that in the standard parameter case $m_U$ is closely related to the excitation wavenumber, $k_e/(2\upi)=6$, and that $m_U$ differs from the Ozmidov wavenumber, $k_O/(2\upi)\approx 56$, and from the buoyancy wavenumber, $k_b/(2\upi)\approx 10$. 

The time evolution of $U$ is shown in figure \ref{fig:NL_Phenom_Hovmollers} (a). 
The VSHF forms by $t\approx15$ and persists until the end of the integration. 
Figure \ref{fig:NL_Phenom_Energy_timeseries} shows the time evolution of the kinetic energy of the VSHF, $\overline{K}$, and of the perturbations, $K'$, where these energies are defined as
\begin{align}
\overline{K}=[\frac{1}{2}U^2], && K'=[\frac{1}{2}\boldsymbol{u' \cdot u'}].
\end{align}
The VSHF is the energetically dominant feature of the statistically steady flow, containing approximately six times more kinetic energy than the perturbations. 
In the statistical equilibrium state, the kinetic energy that is injected into the perturbation field by the stochastic excitation is transferred both into the mean flow, thereby maintaining the VSHF, and into the buoyancy field. 
Energetic balance is maintained by dissipation of the mean and perturbation energies at large scales by Rayleigh drag, with viscosity contributing only weakly to the total dissipation. %playing only a minror role. 

Although the phenomenon of VSHF emergence in stratified turbulence is well-known, the concurrent development of coherent horizontal mean structure in the buoyancy field has not been emphasized in the literature. 
Figure \ref{fig:NL_Phenom_Hovmollers} (b) shows the time evolution of the horizontal mean stratification $\overline{N^2}=N_0^2+\partial_z B$. 
Although $\overline{N^2}$ exhibits more temporal variability than $U$, it is clear that for these parameter values the turbulent fluxes systematically weaken the stratification ($\overline{N^2}<N_0^2$) in the shear regions of the VSHF. 
Association of mean stratification anomalies with the mean shear produces a vertical wavenumber in $\overline{N^2}$ of $m_{B}/2\upi=12$, twice that of the $m_U/2\upi=6$ structure of the VSHF.

\begin{figure}
		\centerline{\includegraphics{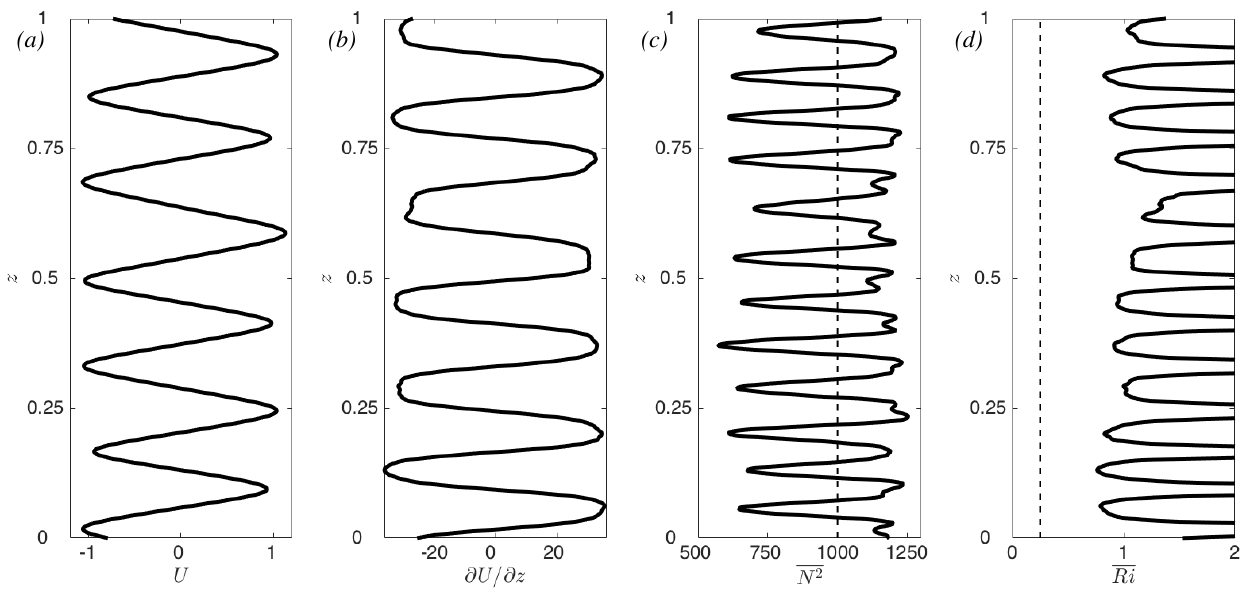}}
  \caption{Vertical structure of the time average horizontal mean state in the standard case NL simulation. (a) Mean flow, $U$. (b) Mean shear, $\partial U/\partial z$. (c) Mean stratification, $\overline{N^2}$. The vertical dashed line indicates $N_0^2$. (d) Mean Richardson number, $\overline{Ri}=\overline{N^2}/(\partial U/\partial z)^2$. The vertical dashed line indicates $\overline{Ri}=1/4$. Profiles are time averages over $t\in[30,60]$ of the structures shown in figure \ref{fig:NL_Phenom_Hovmollers}.}
\label{fig:4panel_USNsqRi_meanplot}
\end{figure}

The statistical equilibrium horizontal mean state, obtained by averaging the flow subsequent to a spin-up period of 30 time units, is shown in figure \ref{fig:4panel_USNsqRi_meanplot}. 
Panels (a) and (b) show that, for these parameters, the VSHF has a vertical structure that deviates somewhat from harmonic, with flattened shear regions resulting in a profile resembling a sawtooth structure. 
Comparison of panels (b) and (c) reveals that the shear extrema coincide with the minima of $\overline{N^2}$. These $\overline{N^2}$ minima correspond to narrow density layers in which $\overline{N^2}$ is reduced by approximately $40\%$ relative to $N_0^2$. Similar density layers have been reported in observations and simulations of the EDJs \citep{Menesguen:2009uu}. 
As the vertical integral of $\overline{N^2}-N_0^2$ must vanish due to the periodicity of the boundary conditions in the vertical direction by (\ref{eq:NL2}), the narrow density layers are compensated by regions of enhanced stratification.  
These regions of enhanced stratification have a characteristic structure in which the $\overline{N^2}$ maxima occur just outside the extrema of $U$, with weak local minima of $\overline{N^2}$ occurring at the locations of the VSHF peaks. 

The locations of strongest shear and weakest stratification correspond to the local minima of the horizontal mean Richardson number, $\overline{Ri}=\overline{N^2}/(\partial_z U)^2$, as shown in figure \ref{fig:4panel_USNsqRi_meanplot} (d). 
The minimum value of $\overline{Ri}$ is near $\overline{Ri}\approx 0.8>0.25$, indicating that the time mean VSHF structure would be free of modal instabilities in the absence of excitation and dissipation by the Miles-Howard (MH) criterion.  
Although the MH criterion is formally valid only for steady unforced inviscid flows, it remains useful in our stochastically maintained turbulent flow to guide intuition about the maximum stable shear attainable by the VSHF for a given stratification. 
We note that this usage of the MH criterion differs from an alternate usage in which $Ri$ is used to distinguish between regions of a flow that are likely to become laminar and regions that are likely to maintain turbulence. 
This alternative interpretation of the implication of $Ri$ is based on the fact that large perturbation growth is obtained by optimal perturbations in shear flows for which $Ri>1/4$ although modal instability is not permitted \citep{Farrell:1993ue}. 
In accord with this result turbulence is observed to be supported in shear flows with $Ri>1$ \citep{Galperin:2007dk}.

To demonstrate that VSHF formation is robust to changes in the control parameters, we show in figure \ref{fig:newfig1} the time evolution of $U$ in four additional cases. 
Panels (a) and (b) show the response of the system to changes in the dissipation parameters.  
Panel (a) shows the development of $U$ when the Rayleigh drag on the mean fields is increased by a factor of five ($r_m=0.5$). 
The mean fields in this case are damped half as rapidly as the perturbations, rather than ten times less rapidly as in the standard case. 
The excitation strength is $\varepsilon=0.5$ and other parameters are as in the standard case, so that $Re_b = 20.8$ and $Fr = 0.84$. 
The VSHF has $m_U/(2\upi)=7$ and is similar to that seen in the standard case in figure \ref{fig:NL_Phenom_Hovmollers} (a). 
In panel (b) is shown the effect of removing Rayleigh drag entirely ($r=r_m=0)$, so that all dissipation is provided by diffusion. 
In this case, some ambiguity arises regarding how the other parameters should be set, as we nondimensionalize time by the perturbation damping time, $1/r$, in examples other than this figure. 
For simplicity we choose to retain all parameters as they are set in the standard case as if Rayleigh drag were still present with $r=1$, which gives $Re_b = 10.4$ and $Fr = 0.23$, where for this example only we use the definition $Fr=(\varepsilon k_e^2)^{1/3}/N_0$ due to the absence of Rayleigh drag. 
The VSHF in this example initially emerges with $m_U/(2\upi)\approx6$ before transitioning to larger scale (smaller $m_U$) as the integration is continued. 
Transition of the VSHF to smaller values of $m_U$ for weaker damping or stronger excitation is consistent with previous studies of VSHF emergence \citep{Herring:1989fx,Smith:2001uo,Smith:2002wg} and is expected on the basis of analysis of the SSD system in the case of strong excitation, as we show in \textsection \ref{sec:equilibration}. 
%For this reason we attribute the development of larger scale VSHFs in this example to the weakness of the dissipation relative to the standard case, rather than to the difference in the mathematical form of the dissipation term between this example and the standard case. 
%%
Panels \ref{fig:newfig1} (c) and (d) show the response of the system to reductions in stratification. 
In these examples we reduce the excitation strength to $\varepsilon=1.5\times 10^{-2}$ for ease of comparison because VSHFs form more rapidly at these stratification values than they do in the standard case.   
Panel (c) shows the development of $U$ when the stratification is reduced by a factor of ten relative to the standard case ($N_0^2=100$ rather than $N_0^2=1000$, corresponding to $Re_b=6.3$, $Fr=.05$) and panel (d) shows the effect of reducing the stratification by a factor of 25 relative to the standard case ($N_0^2=40$, $Re_b=15.6$, $Fr=.12$). 
%The buoyancy Reynolds numbers in each case are $Re_b\approx 6.3$ (panel (c)) and $Re_b\approx 15.6$ (panel (d)). 
As in the case of modified dissipation, the VSHFs in these examples develop with similar structures as in the standard case shown in figure \ref{fig:NL_Phenom_Hovmollers} (a). % jets form spontaneously with similar behaviour. 
We note (not shown) that VSHF formation ceases for sufficiently weak stratification \citep{Smith:2001uo,Kumar:2017ie}. 
We return to the dependence of the VSHF on stratification in \textsection \ref{sec:scaleselection}. 

\section{Mechanism of Horizontal Mean Structure Formation}\label{sec:testfunction}

In a statistically steady state the VSHF, $U$, and the associated buoyancy structure, $B$, must be supported against dissipation by perturbation fluxes of momentum and buoyancy as expressed in (\ref{eq:NL1})-(\ref{eq:NL2}). In the absence of any horizontal mean structure (\emph{i.e.,} if $U=B=0$), isotropy of the stochastic excitation implies that the statistical mean perturbation momentum flux vanishes ($\langle \overline{u'w'}\rangle=0$, where angle brackets indicate the ensemble average over realizations of the stochastic excitation) and that the statistical mean perturbation buoyancy flux is constant ($-\partial_z \langle \overline{w'b'}\rangle=0$). For the observed horizontal mean structures to emerge and persist, their presence must modify the fluxes so that the fluxes reinforce these structures. In this section we analyze the interaction between the turbulence and the horizontal mean state and demonstrate that the horizontal mean structures do influence the turbulent fluxes in this way. 

We analyze turbulence-mean state interactions by applying two modifications to (\ref{eq:NL1})-(\ref{eq:NL4}). 
The first modification is to hold the mean fields constant as $U=U_{\text{test}}$, $B=B_{\text{test}}$. The second modification is to discard the perturbation-perturbation nonlinear terms $[J(\psi',\Delta \psi')-\overline{J(\psi',\Delta \psi')}]$ and $[J(\psi',b')-\overline{J(\psi',b')}]$ from equations  (\ref{eq:NL3})-(\ref{eq:NL4}). 
The resulting equations are
\begin{eqnarray}
\frac{\partial \Delta \psi' }{\partial t} &=& - U_{\text{test}} \frac{\partial \Delta \psi'}{\partial x} +w'\frac{\partial^2 U_{\text{test}}}{\partial z^2}+ \frac{\partial b'}{\partial x} -\Delta \psi' +\nu \Delta^2 \psi'  +\sqrt{\varepsilon}S, \label{eq:testfn1} \\
\frac{\partial b'}{\partial t} &=& -U_{\text{test}} \frac{\partial b'}{\partial x}-w'\overline{N^2}_{\text{test}}-b'+\nu \Delta b' ,\label{eq:testfn2}
\end{eqnarray}
in which $\overline{N^2}_{\text{test}}=N_0^2  + \partial_z B_{\text{test}}$. Equations (\ref{eq:testfn1})-(\ref{eq:testfn2}) are a system of linear differential equations for the perturbation fields. 
For this system the time mean fluxes are identical to the ensemble mean fluxes averaged over noise realizations and either method of averaging can be used to calculate the average fluxes in the presence of the imposed horizontal mean state ($U=U_{\text{test}}$ and $\overline{N^2}=\overline{N^2}_{\text{test}}$). 
We refer to the calculation of perturbation fluxes from (\ref{eq:testfn1})-(\ref{eq:testfn2}) as test function analysis, as it allows us to probe the turbulent dynamics by imposing chosen test functions for the mean flow and buoyancy, $U_{\text{test}}$ and $\overline{N^2}_{\text{test}}$. 
This approach has been applied to estimate perturbation fluxes in the midlatitude atmosphere \citep{Farrell:1993wf} and in wall-bounded shear flows \citep{Farrell:2012jm,Farrell:2017dx} and we will evaluate its effectiveness in the 2D Boussinesq system in \textsection \ref{sec:modelcomparison}. 

That the modified perturbation equations are capable of producing realistic perturbation fluxes given the observed mean flow is related to the non-normality of the perturbation dynamics in the presence of shear \citep{Farrell:1996jj}. 
The modified equations correctly capture the non-normal dynamics, which produce both the positive and negative energetic perturbation-mean flow interactions.
The non-normal dynamics of perturbations in stratified shear flow have been analyzed in 2D \citep{Farrell:1993ue} and in 3D \citep{Bakas:2001fk,Kaminski:2014us}.

\begin{figure}
               \centerline{\includegraphics{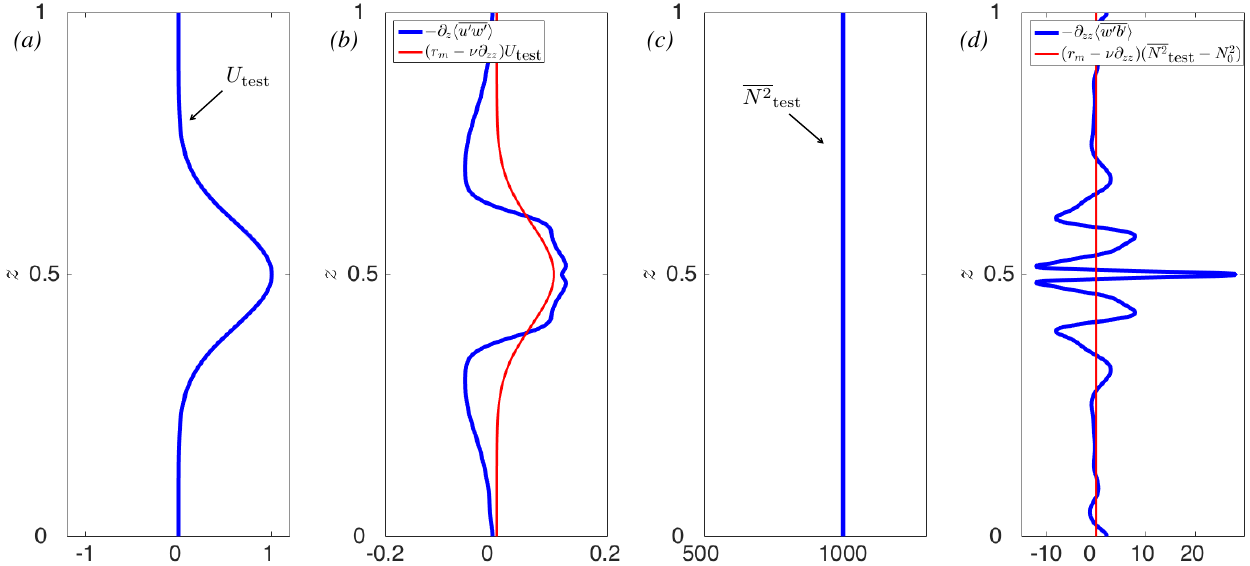}}
 \caption{Test function analysis showing the perturbation flux divergences that develop in response to an imposed horizontal mean state consisting of a Gaussian jet and an unmodified background stratification. (a) Imposed jet, $U_{\text{test}}$. (b) The resulting ensemble mean perturbation momentum flux divergence, $-\partial_z \langle \overline{u'w'} \rangle$, and the negative of the dissipation of the jet, $(r_m- \nu\partial_{zz})U_{\text{test}}$. (c) Imposed stratification, $\overline{N^2}_{\text{test}}$, which is equal to $N_0^2$ in this example. (d) Ensemble mean driving by perturbation fluxes of the stratification anomaly, $-\partial_{zz}\langle \overline{w'b'}\rangle$, and the negative of the dissipation of the stratification anomaly, $(r_m-\nu \partial_{zz})(\overline{N^2}_{\text{test}}-N_0^2)$, which is zero in this example. 
This example shows that a Gaussian jet organizes the turbulence so that the perturbation momentum fluxes generally accelerate the jet. 
The buoyancy fluxes are also organized by the jet in such a way as to drive a stratification anomaly with a complex vertical structure. 
Parameters are as in figure \ref{fig:NL_Phenom_FlowSnapshots}.}
\label{fig:TestFunction_Gaussian}
\end{figure}

As an illustrative example we show in figure \ref{fig:TestFunction_Gaussian} the results of test function analysis in the case of an imposed mean state comprised of a Gaussian jet peaked in the centre of the domain, $U_{\text{test}}=\exp(-50(z-\frac12)^2)$, and an unmodified background stratification, $\overline{N^2}_{\text{test}}=N_0^2=10^3$. 
Panel (a) shows the imposed jet, $U_{\text{test}}$, while panel (b) shows the induced perturbation momentum flux divergence, $-\partial_z \langle \overline{u'w'}\rangle$, alongside the negative of the jet dissipation, $(r_m-\nu \partial_{zz})U_{\text{test}}$. 
The core of the jet is clearly being supported against dissipation by the perturbation momentum fluxes resulting from its modification of the turbulence.
This organization of turbulence producing up-gradient momentum fluxes in the presence of a background shear flow is the essential mechanism of VSHF emergence: an initially perturbative VSHF that arises randomly from turbulent fluctuations modifies the turbulence to produce fluxes reinforcing the initial VSHF. 
This wave-mean flow mechanism is consistent with the results of rapid distortion theory for stratified shear flow \citep{Galmiche:2002cw} and has been identified in simulations of decaying sheared and stratified turbulence \citep{Galmiche:2002gh}. 
Wave-mean flow interaction has also been hypothesized to be the mechanism responsible for the formation and maintenance of the EDJs \citep{Muench:1999dy,Ascani:2015dd}.

Consistent with the results of the NL system shown in \textsection \ref{sec:NLphenom}, the buoyancy fluxes are also modified by imposing a test function horizontal mean state. 
Figure \ref{fig:TestFunction_Gaussian} (c) shows the imposed stratification, $\overline{N^2}_{\text{test}}$, which is equal to $N_0^2$ in this example. 
Figure \ref{fig:TestFunction_Gaussian} (d) shows the driving by perturbation fluxes of the stratification anomaly, $-\partial_{zz} \langle \overline{w'b'}\rangle$, alongside the negative of the dissipation of the stratification anomaly, $(r_m-\nu\partial_{zz})(\overline{N^2}_{\text{test}}-N_0^2)$, which is zero in this Gaussian jet example as $\overline{N^2}=N_0^2$. 
The vertical structure of $-\partial_{zz} \langle\overline{w'b'}\rangle$ is complex. 
For these parameter values the fluxes act to enhance $\overline{N^2}$ most strongly at the jet maximum, which departs from the NL results in which $\overline{N^2}$ has weak local minima at the locations of the VSHF peaks.

This simple example demonstrates the general physical mechanism of horizontal mean structures modifying turbulent fluxes so as to modify the mean state. 
However, the results of this example indicate that a Gaussian jet together with an unmodified background stratification does not constitute a steady state, as neither the jet acceleration nor the driving of the stratification anomaly due to the perturbation fluxes reflect the specific structure of the imposed mean state ($U=U_{\text{test}}$ and $\overline{N^2}=\overline{N^2}_{\text{test}}$). 
Although the perturbation fluxes generally act to strengthen $U$, they also distort its structure by sharpening the jet core and driving retrograde jets on the flanks. Similarly, the $\overline{N^2}=N_0^2$ structure is not in equilibrium with the buoyancy fluxes. 
To maintain a statistically steady mean state as seen in the NL simulations, the turbulence and the mean state must be adjusted by their interaction to produce mean structure which the corresponding fluxes precisely support against dissipation. 

\begin{figure}
                    		\centerline{\includegraphics{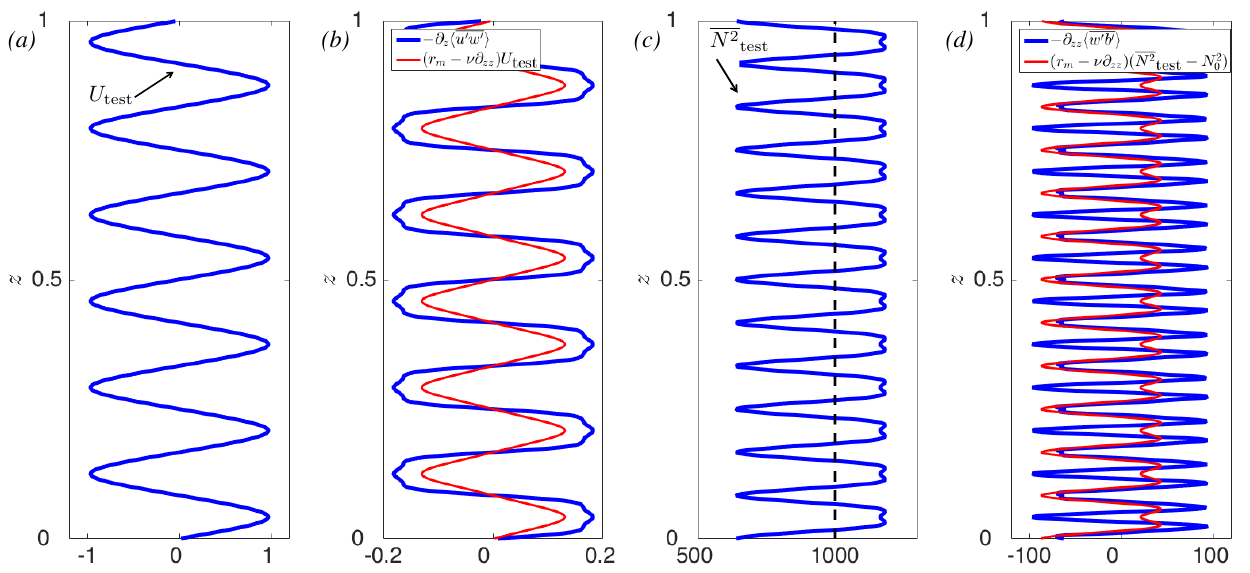}}
  \caption{Test function analysis showing the perturbation flux divergences that develop in response to an imposed horizontal mean state corresponding to that which emerges in the standard case NL simulation shown in \textsection \ref{sec:NLphenom}, with $U_{\text{test}}$ and $\overline{N^2}_{\text{test}}$ smoothed and symmetrized. Panels are as in figure \ref{fig:TestFunction_Gaussian}, with the additional vertical dashed line in panel (c) indicating $N_0^2$. This example shows that the horizontal mean structure that emerges in the NL system, consisting of the VSHF and associated density layers, organizes the turbulent fluxes so that these fluxes support the specific structure of the horizontal mean state against dissipation. Parameters are as in figure \ref{fig:NL_Phenom_FlowSnapshots}.}
\label{fig:TestFunction_Ufinal}
\end{figure}

To demonstrate how such cooperative equilibria are established, we show in figure \ref{fig:TestFunction_Ufinal} the results of test function analysis applied to the case in which $U_{\text{test}}$ (panel (a)) and $\overline{N^2}_{\text{test}}$ (panel (c)) are taken to be the time average profiles from the standard case NL integration discussed in \textsection \ref{sec:NLphenom}, smoothed and symmetrized so that the sixfold symmetry of the VSHF and twelvefold symmetry of $\overline{N^2}$ are made exact. 
As in the Gaussian jet example, the perturbation momentum fluxes support the jet against dissipation (panel (b)). 
However, unlike the results obtained in the case of a Gaussian jet, the approximately harmonic VSHF that emerges in the NL system leads to flux divergences that are precisely in phase with $U$. This provides an explanation for the structure of the emergent VSHF: its approximately harmonic $U$ profile is a structure which the associated statistical equilibrium fluxes precisely support. 
Similarly, the structure of $\overline{N^2}$ is supported against dissipation by the perturbation buoyancy fluxes. 
Some differences between the structure of the perturbation driving and that of the dissipation are seen in panels (b) and (d). In particular, the perturbation driving of the jet is slightly too strong, and the perturbation driving of the stratification anomaly is too strongly negative at the local stratification minima that coincide with the VSHF extrema. These differences arise because the VSHF and horizontal mean stratification anomaly tend to strengthen when perturbation-perturbation nonlinearities are discarded (see \textsection \ref{sec:modelcomparison}) and also because the smoothed and symmetrized stratification anomaly has a somewhat weaker local minimum than that which is found in snapshots of the NL system. 

This analysis demonstrates that the linear dynamics of the stochastically excited Boussinesq equations produces fluxes consistent with the emergent VSHF and density layers seen in the NL system. In this sense, test function analysis provides a `mechanism denial study' that demonstrates that spectrally local perturbation-perturbation interactions associated with a cascade of energy to large scales are not required to produce VSHFs in stratified turbulence. 
That VSHF formation does not occur via such a cascade has previously been noted by \citet[]{Smith:2002wg}. 
The analysis in this section has been conducted using an imposed, constant horizontal mean state. In the next section we extend (\ref{eq:testfn1})-(\ref{eq:testfn2}) by coupling the dynamics of the mean fields to the linearized perturbation equations to formulate the S3T implementation of SSD for this system.
 
\section{Formulating the QL and S3T Equations of Motion}\label{sec:QL_and_SSD_formulation}

The QL system is obtained by combining the perturbation equations (\ref{eq:testfn1})-(\ref{eq:testfn2}) with the NL equations for the horizontal mean state (\ref{eq:NL1})-(\ref{eq:NL2}). The resulting QL equations of motion are 
\begin{eqnarray}
\frac{\partial U}{\partial t} &=& -\frac{\partial}{\partial z}\overline{u'w'}-r_m U + \nu \frac{\partial^2 U}{\partial z^2}, \label{eq:QL1} \\
\frac{\partial B}{\partial t} &=&  -\frac{\partial}{\partial z}\overline{w'b'}-r_m B + \nu \frac{\partial^2 B}{\partial z^2}, \label{eq:QL2}\\
\frac{\partial \Delta \psi '}{\partial t} &=& - U \frac{\partial \Delta \psi'}{\partial x} +w' \frac{\partial^2 U}{\partial z^2} + \frac{\partial b'}{\partial x} -\Delta \psi' +\nu \Delta^2 \psi'  + \sqrt{\varepsilon}S, \label{eq:QL3} \\
\frac{\partial b'}{\partial t} &=& -U \frac{\partial b'}{\partial x}-w'\left(N_0^2+\frac{\partial B}{\partial z}\right)-b'+\nu \Delta b'.\label{eq:QL4}
\end{eqnarray}
This system can also be obtained directly from the NL system (\ref{eq:NL1})-(\ref{eq:NL4}) by discarding the perturbation-perturbation nonlinearities $[J(\psi',\Delta \psi')-\overline{J(\psi',\Delta \psi')}]$ and $[J(\psi',b')-\overline{J(\psi',b')}]$. 
The QL dynamics is a coupled system that while simplified retains the dynamics of the consistent evolution of the horizontal mean state together with the stochastically excited turbulence. 
The 2D Boussinesq equations in the QL approximation have previously been applied to analyze mean flow formation in the case of an unstable background stratification \citep{Fitzgerald:2014wz}.

Because (\ref{eq:QL3})-(\ref{eq:QL4}) are linear in perturbation quantities, the QL system does not retain the transfer by perturbation-perturbation interaction of perturbation energy into horizontal wavenumber components that are not stochastically excited. 
We choose to excite only global horizontal wavenumbers $1-8$. 
The QL system will therefore not exhibit the full range of small scale motions seen in the NL system. However, in \textsection \ref{sec:modelcomparison} we compare the results of QL simulations with those of the NL system, and show that the QL system reproduces the large-scale structure formation observed in the NL system. 
This implies that the small scale structures produced by perturbation-perturbation interaction in the NL system do not strongly influence the horizontal mean state and that a faithful representation of the turbulence at all scales is inessential for understanding the statistical structure of the turbulence to second order.

The energetics of the QL system, with respect to both the mean and perturbation kinetic and potential energies, is identical to that of the NL system, with the exception that the terms originating from perturbation-perturbation interaction, which redistribute energy within the perturbation field but do not change the domain averaged kinetic or potential energies, are not retained in the QL system. 
The QL system thus possesses identical energetics to the NL system in the domain averaged sense. 

Although the QL system constitutes a substantial mathematical and conceptual simplification compared to the NL system, QL dynamics remains stochastic and exhibits significant turbulent fluctuations. These fluctuations obscure the statistical relationships between the horizontal mean structure and the turbulent fluxes discussed in \textsection \ref{sec:testfunction}. 
To understand the mechanism underlying these statistical relationships it is useful to formulate a dynamics directly in terms of statistical quantities, which we refer to as a statistical state dynamics (SSD). 
We now formulate the S3T dynamics, which is the SSD we use to study our system. 
S3T is a closure that retains the interactions between the horizontal mean state and the ensemble mean two-point covariance functions of the perturbation fields which determine the turbulent fluxes. For readers unfamiliar with S3T, Appendix \ref{sec:appendixA} provides a derivation of the S3T equations for a reduced model of stratified turbulence illustrating the conceptual utility of this closure in the context of this reduced model. 

Derivation of the S3T dynamics begins with the QL equations (\ref{eq:QL1})-(\ref{eq:QL4}). We expand the perturbation fields in horizontal Fourier series as
\begin{align}
\psi ' (x,z,t) &= \Real \left[ \sum_{n=1}^{N_k} \tilde{\psi}_n(z,t) e^{\text{i}k_n x} \right], \\
b ' (x,z,t) &= \Real \left[ \sum_{n=1}^{N_k} \tilde{b}_n(z,t) e^{\text{i}k_n x} \right].
\end{align}
Here $N_k$ is the number of retained Fourier modes ($N_k=8$ for our choice of stochastic excitation) and $k_n=2\upi n$. Considering the Fourier coefficients as vectors in the discretized numerical system (\emph{e.g.}, $\tilde{\psi}_n(z,t)\to \boldsymbol{\psi}_n(t)$), the QL equations (\ref{eq:QL3})-(\ref{eq:QL4}) can be combined into the vector equation 
\begin{eqnarray}
\frac{d}{d t}\left( \begin{array}{c} \boldsymbol{\psi}_n \\ \boldsymbol{b}_n \end{array} \right) = \mathsfbi{A}_n (\boldsymbol{U},\boldsymbol{B}) \left( \begin{array}{c} \boldsymbol{\psi}_n \\ \boldsymbol{b}_n \end{array} \right) + \left( \begin{array}{c} \sqrt{\varepsilon} \boldsymbol{\xi}_n \\ 0 \end{array} \right), \label{eq:QLmatrixform}
\end{eqnarray}
where $\boldsymbol{\xi}_n=\mathsfbi{\Delta}_n^{-1}\boldsymbol{S}_n$ is the $n$th horizontal Fourier component of the stochastic excitation of the streamfunction. Here $\mathsfbi{\Delta}_n=-k_n^2\mathsfbi{I}+\mathsfbi{D}^2$ in which $\mathsfbi{I}$ is the identity matrix and $\mathsfbi{D}$ is the discretized vertical derivative operator. 
The linear dynamical operator $\mathsfbi{A}_n$ is given by the expression
\begin{multline}
\mathsfbi{A}_n (\boldsymbol{U},\boldsymbol{B}) = \\
\left( \begin{array}{cc} 
-\text{i}k_n \mathsfbi{\Delta}_n^{-1} \text{diag}(\boldsymbol{U}) \mathsfbi{\Delta}_n + \text{i}k_n\mathsfbi{\Delta}_n^{-1}\text{diag}(\mathsfbi{D}^2\boldsymbol{U})-\mathsfbi{I}+\nu \mathsfbi{\Delta}_n& \text{i}k_n\mathsfbi{\Delta}_n^{-1} \\ 
-\text{i}k_nN_0^2\mathsfbi{I}-\text{i}k_n\text{diag}(\mathsfbi{D}\boldsymbol{B}) & -\text{i}k_n\text{diag}(\boldsymbol{U})-\mathsfbi{I}+\nu \mathsfbi{\Delta}_n \end{array} \right), \label{eq:Adefn}
\end{multline}
in which diag$(\boldsymbol{v})$ denotes the diagonal matrix for which the nonzero elements are given by the entries of the column vector $\boldsymbol{v}$.

We now make use of the ergodic assumption that horizontal averages and ensemble averages are equal, so that, for example, $U=\overline{u}=\langle u \rangle$ and $\overline{u'w'}=\langle u'w' \rangle$. 
For our system, which is statistically horizontally uniform this assumption is justified in a domain large enough so that several approximately independent perturbation structures are found at each height as seen, \emph{e.g.}, in figure \ref{fig:NL_Phenom_FlowSnapshots} (b). 
It can then be shown (using the fact that $\sqrt{\varepsilon}S$ is delta-correlated in time) that the ensemble mean covariance matrix, defined as
\begin{equation}
\mathsfbi{C}_n = 
\left\langle \left( \begin{array}{c} \boldsymbol{\psi}_n \\ \boldsymbol{b}_n \end{array} \right)  \left( \begin{array}{cc} \boldsymbol{\psi}_n^{\dagger} & \boldsymbol{b}_n^{\dagger} \end{array} \right) \right \rangle
= \left( \begin{array}{cc} \langle \boldsymbol{\psi}_n \boldsymbol{\psi}_n^{\dagger} \rangle & \langle \boldsymbol{\psi}_n \boldsymbol{b}_n^{\dagger} \rangle \\ \langle \boldsymbol{b}_n \boldsymbol{\psi}_n^{\dagger} \rangle & \langle \boldsymbol{b}_n \boldsymbol{b}_n^{\dagger} \rangle \end{array} \right) = \left( \begin{array}{cc} \mathsfbi{C}_{\psi \psi , n} & \mathsfbi{C}_{\psi b,n} \\ \mathsfbi{C}_{\psi b,n}^{\dagger} & \mathsfbi{C}_{b b,n} \end{array} \right),
\end{equation}
in which daggers indicate Hermitian conjugation, evolves according to the time-dependent Lyapunov equation
\begin{eqnarray}
\frac{d}{d t}\mathsfbi{C}_n &=& \mathsfbi{A}_n (\boldsymbol{U},\boldsymbol{B}) \mathsfbi{C}_n + \mathsfbi{C}_n \mathsfbi{A}_n (\boldsymbol{U},\boldsymbol{B})^{\dagger} + \varepsilon \mathsfbi{Q}_n, \label{eq:S3T1} \\
\mathsfbi{Q}_n &=& \left[ \begin{array}{cc} \langle \boldsymbol{\xi}_n \boldsymbol{\xi}_n^{\dagger} \rangle & 0 \\ 0  &0 \end{array}\right],
\end{eqnarray}
where $\mathsfbi{Q}_n$ is the ensemble mean covariance matrix of the stochastic excitation and has nonzero entries only in the upper left block matrix because we apply excitation only to the vorticity field. Equation (\ref{eq:S3T1}) constitutes the perturbation dynamics of the S3T system and is the S3T analog of the QL equations (\ref{eq:QL3})-(\ref{eq:QL4}).

To complete the derivation of the S3T system it remains to write the mean equations (\ref{eq:QL1})-(\ref{eq:QL2}) in terms of the covariance matrix. The ensemble mean perturbation flux divergences can be written as functions of the covariance matrix as
\begin{eqnarray}
-\frac{\partial}{\partial z}\langle \overline{u'w'}\rangle &=& \sum_{n=1}^{N_k} \frac{k_n}{2} \mbox{Im} \left[ \text{vecd} \left( \mathsfbi{\Delta}_n \mathsfbi{C}_{\psi \psi,n} \right) \right] , \\
-\frac{\partial}{\partial z}\langle \overline{w'b'}\rangle &=& \sum_{n=1}^{N_k} \frac{k_n}{2} \mbox{Im} \left[ \text{vecd} \left( \mathsfbi{D} \mathsfbi{C}_{\psi b,n} \right) \right] ,
\end{eqnarray}
in which vecd$(\mathsfbi{M})$ denotes the vector comprised of the diagonal elements of the matrix $\mathsfbi{M}$. The mean state dynamics then become
\begin{eqnarray}
\frac{d}{d t}\boldsymbol{U} &=& \sum_{n=1}^{N_k} \frac{k_n}{2} \mbox{Im} \left[ \text{vecd} \left( \mathsfbi{\Delta}_n \mathsfbi{C}_{\psi \psi,n} \right) \right] -r_m \boldsymbol{U} + \nu \mathsfbi{D}^2 \boldsymbol{U} , \label{eq:S3T2} \\
\frac{d}{d t}\boldsymbol{B} &=& \sum_{n=1}^{N_k} \frac{k_n}{2} \mbox{Im} \left[ \text{vecd} \left( \mathsfbi{D} \mathsfbi{C}_{\psi b,n} \right) \right]  -r_m \boldsymbol{B} + \nu \mathsfbi{D}^2 \boldsymbol{B} . \label{eq:S3T3}
\end{eqnarray}
Equations (\ref{eq:S3T1}), (\ref{eq:S3T2}), and (\ref{eq:S3T3}) together constitute the S3T SSD closed at second order. 

The S3T system is deterministic and autonomous and so provides an analytic description of the evolving relationships between the statistical quantities of the turbulence up to second order, including fluxes and horizontal mean structures, without the turbulent fluctuations inherent in the dynamics of particular realizations of turbulence, such as those present in the NL and QL systems. 
Although some previous attempts to formulate turbulence closures have been found to have inconsistent energetics \citep{KRAICHNAN:1957ty,OGURA:1963wp}, the dynamics of the S3T system are QL and so S3T inherits the consistent energetics of the QL system. % closure attempts 
That the second-order S3T closure is capable of capturing the dynamics of VSHF formation, as we demonstrate in \textsection \ref{sec:modelcomparison}, is due to the appropriate choice of mean. 
In the present formulation of S3T, we have chosen the mean to be the horizontal average structure so that the second-order closure retains the QL mean-perturbation interactions that account for the mechanism of VSHF formation in turbulence.

Before proceeding to analysis of the QL and S3T systems, we wish to make two remarks regarding the mathematical structure and physical basis of S3T.  
First, we note that the ergodic assumption used in deriving S3T is formally valid in the limit that the horizontal extent of the domain tends to infinity and the number of independent perturbation structures at each height correspondingly tends to infinity. 
In this ideal limit described by S3T, the statistical homogeneity of the turbulence is only broken by the initial state of the horizontal mean structure (which in the examples is perturbatively small), which then determines the phase of the emergent VSHF in the vertical direction. 
In simulations of the QL and NL systems in a finite domain, the initial mean structure instead results from random Reynolds stresses arising from fluctuations in the perturbation fields. 
Second, we note that S3T is a canonical closure of the turbulence problem at second order, in that it is a truncation of the cumulant expansion at second order achieved by setting the third cumulant to zero. 
The mathematical structure of the cumulant expansion determines which nonlinearities are retained and discarded in the QL system, which is a stochastic approximation to the ideal S3T closure. 
Wave-mean flow coupling enters the equations through second order cumulants and so is retained, while perturbation-perturbation nonlinearities enter as third-order cumulants and so are not retained. 

In the next section we demonstrate that the QL and S3T systems reproduce the major statistical phenomena observed in the NL system. 

 \section{Comparison of the NL, QL, and S3T Systems}\label{sec:modelcomparison}
 
The most striking feature of the standard case NL simulation discussed in \textsection \ref{sec:NLphenom} is the spontaneous development of a VSHF, $U$, with $\text{max}(U)\approx1$ and vertical wavenumber $m_U/2\upi=6$. The horizontal mean stratification, $\overline{N^2}$, is also modified by the turbulence and develops a structure with vertical wavenumber $m_B/2\upi=12$ in phase with $U$ such that weakly stratified density layers develop in the regions of strongest mean shear. The VSHF in the NL system is approximately steady in time, while the horizontal mean stratification is more variable. In this section we compare these NL results to the behaviour of the QL and S3T systems for the same parameter values. We initialize the QL system from rest, matching the procedure used for the NL system. 
%We initialize the S3T system with $\mathsfbi{C}_n$ corresponding to homogeneous turbulence together with a weak, slightly perturbed VSHF with $m_U/2\upi=6$.  
We initialize the S3T system with $\mathsfbi{C}_n$ corresponding to homogeneous turbulence together with a small VSHF perturbation (amplitude $0.1$) with $m_U/2\upi=6$ that is slightly modified by additional small perturbations (amplitude $.005$). 
We note that the details of the S3T initialization are unimportant in this example because, as we will show in \textsection \ref{sec:scaleselection}, the VSHF emerges via a linear instability of the homogeneous turbulence and so any sufficiently small initial perturbation to the S3T system will evolve into a $m_U/2\upi=6$ VSHF for these parameter values. %s an exponentially growing mode associated with a linear instability, and as such the VSHF emerges t
	
\begin{figure}
       \centerline{\includegraphics{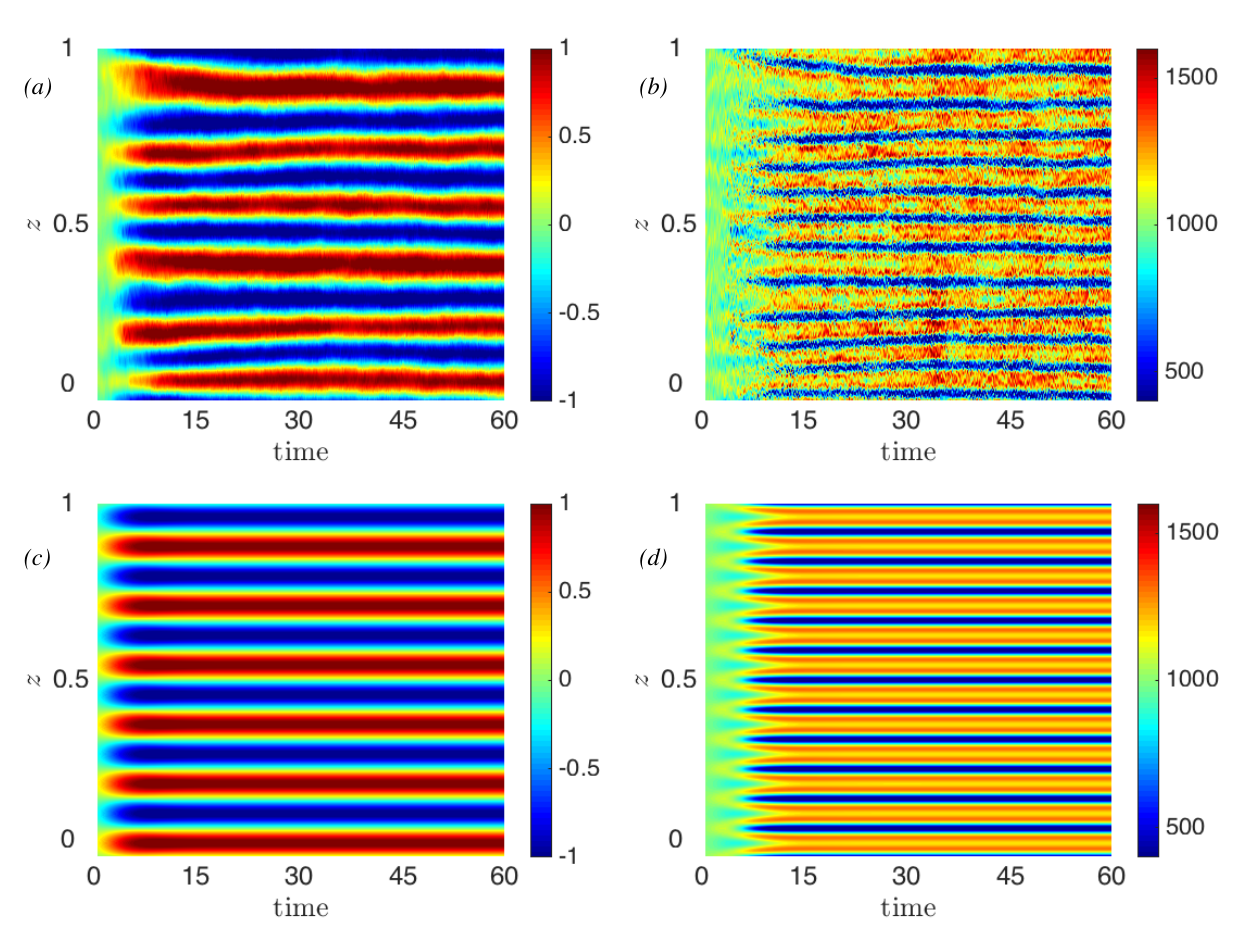}} 
  \caption{Development of the VSHF and associated density layers in the QL and S3T systems. Panels show the time evolution of (a) $U$ in the QL system, (b) $\overline{N^2}$ in the QL system, (c) $U$ in the S3T system, and (d) $\overline{N^2}$ in the S3T system. This figure demonstrates that the QL and S3T systems reproduce the phenomenon of spontaneous VSHF and density layer formation shown in figure \ref{fig:NL_Phenom_Hovmollers} for the NL system. Parameters are as in figure \ref{fig:NL_Phenom_FlowSnapshots}.}
\label{fig:ModelComparison_Hovmollers}
\end{figure}

\begin{figure}
            \centerline{\includegraphics{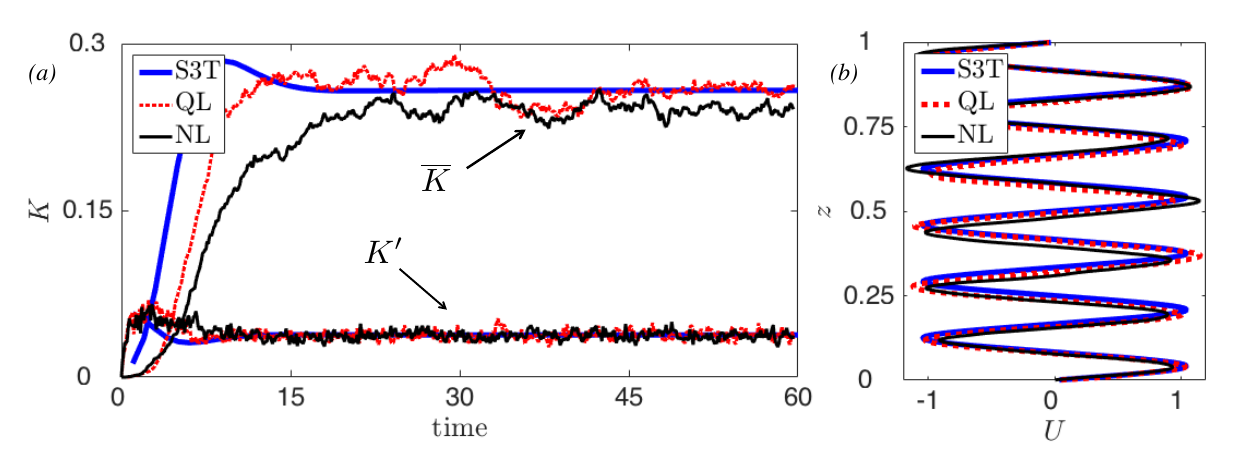}} 
  \caption{Comparison of the kinetic energy evolution and equilibrium VSHF profiles in the NL, QL, and S3T systems. (a) Mean and perturbation kinetic energy evolution. (b) Aligned VSHF profiles. The NL and QL profiles are averaged over $t\in[30,60]$ and the S3T profile is taken to be the state after the S3T system reaches a fixed point. This figure demonstrates that VSHF emergence in the S3T and QL systems occurs with similar structure and energy evolution to that which occurs in the NL system. Parameters are as in figure \ref{fig:NL_Phenom_FlowSnapshots}.}
\label{fig:ModelComparison_Energy_and_Uprofiles}
\end{figure}

Figures \ref{fig:ModelComparison_Hovmollers} (a) and (c) show the time evolution of the VSHF in the QL and S3T systems (see figure \ref{fig:NL_Phenom_Hovmollers} for the corresponding evolution in the NL system). 
The QL and S3T systems develop VSHF structures with $m_U/2\upi=6$ and the $U$ profiles in the NL, QL, and S3T systems are compared in figure \ref{fig:ModelComparison_Energy_and_Uprofiles} (b). 
For the NL and QL systems the profiles are time averaged over $t\in[30,60]$, while for the S3T system we show the $U$ state after the S3T system has reached a fixed point. 
The aligned VSHF structures agree well across the three systems. 
The time evolution of the horizontal mean stratification, $\overline{N^2}$, in the QL and S3T systems is shown in figures \ref{fig:ModelComparison_Hovmollers} (b) and (d). Like the NL stratification, the QL profile of $\overline{N^2}$ develops a $m_B/2\upi=12$ structure that is more variable in time than $U$ and is phase-aligned with $U$ so that $\overline{N^2}$ is weakest in the regions of strongest shear. The S3T system behaves similarly but is free of fluctuations. The evolution of $\overline{N^2}$ in the S3T system also reveals that the vertical structure of $\overline{N^2}$ changes over time. During the development of the VSHF ($t\lesssim8$), the stratification is enhanced in the regions of strongest shear. As the VSHF begins to equilibrate at finite amplitude, the $\overline{N^2}$ profile reorganizes such that the shear regions are the most weakly stratified. Such reorganization may also occur in the NL and QL systems but is difficult to identify due to the fluctuations present in these systems.
		
Figure \ref{fig:ModelComparison_Energy_and_Uprofiles} (a) shows the evolution of the mean and perturbation kinetic energies of the NL, QL and S3T systems. The growth rate of mean kinetic energy is similar in these three systems. 
The equilibrium mean energies differ somewhat among the systems, with the VSHFs in the S3T and QL systems having more energy than the NL VSHF. The relative weakness of the NL VSHF is consistent with the scattering of perturbation energy to small scales by the perturbation-perturbation advection terms that are included in NL but not in QL or S3T. 
The temporal variability of the NL and QL VSHFs, as indicated by the fluctuations in $\overline{K}$, is similar in the stochastic NL and QL systems. 
The VSHF in S3T is time-independent once equilibrium has been reached as the S3T VSHF corresponds to a fixed point of the S3T dynamics.	

\begin{figure}
\centerline{\includegraphics{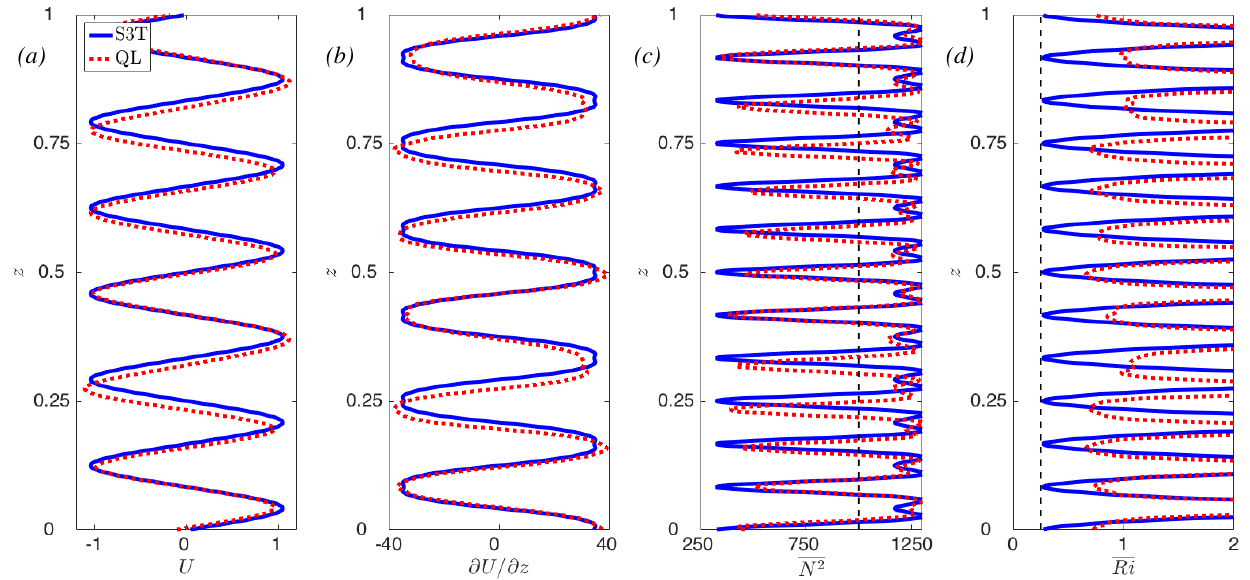}}
  \caption{Vertical structure of the horizontal mean states of the QL and S3T systems. Panels are as in figure \ref{fig:4panel_USNsqRi_meanplot} with solid lines showing the S3T state and dotted lines showing the QL state. This figure demonstrates that the QL and S3T systems capture the structure of the horizontal mean state in the NL system, including the phase relationship between $U$ and $\overline{N^2}$. Parameters are as in figure \ref{fig:NL_Phenom_FlowSnapshots}.}
\label{fig:New_ModelComparison_QLS3T_4panelplots}
\end{figure}

The relationship between the $U$ and $\overline{N^2}$ structures is shown in figure \ref{fig:New_ModelComparison_QLS3T_4panelplots} for the QL (dotted curves) and S3T (solid curves) systems (see figure \ref{fig:4panel_USNsqRi_meanplot} for the corresponding structures in the NL system). The equilibrium horizontal mean structures in the QL and S3T systems agree well with those of the NL system. The $U$ profiles (panel (a)) are approximately harmonic with somewhat flattened shear regions and, remarkably, the detailed structure of $\overline{N^2}$ seen in the NL integration is reproduced by the QL and S3T systems (panel (c)), which discard perturbation-perturbation nonlinear interactions. In particular, the presence of weak local stratification minima at the locations of the VSHF peaks is captured by the QL and S3T systems. 
	
\begin{figure}
              \centerline{\includegraphics{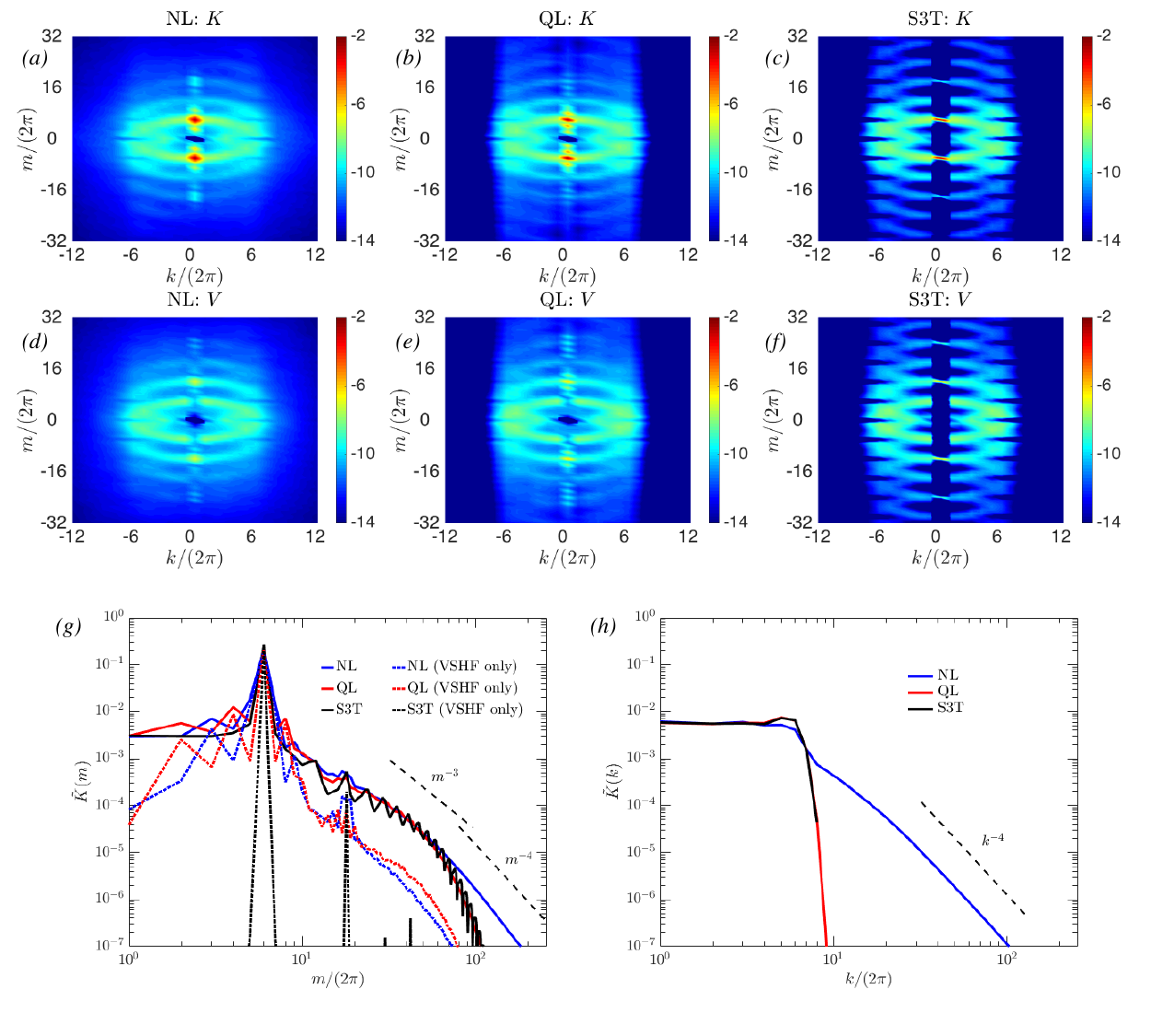}} 
  \caption{Comparison of the wavenumber power spectra of kinetic ($K$) and potential ($V$) energy in the NL, QL, and S3T systems. The top row shows the 2D $K$ spectra of the (a) NL, (b) QL, and (c) S3T systems as functions of $(k,m)$, while the middle row (d)-(f) shows the 2D $V$ spectra of those systems.  2D spectra are shown in terms of their natural logarithms and no normalization is performed. The bottom row shows the kinetic energy spectra in the conventional 1D form as functions of (g) vertical wavenumber, $m$, and (h) horizontal wavenumber, $k$. In panel (g), the contributions to the spectra from the VSHFs in each system are also shown. This figure demonstrates that the QL and S3T systems reproduce structural details of the turbulence beyond the horizontal mean state, including the wavenumber distribution of perturbation energy at large scales. Parameters are as in figure \ref{fig:NL_Phenom_FlowSnapshots}.}
\label{fig:New_ModelComparison_Spectra}
\end{figure}

The above comparisons demonstrate that the horizontal mean structures and domain mean kinetic energies of the QL and S3T systems show good agreement with those of the NL system. 
Figure \ref{fig:New_ModelComparison_Spectra} compares the energy spectra in the three systems. 
In panels (a)-(f), the 2D spectra of kinetic and potential energy are compared as functions of $(k,m)$, while in panels (g)-(h) the kinetic energy spectra are compared in the more traditional 1D integrated forms as functions of $k$ and $m$ separately. 
Panel (a) shows the kinetic energy spectrum of the NL system. 
The dominant and most important feature of the $K$ spectrum is the concentration of energy at $(k,m)=2\upi(0,6)$ which corresponds to the $m_U/2\upi=6$ VSHF structure. 
This feature is also evident in panel (g), which shows that the peak in the vertical wavenumber spectrum of NL kinetic energy is dominated by the VSHF component of the flow. %kinetic energy spectrum . 
The energy of the VSHF is also spread across the neighbouring vertical wavenumbers, reflecting both the deviation of the structure of the VSHF from a pure harmonic and also that fluctuations in the VSHF structure project onto nearby vertical wavenumbers. 
Away from the $k=0$ axis, the $K$ spectrum reveals the expected concentration of energy on the ring of excited wavenumbers $k^2+m^2=k_e^2$, and the spread of this ring to higher $m$ values. 
This spread is due to the shearing of the ring by the $m_U/2\upi=6$ VSHF, which produces the sum and difference wavenumber components.
The quantitative structure of the spectrum associated with this spreading can be seen more clearly in panel (g), in which selected power law slopes are provided for reference. 

The most important features of the 2D $K$ spectrum of the NL system are captured by the QL and S3T systems. The QL $K$ spectrum (figure \ref{fig:New_ModelComparison_Spectra} (b)) reproduces the primary feature of the energetic dominance of the VSHF over the perturbation field, as well as the minor features of concentration of energy at $k_e$ and the spread of the excited ring structure to higher vertical wavenumbers. 
The 1D spectrum in panel (g) shows that the QL system quantitatively captures the spectrum of perturbation kinetic energy in the wavenumber range $m/(2\upi)\lesssim 80$. 
We note that the stochastic excitation directly influences the energy spectrum only near $m/(2\upi)\approx 6$, and so the agreement seen in panel (g) is not a direct result of the structure of the excitation. 
The primary difference between the $K$ spectra of the NL and QL systems is that the NL system scatters some kinetic energy into the unexcited part of the horizontal wavenumber spectrum ($|k|/2\upi>8$), whereas these unexcited wavenumber components have no energy in the QL system, as can also be seen in panel (h). 
The vertical wavenumber spectrum of $K$ in the NL system, shown in panel (g), also contains small scale structure for $m\gtrsim 80$ that is not present in the QL system and so can be attributed to perturbation-perturbation nonlinearity. %  , and that the vertical  
%This difference can be seen especially clearly in panel (h). %, which shows that the small scales in the NL system beyond $k/(\2pi)=8$ do not exist in the QL system. 
The S3T $K$ spectrum (figure \ref{fig:New_ModelComparison_Spectra} (c)) also captures the most important features of the NL spectrum, but some differences between the S3T spectrum and those of the NL and QL systems are also visible. In the S3T system the VSHF energy is more strongly concentrated in the $m_U/2\upi=6$ harmonic than it is in the NL and QL systems. 
Additionally, the concentration of energy at the excited ring and the spread of energy to higher $m$ are more distinct in the S3T system than in the NL and QL systems, in which the gaps are filled in by a broad background spectrum. 
These features are also visible in panel (g). 
These minor differences between the spectrum of S3T and those of the NL and QL systems are due in part to the absence of fluctuations in the S3T system that are present in the stochastic NL and QL systems. Noise in the stochastic systems produces VSHF fluctuations that spread mean flow energy into $k=0$ modes neighbouring the $m_U/2\upi=6$ harmonic. These transient VSHF fluctuations also contribute to producing the broad background spectrum seen in the NL and QL systems by shearing the ring of excited wavenumbers. 
	
The spectrum of potential energy in the NL system is shown in figure \ref{fig:New_ModelComparison_Spectra} (d). Unlike the $K$ spectrum, which is dominated by the horizontal mean flow, $U$, the $V$ spectrum is not dominated by the horizontal mean buoyancy, $B$, although a peak is evident at the $m_B/2\upi=12$ component. 
In this sense, the VSHF is a `manifest' structure, whereas the horizontal mean density layers are `latent' structures \citep{Berloff:2009ct}. 
Other features of the spectrum are the expected concentration of potential energy at the ring wavenumber and the spread of the ring to higher vertical wavenumbers as was found for the $K$ spectrum. The $V$ spectra for the QL and S3T systems are shown in figures \ref{fig:New_ModelComparison_Spectra} (e) and (f). The QL and S3T spectra capture the peak associated with the horizontal mean buoyancy layers, the concentration of potential energy at the excitation scale, and the spread of the ring to higher $m$. The differences between the three $V$ spectra are similar to those identified when comparing the three $K$ spectra.

Agreement between the NL, QL, and S3T systems indicates that the QL dynamics of the horizontal mean state interacting with the perturbation field accounts for the physical mechanisms responsible for determining the most important aspects of the energy spectra. 	
%The comparisons presented in this section indicate that the approximations made in developing the QL and S3T systems do not strongly modify the essential statistical properties of the turbulence up to second order.  
We emphasize that the QL and S3T systems involve no free parameters, and that the demonstrated agreement between the three systems is not the result of parameter tuning. 
%, including the dominance of the VSHF and the distribution of spectral distribution of perturbation energy at large scales. 
%%
% First try at response to one of Galperin's points:
Because the QL and S3T systems do not contain the perturbation-perturbation interactions required to produce a turbulent cascade, agreement between these systems and the NL system indicates that, in the present model configuration, such a cascade is not essentially involved in determining the equilibrium turbulent state, including the large-scale spectrum.
Nonlinear cascades in the NL system only weakly influence the large-scale dynamics and energetics, and we note that for this reason the parameter $\varepsilon$, which is the energy injection rate of the stochastic excitation, should not be interpreted as a rate of turbulent dissipation in the sense of a classical cascade.   
In the NL system we employ to model VSHF formation the turbulent dissipation rate in the classical cascade sense is close to zero and it is exactly zero in the QL and S3T systems. %, which do not permit cascades. 
%Instead, the statistical equilibrium state of the turbulence, including the structure of the spectrum, is primarily determined by spectrally nonlocal interaction between the perturbations and the mean state. % interactions 

Similar results have been obtained for barotropic turbulence characterized by strong zonal jets. 
In the presence of such jets the meridional wavenumber ($\ell$) spectrum of the zonal flow obtains a well-known $\ell^{-5}$ structure at large scales \citep{Huang:2001im,Galperin:2010ix}. 
\citet{Constantinou:2015ti} showed that the barotropic S3T system, with underlying QL dynamics, captures this $\ell^{-5}$ spectrum, indicating that the large-scale structure of the spectrum in barotropic turbulence with strong zonal jets is primarily determined by perturbation-mean interaction rather than by a turbulent cascade. %, indicating that the turbulent cascade is inessential to the large scale spectrum. %structure in this system as well. 

QL dynamics does not account for the spectrum at very small scales, which is produced by perturbation-perturbation nonlinearity and is inessential to the dynamics of VSHFs. 
We note that these small scale features may be important to the stirring of passive tracers at small scales \citep{Sukoriansky:2009dz}, which the QL and S3T systems would not be expected to accurately capture in the present model turbulence.  
%As a result, the QL and S3T systems would not be expected to accurately capture such stirring in the present model turbulence. %and so the QL and S3T systems would not expected to accuratley capture such stirring 

Motivated by these results we proceed in the rest of this paper to exploit the S3T system to analyze the mechanisms underlying the organization of structure in stratified turbulence. 
	
\section{Linear Stability Analysis of the S3T System}\label{sec:scaleselection}

In the previous section we showed that the S3T system reproduces the essential statistical features, up to second order, of the NL system, including both the structure of the horizontal mean state as well as the spectral characteristics of the perturbation field. The S3T system can be understood and analyzed with much greater clarity than the NL system because the S3T system is a deterministic and autonomous dynamical system and is amenable to the usual techniques of dynamical systems analysis. 
In this section we show that the emergence of VSHFs in 2D stratified turbulence can be traced to a linear instability in the SSD of the stationary state of homogeneous turbulence that has analytic expression in the S3T SSD while lacking analytic expression in the dynamics of single realizations. 
To determine the properties of this instability, and in particular to understand how the vertical scale of the initially emergent VSHF is selected, we now perform a linear stability analysis of the S3T system. 

Before linearizing the S3T system it is first necessary to obtain the fixed point statistical state that is unstable to VSHF formation. As shown in \textsection \ref{sec:modelcomparison}, the equilibrium state with a finite-amplitude VSHF and modified horizontal mean stratification is a fixed point of the S3T system. However, the fixed point of the S3T system whose stability we wish to analyze is the state of homogeneous turbulence that is excited by the stochastic excitation and equilibrated by dissipation. 
This homogeneous state is obscured in the NL and QL systems, both by noise fluctuations and (in examples for which it is SSD unstable) by the development of a VSHF, but roughly corresponds to the interval of nearly constant perturbation kinetic energy at early times ($t\lesssim 5$) in figure \ref{fig:ModelComparison_Energy_and_Uprofiles} (a). If homogeneous turbulence is unstable we obtain an explanation for the observed VSHF formation, since the alternative possibility of sustained homogeneous turbulence is not possible in the presence of small perturbations.

For homogeneous turbulence $\boldsymbol{U}=\boldsymbol{B}=0$ and from (\ref{eq:S3T1}) the steady-state perturbation covariance matrix at wavenumber $k_n$ obeys
\begin{equation}
\mathsfbi{A}_n^{\star}\mathsfbi{C}_n^{\star}+\mathsfbi{C}_n^{\star}\mathsfbi{A}_n^{\star,\dagger}+\varepsilon \mathsfbi{Q}_n=0, \label{eq:S3TFP1}
\end{equation}
where the $\mathsfbi{A}_n^{\star}$ operator is given by
\begin{equation}
\mathsfbi{A}_n^{\star} = \left( \begin{array}{cc} -\mathsfbi{I}+\nu\mathsfbi{\Delta}_n & \text{i}k_n \mathsfbi{\Delta}_n^{-1} \\ -\text{i}k_n N_0^2 \mathsfbi{I} & -\mathsfbi{I}+\nu \mathsfbi{\Delta}_n \end{array} \right). \label{eq:S3TAstar}
\end{equation}
Equation (\ref{eq:S3TFP1}) can be solved analytically for $\mathsfbi{C}_n^{\star}$, and we show details of the solution in Appendix \ref{sec:appendixB}. Figure \ref{fig:Homogeneous_Spectrum_Plot} shows the kinetic and potential energy spectra for this fixed point homogeneous turbulent state.

\begin{figure}
        \centerline{\includegraphics{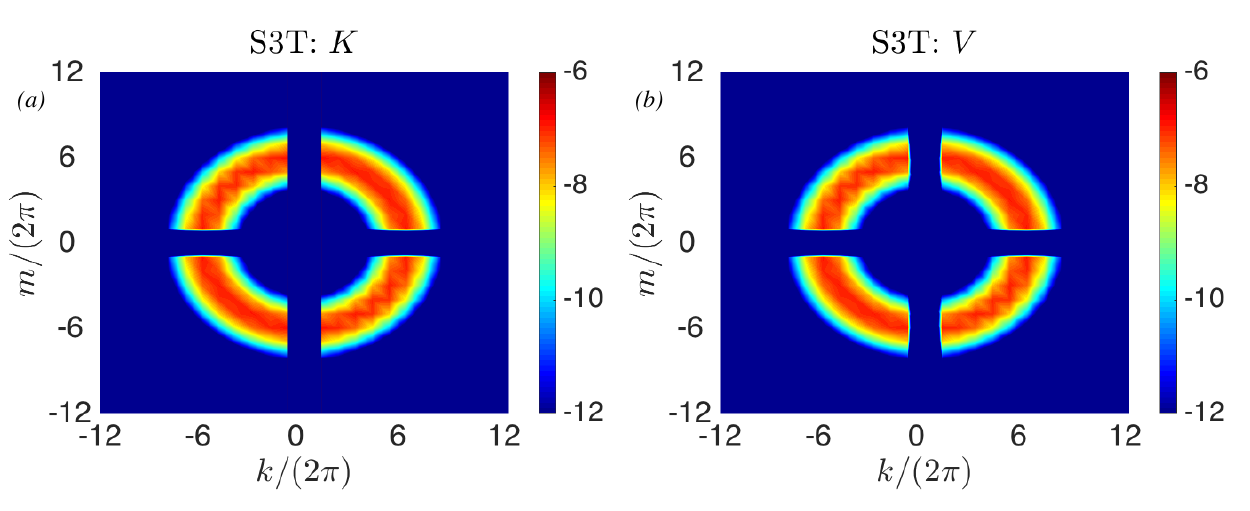}} 
        \caption{Spectral structure of the homogeneous S3T fixed point. (a) Kinetic energy ($K$) spectrum. (b) Potential energy ($V$) spectrum. The spectra are shown in terms of their natural logarithms and no normalization is performed. The $K$ and $V$ spectra are nearly identical to one another, even though only the vorticity field is stochastically excited, due to the strong stratification. This figure shows that the homogeneous turbulence from which the VSHF emerges inherits its structure directly from the stochastic excitation whose structure is shown in figure \ref{fig:Z_Forcing_Plot}. Parameters are as in figure \ref{fig:NL_Phenom_FlowSnapshots}.}
\label{fig:Homogeneous_Spectrum_Plot}
\end{figure}

To analyze the linear stability of this homogeneous turbulent state we perturb the S3T state, $(\mathsfbi{C}_n,\boldsymbol{U},\boldsymbol{B})$, about the fixed point, $(\mathsfbi{C}_n^{\star},0,0)$, as
\begin{align}
\mathsfbi{C}_n = \mathsfbi{C}_n^{\star} + \delta \mathsfbi{C}_n, &&
\boldsymbol{U} = \delta \boldsymbol{U}, &&
\boldsymbol{B} = \delta \boldsymbol{B},
\end{align}
where the $\delta$ notation indicates that the first order terms are treated as infinitesimal perturbations. The operator $\mathsfbi{A}_n$ in (\ref{eq:Adefn}) may then be written as 
\begin{equation}
\mathsfbi{A}_n = \mathsfbi{A}_n^{\star}+\delta \mathsfbi{A}_n,
\end{equation}
where $\mathsfbi{A}_n^{\star}$ given by (\ref{eq:S3TAstar}) and 
\begin{equation}
\mathsfbi{\delta A}_ n = \left( \begin{array}{cc} -\text{i}k_n\mathsfbi{\Delta}_n^{-1}\text{diag}(\delta\boldsymbol{U})\mathsfbi{\Delta}_n+\text{i}k_n\mathsfbi{\Delta}^{-1}_n\text{diag}(\mathsfbi{D}^2\delta\boldsymbol{U}) & 0 \\ -\text{i}k_n\text{diag}(\mathsfbi{D}\delta \boldsymbol{B}) & -\text{i}k_n\text{diag}(\delta \boldsymbol{U}) \end{array}\right).
\end{equation}
The linearized equations of motion are
\begin{eqnarray}
\frac{d}{dt} \delta \boldsymbol{U} &=& \sum_{n=1}^{N_k} \frac{k_n}{2} \mbox{Im} \left[ \text{vecd}\left(\mathsfbi{\Delta}_n \delta\mathsfbi{C}_{\psi \psi,n} \right) \right] -r_m \delta \boldsymbol{U} + \nu \mathsfbi{D}^2 \delta  \boldsymbol{U} , \label{eq:S3TLIN1} \\
\frac{d}{dt} \delta \boldsymbol{B} &=& \sum_{n=1}^{N_k} \frac{k_n}{2} \mbox{Im} \left[ \text{vecd}\left(\mathsfbi{D}\delta \mathsfbi{C}_{\psi b,n} \right) \right] -r_m \delta \boldsymbol{B} + \nu \mathsfbi{D}^2 \delta \boldsymbol{B} , \label{eq:S3TLIN2} \\
\frac{d}{dt} \delta \mathsfbi{C}_n &=& \mathsfbi{A}_n^{\star} \delta \mathsfbi{C}_n + \delta \mathsfbi{C}_n \mathsfbi{A}_n^{\star,\dagger} + \delta \mathsfbi{A}_n \mathsfbi{C}_n^{\star} + \mathsfbi{C}_n^{\star} \delta \mathsfbi{A}_n^{\dagger} .\label{eq:S3TLIN3}
\end{eqnarray}

As usual in linear stability analysis, we express the solutions of (\ref{eq:S3TLIN1})-(\ref{eq:S3TLIN3}) in terms of the eigenvectors and eigenvalues of the system. The natural matrix form of the S3T equations obscures the operator-vector structure of the linearized system. The most direct technique for conducting the eigenanalysis is to rewrite the equations in superoperator form by unfolding the matrices $\mathsfbi{\delta C}_n$ \citep{FARRELL:2002tz}. This technique results in linear operators of very high dimension for which eigenanalysis is expensive. We use an alternate method to obtain the eigenstructures in which the linearized equations are rewritten as coupled Sylvester equations \citep[see Appendix B in][]{Constantinou:2014fh}.

We note that, for our choice of stochastic excitation, equations (\ref{eq:S3TLIN1})-(\ref{eq:S3TLIN3}) decouple into two separate eigenproblems: one determining the eigenmodes involving mean flow perturbations $\delta \boldsymbol{U}$, which have $\delta \boldsymbol{B}=0$, and a separate eigenproblem determining the eigenmodes involving mean buoyancy perturbations $\delta \boldsymbol{B}$, which have $\delta \boldsymbol{U}=0$. The eigenproblem involving $\delta \boldsymbol{U}$ gives unstable eigenmodes associated with growing VSHFs for the parameter regime we address in this work, while the mean buoyancy eigenproblem has only stable eigenmodes in our parameter regime. The mean buoyancy eigenproblem is therefore irrelevant, in our parameter range, to VSHF formation and we focus on the eigenproblem concerning $\delta \boldsymbol{U}$.

We now describe the results of the eigenanalysis of equations (\ref{eq:S3TLIN1})-(\ref{eq:S3TLIN3}). As the fixed point underlying the linearization corresponds to homogeneous turbulence, the eigenfunctions have harmonic structure in $z$ so that $\delta \boldsymbol{U}$ and $\delta \langle \overline{u'w'} \rangle$ are both proportional to $e^{st}e^{\text{i}m_Uz}$. For each $m_U$ permitted by the periodic domain there is a dominant eigenmode with eigenvalue $s(m_U)$. For the parameter range we study, we find that these eigenvalues are real, corresponding to structures for which the perturbation momentum flux divergence, $-\partial_z \delta \langle \overline{u'w'} \rangle$, and the mean flow, $\delta \boldsymbol{U}$, are aligned in phase.

\begin{figure}
       \centerline{\includegraphics{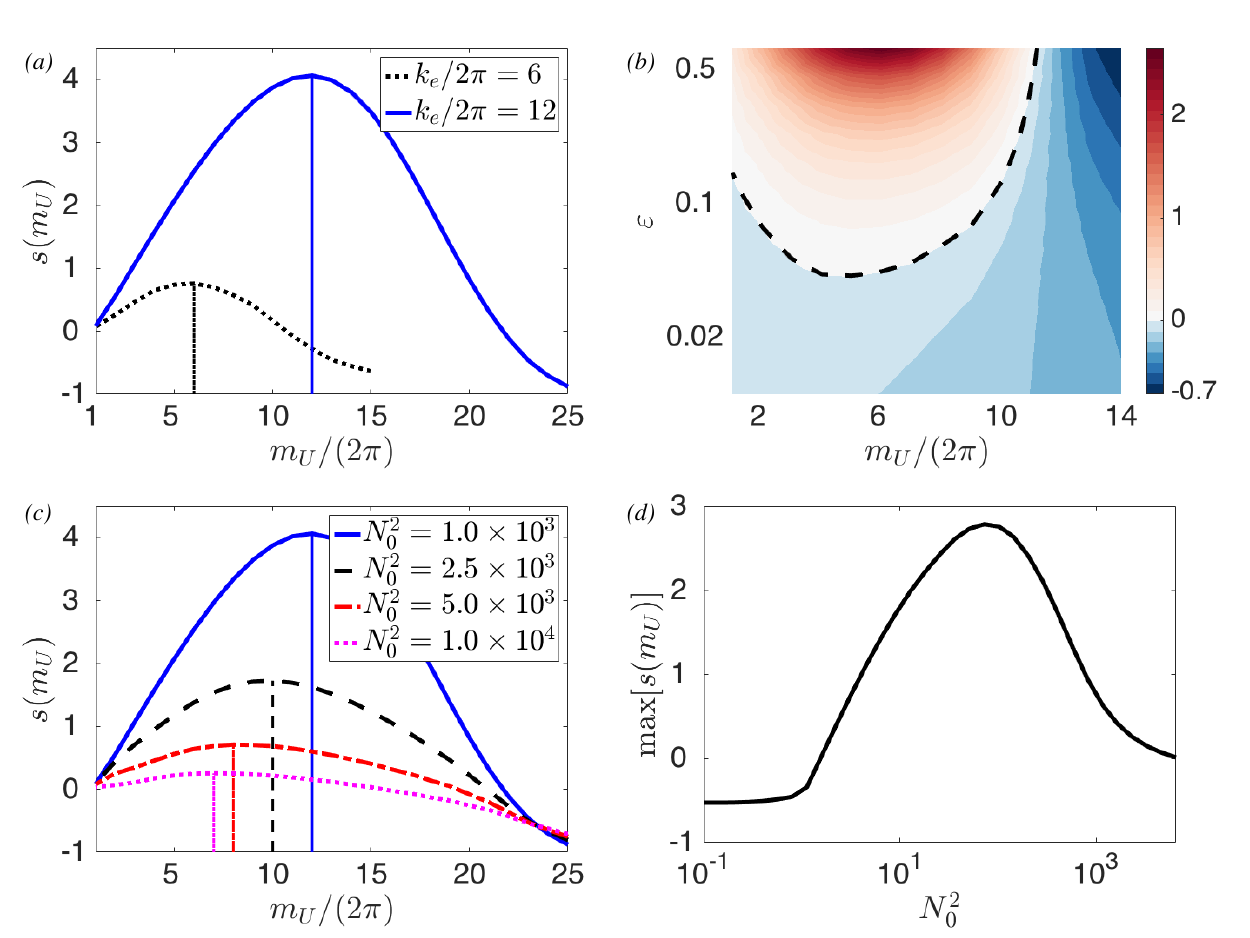}} 
  \caption{Growth rates of the eigenmodes responsible for VSHF formation in the S3T system. (a) Growth rate as a function of the VSHF wavenumber $m_U$ for $\varepsilon=0.25$ and two different excitation structures: $k_e/2\upi=6$ (dotted, $Fr=0.6$) and $k_e/2\upi=12$ (solid, $Fr=1.2$). (b) Growth rate as a function of $m_U$ and $\varepsilon$ for $k_e/2\upi=6$. Note the logarithmic $\varepsilon$ axis. (c) Growth rate as a function of $m_U$ for $k_e/2\upi=12$ and four values of $N_0^2$. (d) Growth rate of the fastest growing VSHF structure as a function of $N_0^2$ for $k_e/2\upi=6$. This figure shows that the vertical wavenumber, $m_U$, of the initially emergent VSHF is very sensitive to changes in the spectral structure of the excitation, and also that $m_U\to0$ as $N_0^2\to \infty$ so that the initially emergent VSHF takes on the largest scale permitted by the domain if the stratification is sufficiently strong. Unless otherwise specified, parameters are as in figure \ref{fig:NL_Phenom_FlowSnapshots}.}
\label{fig:ScaleSelectionPlot_Combined}
\end{figure}

Figure \ref{fig:ScaleSelectionPlot_Combined} summarizes how the dominant eigenvalue, $s$, which is the VSHF growth rate, depends on $m_U$ and on the parameters $k_e$, $N_0^2$, and $\varepsilon$. The dotted curve in panel (a) shows the VSHF growth rate as a function of $m_U$ for the standard parameter case with $\varepsilon=0.25$ ($Fr=0.6$). VSHFs with $1\le m_U/2\upi \le 10$ have positive growth rates, with the $m_U/2\upi=6$ structure having the fastest growth rate. This eigenvalue problem thus predicts that a VSHF with vertical wavenumber $m_U/2\upi=6$ will initially emerge from the turbulence, consistent with the structure of the VSHF discussed in \textsection \ref{sec:NLphenom} and \textsection \ref{sec:modelcomparison}. 
The solid curve in panel (a) shows the VSHF growth rates for the standard parameter case except with $k_e/2\upi=12$ ($Fr=1.2$) so that the excitation is at a smaller scale. Increasing $k_e$ shifts the peak of the growth rate curve toward larger $m_U$ values, resulting in smaller scale VSHFs, and also enhances the VSHF growth rates.

Figure \ref{fig:ScaleSelectionPlot_Combined} (b) shows the dependence of $s$ on $m_U$ and $\varepsilon$ for the standard parameter case with $k_e/2\upi=6$. For VSHFs with $1\le m_U/2\upi \le 11$ the growth rate becomes positive for sufficiently large $\varepsilon$. For VSHFs in this wavenumber band, the perturbation fluxes reinforce the infinitesimal VSHF and $s$ increases with increasing $\varepsilon$. For VSHFs outside this band, with $m_U/2\upi \ge 12$, the perturbation fluxes oppose the VSHF so that the growth rate becomes increasingly negative as $\varepsilon$ increases. The dashed line shows the stability boundary, $s=0$. The homogeneous turbulence first becomes unstable near $\varepsilon\approx.042$ ($Fr=.25$) to a VSHF with $m_U/2\upi=5$. As $\varepsilon$ increases, the growth rate of the $m_U/2\upi=6$ VSHF structure exceeds that of the $m_U/2\upi=5$ VSHF so that for the standard parameter case with $\varepsilon=0.25$ ($Fr=0.6$) a VSHF with vertical wavenumber $m_U/2\upi=6$ initially emerges in the turbulence. 
We note that the emergent VSHF wavenumber varies between NL simulations depending on the particular realization of the stochastic excitation, with the $m_U/(2\upi)=7$ structure occurring somewhat more frequently than $m_U/(2\upi)=6$. 
The existence of multiple turbulent equilibria in this system is predicted by S3T as discussed in \textsection \ref{sec:multiple_equilibria}. 
That the NL system often forms a VSHF with a slightly different scale than that predicted by linear stability analysis of the S3T system is likely to due the modification of the background spectrum of turbulence by perturbation-perturbation interactions. 
The influence of perturbation-perturbation interactions on the S3T stability of homogeneous turbulence has been analyzed in detail by \citet[]{Constantinou:2014fh} in the context of barotropic beta-plane turbulence.
%from t has been explained by \citet[]{Constantinou:2014fh}, in the context of barotropic beta-plane turbulence, to result from the modification of the background spectrum of the NL system by the perturbation-perturbation interactions that are retained in the NL system but not in the QL or S3T systems.
%
%The existence of multiple turbulent equilibria in this system is predicted by S3T as discussed in \textsection \ref{sec:multiple_equilibria}. 
%The tendency of the NL system to form a VSHF with a slightly different scale than that which is predicted by linear stability analysis of the S3T system has been explained by \citet[]{Constantinou:2014fh}, in the context of barotropic beta-plane turbulence, to result from the modification of the background spectrum of the NL system by the perturbation-perturbation interactions that are retained in the NL system but not in the QL or S3T systems.
%

Figures \ref{fig:ScaleSelectionPlot_Combined} (c) show how $N_0^2$ influences the scale selection of the initially emergent VSHF. In panel (c) $s$ is shown as a function of $m_U$ for four $N_0^2$ values with $k_e/2\upi=12$ and $\varepsilon=0.25$. 
As $N^2_0$ increases, $s$ decreases and the peak (indicated by the vertical lines) shifts toward smaller $m_U$. 
For very large $N_0^2$ the largest values of $s$ occur for VSHFs at the domain scale with $m_U/2\upi=1$. However, unless $\varepsilon$ is also very large the homogeneous turbulent state will remain stable and a domain-scale VSHF will not emerge, because $s$ decreases as $N_0^2$ becomes large. 
We note that the decrease of the VSHF wavenumber as $N_0^2$ increases demonstrates that $m_U$ is not directly related to either the Ozmidov wavenumber, $k_{O}=(N_0^3/\varepsilon)^{1/2}$, or the buoyancy wavenumber, $k_b=N_0/\sqrt{\varepsilon}$, both of which increase as $N_0^2$ is increased, and also that the S3T prediction of $m_U$ depends on the parameter values and is not always equal to the excitation wavenumber, $k_e$. %is not alwaysrequired to be equal to the excitation wavenumber, $k_e$. %, in all cases. 
As $N^2_0$ is decreased toward moderate and weak stratification (not shown), the wavenumber of the initially emergent VSHF remains near $k_e$, consistent with the results of NL simulations shown in figure \ref{fig:newfig1} (c,d). %, which show 
%We note that this dependence of $m_U$ on $N_0^2$ is inconsistent with 
%

The dependence of $s$ on $N_0^2$ is shown directly in panel (d), which shows $\text{max}[s(m_U)]$, where the maximum is taken over $m_U$, as a function of $N^2_0$. 
For small $N^2_0$, all modes have negative growth rates. 
This result provides an explanation for the frequent observation in numerical simulations that the VSHF ceases to emerge when the stratification becomes sufficiently weak. 
However, this result depends on the details of the stochastic excitation. In Appendix \ref{sec:appendixA} we describe a reduced model in which the excitation is anisotropic and which has the property that $s$ remains positive as $N_0^2\to0$, a result which was also obtained for similarly anisotropic excitation by \citet[]{Bakas:2011bt}. As $N^2_0$ increases from zero $s$ increases to a maximum near $N_0^2\approx10^2$. This increase in growth rate is associated with the strengthening of the feedback between the VSHF and the turbulence described in \textsection \ref{sec:testfunction}. 
The dependence of the S3T wave-mean flow feedback for harmonic mean structures on the parameter that sets the wave restoring force has been explained analytically in terms of wave dynamics by \citet[]{Bakas:2013ft} for the case of barotropic beta-plane turbulence.  
For $N^2_0\gtrsim 10^3$ the growth rate falls off as $\sim1/N^2_0$ and approaches a constant asymptotic value as $N_0^2\to\infty$ that is set by the dissipation parameters.

\begin{figure}
    \centerline{\includegraphics{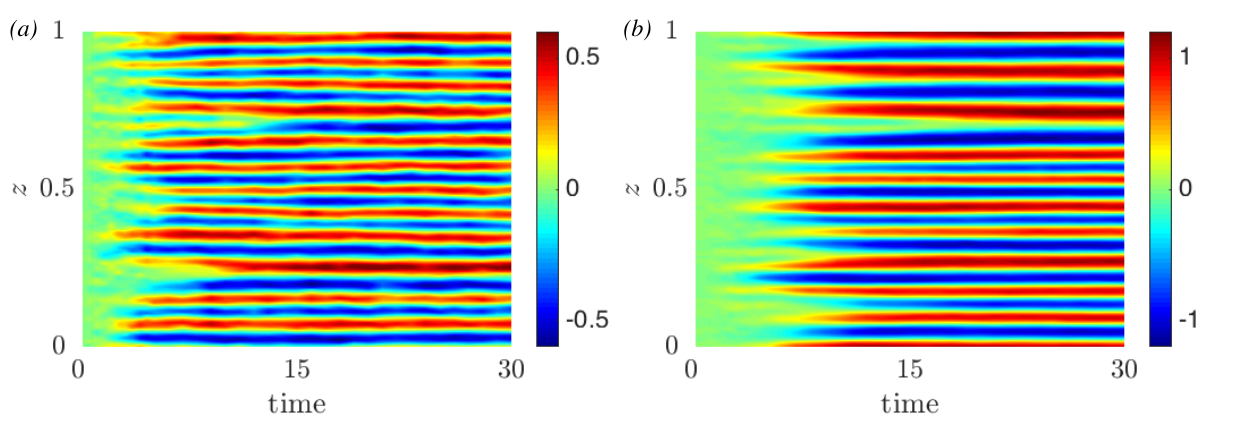}}
    \caption{Time evolution of the VSHF in two example simulations illustrating the correspondence between the behavior of the NL system and the predictions of the linear stability analysis of the S3T system as the parameters are varied. Unless otherwise stated all parameters are as in figure \ref{fig:NL_Phenom_FlowSnapshots}. (a) An example with smaller scale excitation, $k_e/(2\upi)=12$ ($Re_b=10.4$, $Fr=1.2$). The VSHF forms with $m_U/(2\upi)\approx 12$. (b) An example with smaller scale excitation, $k_e/(2\upi)=12$, and also stronger stratification, $N_0^2=5\times10^3$ ($Re_b=2.1$, $Fr=0.53$). The VSHF forms with $m_U/(2\upi)\approx10$. This figure demonstrates that, in the NL system, the VSHF forms at smaller scale when the turbulence is excited at smaller scale and that the VSHF forms at larger scale when the stratification is increased, consistent with the predictions of S3T.} 
\label{fig:New_Jet_Cases}
\end{figure}

Figure \ref{fig:New_Jet_Cases} shows the time evolution of the VSHF, $U$, in two example simulations that illustrate the correspondence of the behavior of the NL system with the predictions of linear stability analysis of the S3T system shown in figure \ref{fig:ScaleSelectionPlot_Combined}. 
Panel (a) shows the VSHF evolution in an example in which the parameters are as in the standard case simulation but with the excitation scale modified to $k_e/(2\upi)=12$ ($Fr=1.2$). 
Consistent with the shift of peak of the VSHF growth rate curve to $m_U/(2\upi)=12$ in figure \ref{fig:ScaleSelectionPlot_Combined} (a), the emergent VSHF in the NL system has $m_U/(2\upi)\approx12$. 
In panel (b) is shown the VSHF evolution in an example with the same parameters as in panel (a), but with the stratification increased to $N_0^2=5\times10^3$ ($Fr=.53$). 
Consistent with the shift of the peak of the VSHF growth rate curve toward lower values of $m_U$ under increased stratification in figure \ref{fig:ScaleSelectionPlot_Combined} (c), the emergent VSHF in the NL system has $m_U/(2\upi)\approx10$. 
Although the peak of $s(m_U)$ in the S3T system occurs at the slightly larger scale $m_U/(2\upi)=8$ for $N_0^2=5\times10^3$, the growth rates of the nearby VSHF wavenumbers are very similar to the peak value, as shown in figure \ref{fig:ScaleSelectionPlot_Combined} (c), so that the VSHF wavenumber that is observed in a given realization of the stochastic system for these parameter values is likely to depend on the particular noise realization. 

Results of this section demonstrate that the scale of the initially emergent VSHF, $m_U$, is strongly influenced by the spectral structure of the perturbation field, which in our problem is set by $k_e$. 
As the stratification becomes very strong, the VSHF scale is modified from the scale set by the excitation and tends toward the largest scale allowed by the domain. 
In realistic turbulence, the implication of this result is that we expect the spectral characteristics of the background turbulence to imprint strongly on the VSHF scale if the turbulence is sufficiently close to the stability boundary and the stratification is not too strong. 
% and is also dependent on the strength of the stratification.
%For the parameter values considered here, the initially emergent VSHF wavenumber is approximately equal to the excitation wavenumber for stratification strengths up to $N_0^2\approx10^3$. 
%As the stratification strength is increased further, the VSHF wavenumber decreases, with $m_U\to0$ as $N_0^2\to\infty$. 
We emphasize that the linear stability analysis conducted in this section provides a prediction of the scale of the initially emergent VSHF, rather than of the scale of the statistical equilibrium VSHF. 
For excitation strengths sufficiently near the stability boundary, the prediction based on linear stability analysis is expected to agree with the equilibrium VSHF structure. %  accurate. 
As the excitation strength is increased beyond the stability boundary, the structure of the VSHF may be modified from the initially emergent structure, as suggested by the NL simulation shown in figure \ref{fig:newfig1} (b). 
Previous studies of VSHF emergence have primarily been conducted using weak dissipation or strong excitation, so that the excitation strength lies well beyond the stability boundary, and these studies have also observed VSHFs with larger scale than that of the excitation. 
For example, \citet[]{Herring:1989fx} obtained a $m_U=6$ VSHF in 3D stratified turbulence maintained with $k_e\approx 11$ excitation, \citet[]{Smith:2001uo} obtained a VSHF with energy concentrated near $m_U\approx 10-15$ in the 2D system using $k_e\approx96$, and \citet[]{Smith:2002wg} obtained a VSHF with energy concentrated near $m_U\approx 9-11$ in the 3D system maintained with $k_e\approx 24$. 
Although precise parameter correspondence between our study and these previous studies is difficult to establish due to differences in model formulation, these previous examples demonstrate that VSHFs with larger scale than that of the excitation are often observed when the system is strongly excited. %pushed well beyond the stability boundary. 
This behavior is expected based on analysis of the S3T system, which we discuss in \textsection \ref{sec:equilibration}. 

\section{Equilibration of Horizontal Mean Structure}\label{sec:equilibration}

In \textsection \ref{sec:modelcomparison} we showed that the S3T system initialized with a perturbative VSHF with $m_U/2\upi=6$ in the standard parameter case evolves into an equilibrium state with the same VSHF wavenumber (figures \ref{fig:ModelComparison_Hovmollers} and \ref{fig:ModelComparison_Energy_and_Uprofiles}). We now analyze how the structure of this finite-amplitude equilibrium depends on the control parameters. 

\begin{figure}
       \centerline{\includegraphics{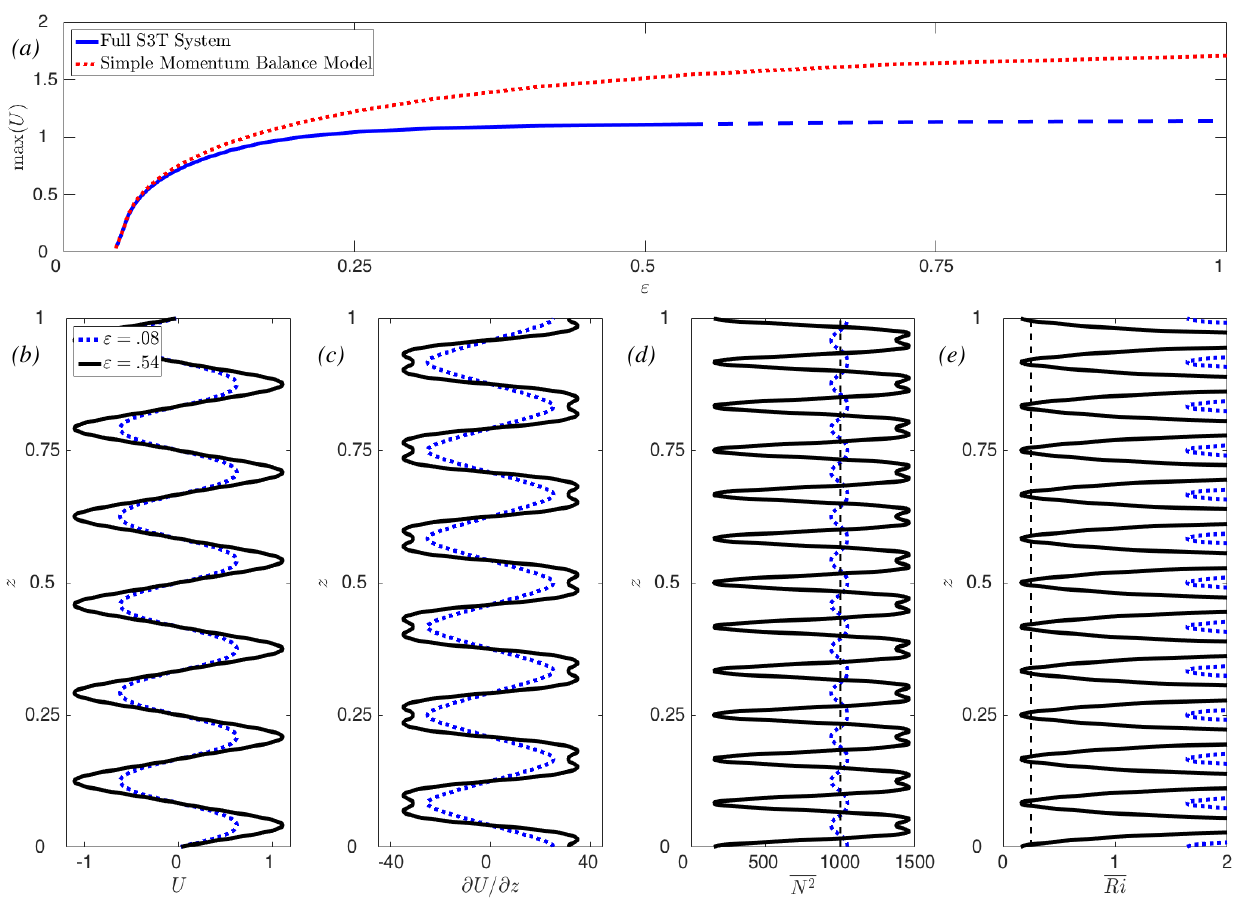}} 
  \caption{Equilibrium structure diagnostics for S3T and the simple momentum balance model in the case of the $m_U/2\upi=6$ horizontal mean state as a function of $\varepsilon$. (a) Maximum value of $U$, maximized over $z$, for the stable S3T fixed point with $m_U/2\upi=6$ (solid). This fixed point becomes secondarily unstable near $\varepsilon=0.55$ ($Fr=0.88$) and the dashed continuation shows the amplitude of the unstable solution. The dotted curve shows the estimate of the amplitude of $U$ from the simple momentum balance model (see text). Panels (b)-(e) show the vertical structure of the horizontal mean state as in figure \ref{fig:4panel_USNsqRi_meanplot} with dotted curves indicating the $\varepsilon=0.08$ ($Fr=.34$) state and solid curves indicating the $\varepsilon=0.54$ ($Fr=.88$) state. This figure shows that weak equilibration of the VSHF is captured by the simple momentum balance model and that the $U$ and $\overline{N^2}$ structures, and their phase relationship to one another, vary as $\varepsilon$ is increased. Unless otherwise specified, parameters are as in figure \ref{fig:NL_Phenom_FlowSnapshots}.}
\label{fig:Drafting_Equilibration_Plot_full}
\end{figure}

Figure \ref{fig:Drafting_Equilibration_Plot_full} (a, solid curve) shows the maximum value of the $m_U/2\upi=6$ equilibrium $U$ structure, maximized over $z$, as a function of $\varepsilon$. The dotted curve shows an estimate of $U$ from a simple momentum balance model which we will explain later in this section. As suggested by the stability analysis in \textsection \ref{sec:scaleselection}, the $m_U/2\upi=6$ VSHF forms near $\varepsilon\approx0.044$ ($Fr=.25$) when the growth rate of the corresponding eigenmode crosses zero. The bifurcation is supercritical, with the VSHF equilibrating at weak amplitude just beyond the bifurcation point. Near the bifurcation point $U$ increases rapidly with $\varepsilon$, with this rate of increase slowing as $\varepsilon$ increases. 

The structure of the horizontal mean state depends on $\varepsilon$. Figures \ref{fig:Drafting_Equilibration_Plot_full} (b)-(e) (dotted curves) show the horizontal mean structure of the marginally supercritical equilibrium at $\varepsilon=0.08$ ($Fr=.34$). The VSHF structure is similar to that of the unstable $m_U/2\upi=6$ harmonic eigenmode. The phase relationship between $U$ and $\overline{N^2}$ differs from that found in the more strongly excited $\varepsilon=0.25$ ($Fr=0.6$) case discussed in \textsection \ref{sec:NLphenom} and \textsection \ref{sec:modelcomparison}. In the $\varepsilon=0.08$ ($Fr=.34$) case of weak equilibration the stratification is enhanced in the shear regions, rather than weakened, and $\overline{Ri}$ is large for all $z$ due to the weak shear. The solid curves in figures \ref{fig:Drafting_Equilibration_Plot_full} (b)-(e) show the horizontal mean structure of the $\varepsilon=0.54$ ($Fr=.88$) equilibrium. For this more strongly supercritical equilibrium, the shear regions are characterized by weakened stratification and $\overline{Ri}<1/4$. This structure is similar to that shown in figure \ref{fig:New_ModelComparison_QLS3T_4panelplots} for the $\varepsilon=0.25$ case, but with stronger shear and smaller $\overline{Ri}$ values. The VSHF remains hydrodynamically stable (\emph{i.e.,} all eigenvalues of $\mathsfbi{A}_n$ have negative real parts) despite having $\overline{Ri}<1/4$ due to the dissipation acting on the perturbation fields. 
When $\varepsilon$ is further increased the $m_U/2\upi=6$ fixed point becomes secondarily unstable, indicated by the dashed continuation of the solid curve in figure \ref{fig:Drafting_Equilibration_Plot_full} (a). 
 
The changing phase relationship between $U$ and $\overline{N^2}$ shown in figure \ref{fig:Drafting_Equilibration_Plot_full} that occurs as a function of $\varepsilon$ mirrors the change in this relationship shown in figure \ref{fig:ModelComparison_Hovmollers} that occurs as a function of time. Comparison of figures \ref{fig:ModelComparison_Hovmollers} (c) and (d) shows that, when the developing VSHF is weak, $\overline{N^2}$ is enhanced where the shear is strongest. When the VSHF becomes strong, the stratification is reorganized by the turbulent fluxes so that $\overline{N^2}$ is weakest where the shear is strongest. 

\begin{figure}
        \centerline{\includegraphics{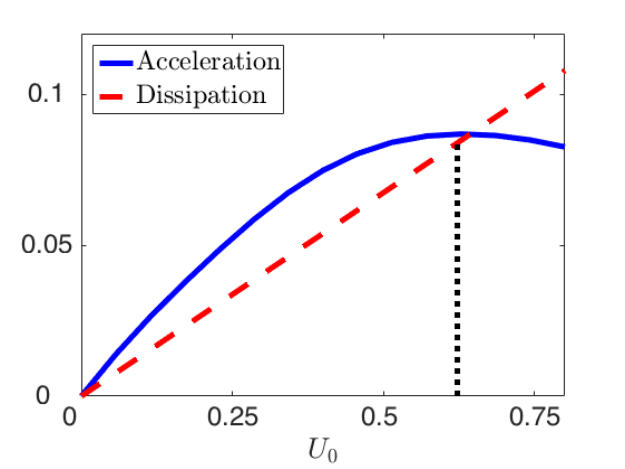}} 
  \caption{Illustration of the simple momentum balance model for weakly supercritical VSHF equilibration for $\varepsilon=0.08$ and $m_U/2\upi=6$, with other parameters as in figure \ref{fig:NL_Phenom_FlowSnapshots}. The solid curve shows the projection of the perturbation momentum flux divergence, calculated using the test function analysis of \textsection\ref{sec:testfunction}, onto the assumed harmonic VSHF structure. The dashed line shows the dissipation acting on the VSHF, given by $(r_m+\nu m_U^2)U_0$. The simple model estimate of the equilibrium VSHF amplitude is the value of $U_0$ at which these terms balance one another. The vertical dotted line indicates the equilibrium VSHF amplitude obtained from the full S3T system. This figure demonstrates that the dynamics of weakly supercritical VSHF equilibration is captured by the simple balance model.}
\label{fig:WeakEquil_SimpleBalance_Plot}
\end{figure}

The mechanism of VSHF equilibration at weak amplitude can be understood using a simple momentum balance model based on the test function analysis of \textsection \ref{sec:testfunction}. To construct the simple model we first approximate the horizontal mean state as $U=U_0 \sin(m_U z)$ and $B=0$, where $U_0$ is the equilibrium VSHF amplitude that we will estimate. We then estimate the acceleration of the VSHF produced by the induced perturbation momentum fluxes as a function of $U_0$ using (\ref{eq:testfn1})-(\ref{eq:testfn2}). Our estimate of the equilibrium VSHF amplitude is the value of $U_0$ for which this acceleration is balanced by dissipation. As $\varepsilon\to\varepsilon_c$ this simple model becomes exact because both $U$ and the perturbation flux divergence become exactly harmonic and $B\to0$. For $\varepsilon > \varepsilon_c$, the structure of $-\partial_z \langle \overline{u'w'} \rangle$ deviates from harmonic and we estimate the equilibrium VSHF amplitude by projecting the acceleration onto the assumed harmonic VSHF structure. 

We illustrate the simple momentum balance model for $\varepsilon=0.08$ ($Fr=.34$) and $m_U/2\upi=6$ in figure \ref{fig:WeakEquil_SimpleBalance_Plot}, which shows the estimated acceleration (solid) and dissipation (dashed) of the VSHF as functions of the VSHF amplitude, $U_0$. The dissipation, $(r_m+\nu m_U^2)U_0$, increases linearly with $U_0$. For small $U_0$ the acceleration is stronger than the dissipation, consistent with spontaneous VSHF formation as a linear instability for these parameters. Due to the negative curvature of the acceleration as a function of $U_0$ the two terms balance near $U_0\approx 0.65$, which gives the simple model estimate of the equilibrium VSHF amplitude. The vertical dotted line indicates the equilibrium VSHF strength in the full S3T system. For this value of $\varepsilon$ the simple model captures the equilibration dynamics, implying that modification of $\overline{N^2}$ and changes in $U$ structure do not play important roles in the weak equilibration process. The simple model estimate of $\text{max}(U)$ as a function of $\varepsilon$ is shown in figure \ref{fig:Drafting_Equilibration_Plot_full} (a) as the dotted curve. The model estimate matches the results of the full calculation as $\varepsilon\to\varepsilon_c$ and diverges from the full solution as $\varepsilon$ increases.

\begin{figure}
\centerline{\includegraphics{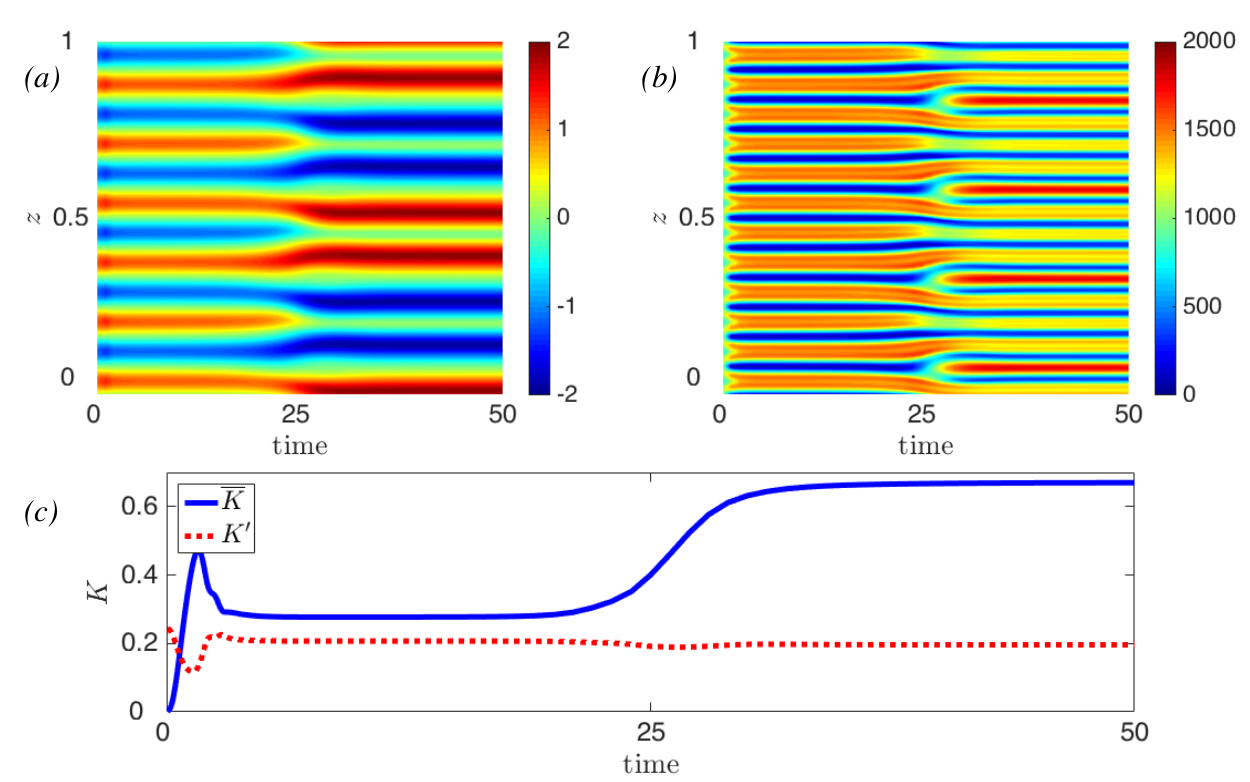}} 
  \caption{Secondary instability of the S3T fixed point corresponding to the $m_U/2\upi=6$ VSHF for $\varepsilon=1$ ($Fr=1.2$), with other parameters as in figure \ref{fig:NL_Phenom_FlowSnapshots}. The upper panels show the time evolution of (a) $U$ and (b) $\overline{N^2}$. The lower panel (c) shows the kinetic energy evolution. This figure shows that for strong excitation the $m_U/2\upi=6$ VSHF state is unstable to the development of a global vertical wavenumber 2 pattern in $U$ that is superimposed on the initial wavenumber 6 pattern, strengthening the VSHF and modifying its structure to produce wider shear regions.}
\label{fig:Developing_HardeqInstFig1_Combined}
\end{figure}

As $\varepsilon$ is increased, the global minimum of $\overline{Ri}$ falls further below $1/4$ and the $m_U/2\upi=6$ state becomes secondarily unstable just beyond $\varepsilon=0.54$ ($Fr=.88$). 
Although this instability occurs when $U$ is near the laminar stability boundary, which is modified from $\overline{Ri}=1/4$ by the presence of dissipation and by our choice of a finite periodic domain which quantizes the permitted perturbation horizontal wavenumbers, we emphasize that this secondary instability is a property of the S3T dynamics, rather than of the perturbation dynamics determined by the operator $\mathsfbi{A}_n$.
In particular, the $m_U/2\upi=6$ state remains hydrodynamically stable at all times during the instability development. Figures \ref{fig:Developing_HardeqInstFig1_Combined} (a) and (b) show the time evolution of $U$ and $\overline{N^2}$ during the development of the secondary instability for $\varepsilon=1$ ($Fr=1.2$). The VSHF structure near $t=30$ reveals that the 6-fold symmetry of the $m_U/2\upi=6$ VSHF is spontaneously broken by the instability. As the instability develops the positive VSHF peaks near $z=0.2$ and $z=0.7$ contract and weaken while their neighbouring negative peaks strengthen and expand. Similarly, the negative VSHF peaks near $z=0.5$ and $z=0.9$ contract and weaken while their neighbouring positive VSHF peaks strengthen and expand. The particular locations of the strengthening and weakening features are the result of the symmetry breaking and so depend on the small perturbations included in the initialization. Figure \ref{fig:Developing_HardeqInstFig1_Combined} (c) shows the evolution of kinetic energy during the instability. The changes in the VSHF structure are associated with an increase in the mean kinetic energy consistent with the broadening of the VSHF pattern allowing $U$ to strengthen while maintaining a hydrodynamically stable shear. 
Similar behavior is shown to occur in the NL system in figure \ref{fig:newfig1} (b), which shows an example in which the effective excitation strength has been increased relative to the standard case integration by removing the Rayleigh drag terms from the dynamics. %, reducing the dissipation. 
In this example, a VSHF with $m_U/(2\upi)\approx 6$ initially emerges from the turbulence and this VSHF transitions to lower wavenumber as the integration is continued. 
Secondary instabilities of finite-amplitude mean shear flows that result in broader shear patterns also occur in the barotropic beta-plane system and have been analyzed using S3T by \citet[]{Constantinou:2014fh}.

\begin{figure}
 % \centerline{\includegraphics[scale=.32]{Fig17_v1p0.png}}   
  % \centerline{\includegraphics{Fig17_v2p0.pdf}}   
     \centerline{\includegraphics{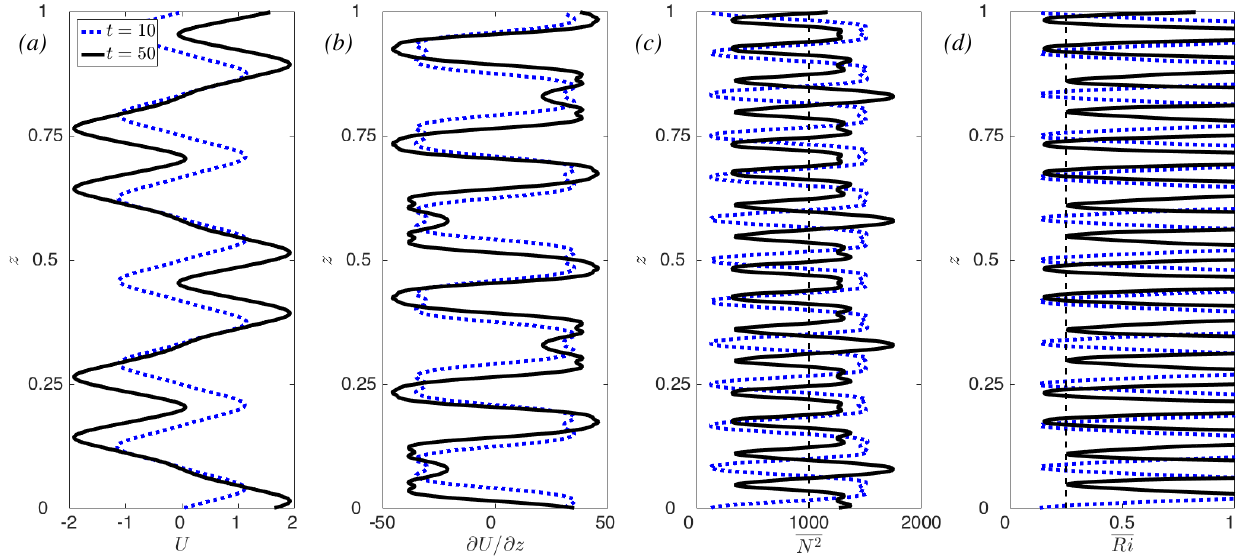}}   
  \caption{Vertical structure of the horizontal mean state in the S3T system before and after the development of the secondary instability for $\varepsilon=1$ ($Fr=1.2$), with other parameters as in figure \ref{fig:NL_Phenom_FlowSnapshots}. Panels are as in figure \ref{fig:4panel_USNsqRi_meanplot} with dotted curves showing the structure for $t=10$ and solid curves showing the structure for $t=50$. This figure shows how the structure of the horizontal mean state is reorganized by the secondary instability. The unstable equilibrium state at $t=10$ has $\overline{Ri}<1/4$ in regions of strongest shear and weakest stratification. The final equilibrium state has shear regions of two different widths in which the broader shear regions have $\overline{Ri}>1/4$ due to enhanced stratification and weakened shear in the cores of the shear regions.}
\label{fig:Developing_HardeqInstFig2_Combined}
\end{figure}

The structure of the horizontal mean state before and after the development of the secondary instability for $\varepsilon=1$ ($Fr=1.2$) is shown in figure \ref{fig:Developing_HardeqInstFig2_Combined}. Prior to the instability development ($t=10$, dotted) the structure is similar to that shown for $\varepsilon=0.54$ ($Fr=.88$) in figures \ref{fig:Drafting_Equilibration_Plot_full} (b)-(e) and is characterized by a VSHF with $m_U/2\upi=6$ and weakened stratification in the shear extrema. The $U$ profile of the final equilibrium structure ($t=50$, solid) contains shear regions with two distinct widths which are associated with distinct phase relationships between $U$ and $\overline{N^2}$. For the wider shear regions, the $U$ profile inflects in the centre of the shear region and $\overline{N^2}$ is locally maximized there, resulting in $\overline{Ri}\gtrsim1/4$. The narrower shear regions are similar to those that precede the secondary instability development and have $\overline{Ri}<1/4$. 

%\section{Existence of Multiple Turbulent Equilibria}\label{sec:multiple_equilibria}
\section{Multiple Turbulent Equilibria in Stratified Turbulence}\label{sec:multiple_equilibria}

In \textsection \ref{sec:scaleselection} we showed an NL simulation in which a $m_U/2\upi=6$ VSHF emerges, corresponding to the eigenmode of the linearized S3T system that has the fastest growth rate in the standard parameter case. However, figure \ref{fig:ScaleSelectionPlot_Combined} (a) shows that, for the standard parameter case, all VSHF structures in the wavenumber band $1\leq m_U/2\upi \leq10$ have positive growth rates. The subdominant eigenmodes (\emph{i.e}, those with $m_U/2\upi \neq 6$) continue to finite-amplitude VSHF equilibria at the corresponding wavenumbers. These equilibria may or may not be stable. In this section we demonstrate that multiple turbulent equilibrium states are possible in 2D Boussinesq turbulence by providing an example of such an alternate stable equilibrium in the S3T and NL systems. 

\begin{figure}
   %\centerline{\includegraphics[scale=.3]{Fig18_v1p0.png}} 
      \centerline{\includegraphics{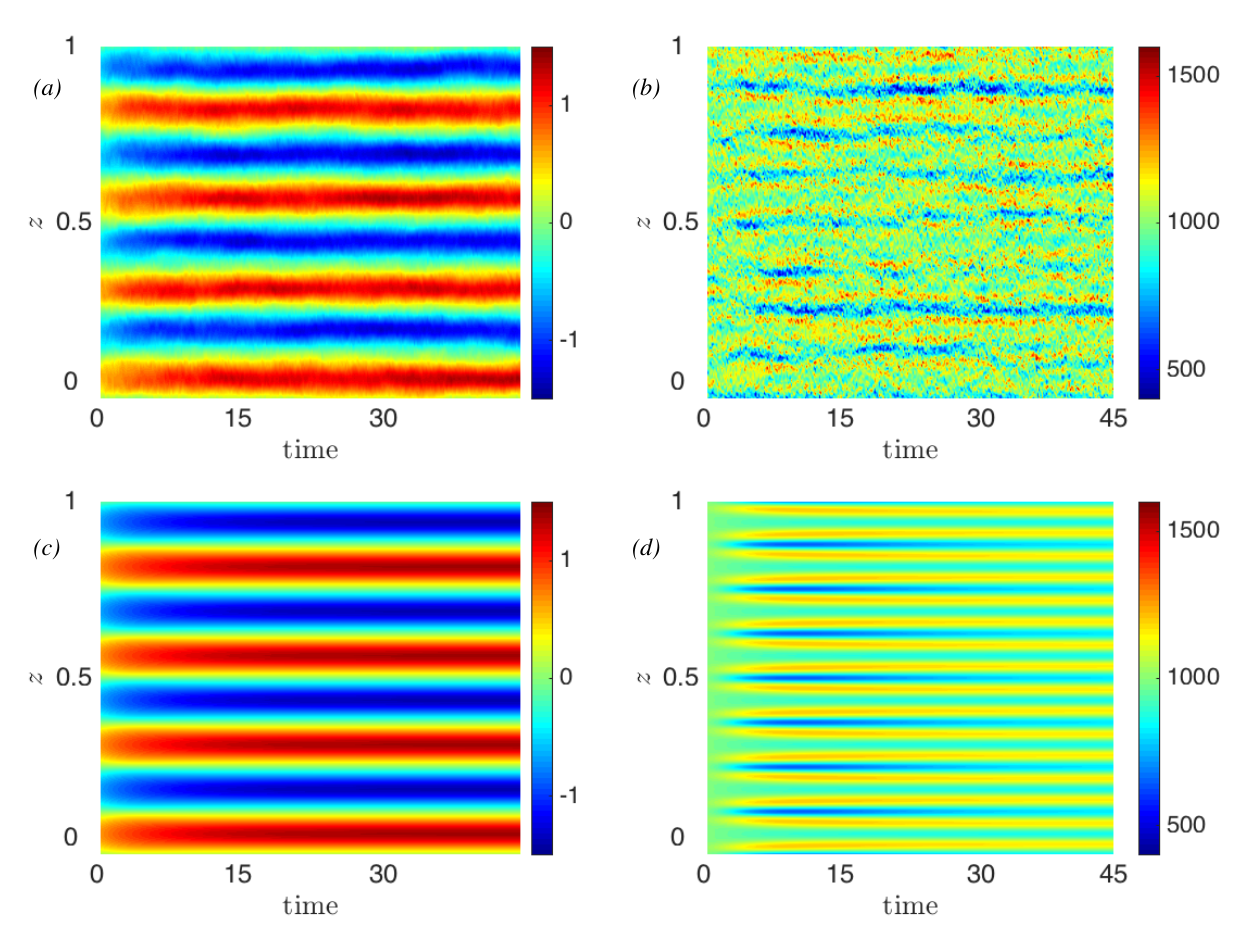}} 
  \caption{Time evolution of the horizontal mean structure of the $m_U/2\upi=4$ equilibrium state in the NL and S3T systems. Panels show the time evolution of (a) $U$ and (b) $\overline{N^2}$ in the NL system and (c) $U$ and (d) $\overline{N^2}$ in the S3T system. This figure shows that when initialized with a finite-amplitude VSHF with wavenumber $m_U/2\upi=4$ the NL system maintains this structure, resulting in a turbulent equilibrium state different from that discussed in \textsection \ref{sec:NLphenom} for the same parameter values, and that this alternate equilibrium state is also a fixed point of the S3T system. Parameters are as in figure \ref{fig:NL_Phenom_FlowSnapshots}.}
\label{fig:New_MultipleEquilibria1}
\end{figure}

\begin{figure}
%    \centerline{\includegraphics[scale=.3]{Fig19_v1p0.png}}
    %\centerline{\includegraphics{Fig19_v2p0.pdf}}  
        \centerline{\includegraphics{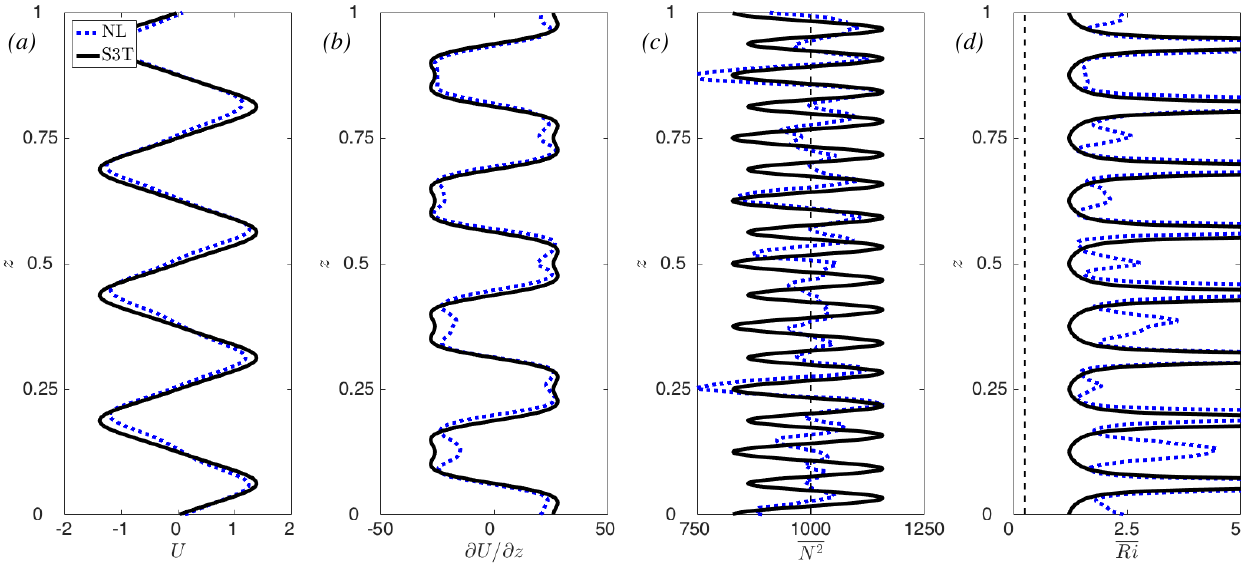}}  
  \caption{Vertical structure of the horizontal mean state of the $m_U/2\upi=4$ equilibrium in the NL and S3T systems. Panels are as in figure \ref{fig:4panel_USNsqRi_meanplot} with dotted curves showing the time-averaged structure over $t\in[22,45]$ for the NL system and solid curves showing the final fixed point structure for the S3T system. Parameters are as in figure \ref{fig:NL_Phenom_FlowSnapshots}.}
\label{fig:New_MultipleEquilibria3}
\end{figure}

In figures \ref{fig:New_MultipleEquilibria1} (a) and (c) we show the development of $U$ in the NL and S3T systems in an example in which the parameters are set to the standard values ($Fr=0.6$ as in figure \ref{fig:NL_Phenom_FlowSnapshots}) but the initial conditions are chosen to initiate a VSHF with wavenumber $m_U/2\upi=4$. The NL system is initialized with a mean flow $U\propto \sin(m_U z)$ for $m_U/2\upi=4$ and the S3T system is initialized with the same $U$ profile and $\mathsfbi{C}_n=\mathsfbi{C}_n^{\star}$. 
In the S3T system this initial condition evolves into a stable $m_U/2\upi=4$ fixed point. 
We note that, as shown in \textsection \ref{sec:scaleselection}, the S3T system will evolve, in the standard parameter case, toward the $m_U/2\upi=6$ fixed point for any sufficiently small initial perturbation, but that in this example the system evolves toward the $m_U/2\upi=4$ fixed point as a result of the finite initial perturbation. % directs the system toward an alternate fixed point. 
In the NL system the $m_U/2\upi=4$ turbulent equilibrium is maintained for the length of the integration. Due to noise in the NL system, the turbulence may eventually transition to another equilibrium state, such as the $m_U/2\upi=6$ state discussed in \textsection \ref{sec:NLphenom}. 
The development of $\overline{N^2}$ for this example is shown in figures \ref{fig:New_MultipleEquilibria1} (b) and (d). As in the previous examples of equilibria, $\overline{N^2}$ has a doubled vertical wavenumber relative to $U$ and is more variable than $U$ in the NL system. The vertical structure of the horizontal mean state is shown in figure \ref{fig:New_MultipleEquilibria3}. The VSHF structure (panels (a,b)) resembles a sawtooth in both the NL (dotted) and S3T (solid) systems. The phase relationship between $U$ and $\overline{N^2}$ (panel (c)) shares some features with that shown in figure \ref{fig:Drafting_Equilibration_Plot_full} for the $m_U/2\upi=6$ equilibrium with $\varepsilon=0.54$ ($Fr=.88$). In particular, the weakest values of $\overline{N^2}$ occur in the centres of the shear regions. The excitation strength $\varepsilon=0.25$ falls in a transitional range for the $m_U/2\upi=4$ equilibrium in which $\overline{N^2}$ has an approximately $m_B/2\upi=16$ structure. As $\varepsilon$ is increased (not shown), the stratification in the shear centres continues to weaken, producing density layers at these locations, and the stratification near the VSHF peaks is enhanced.

\begin{figure}
    %\centerline{\includegraphics[scale=.3]{Fig20_v1p0.png}} 
      %  \centerline{\includegraphics{Fig20_v2p0.pdf}} 
        %      \centerline{\includegraphics{Fig20_v2p1.pdf}} 
                      \centerline{\includegraphics{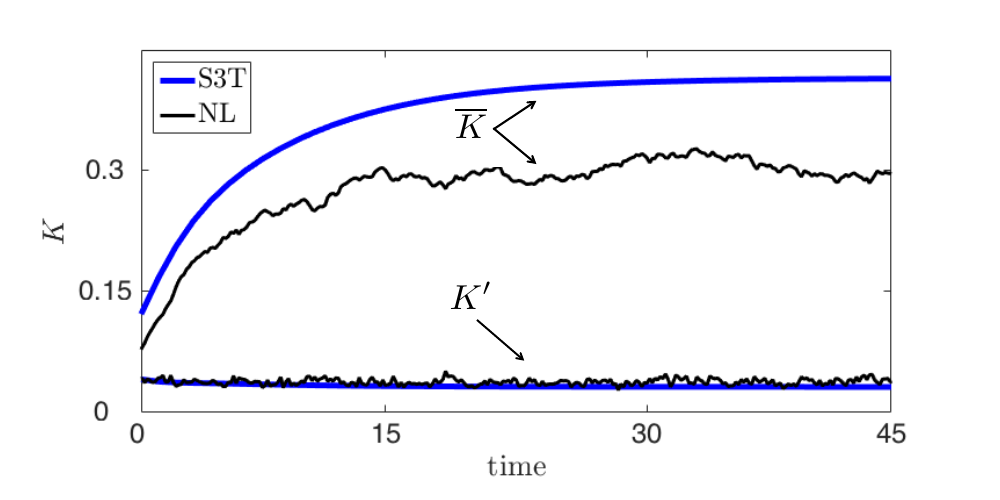}} 
  \caption{Kinetic energy evolution in the NL and S3T systems initialized with a $m_U/2\upi=4$ VSHF. This figure shows that, as for the $m_U/2\upi=6$ equilibrium, the VSHF in the S3T system is more energetic than the VSHF in the NL system, and comparison with figure \ref{fig:ModelComparison_Energy_and_Uprofiles} shows that in both the NL and S3T systems the $m_U/2\upi=4$ VSHF is more energetic than the $m_U/2\upi=6$ VSHF. Parameters are as in figure \ref{fig:NL_Phenom_FlowSnapshots}.}
\label{fig:New_MultipleEquilibria2}
\end{figure}

Figure \ref{fig:New_MultipleEquilibria2} shows the evolution of kinetic energy in the NL and S3T systems. Consistent with the results for the $m_U/2\upi=6$ equilibrium in \textsection \ref{sec:modelcomparison}, the equilibrium value of $\overline{K}$ in the S3T system exceeds that of NL system. In both systems, the broader VSHF in the $m_U/2\upi=4$ equilibrium is more energetic than the VSHF in the $m_U/2\upi=6$ equilibrium. This is consistent with the behaviour shown in figure \ref{fig:Developing_HardeqInstFig1_Combined} (c) in which the broadened VSHF resulting from the secondary instability is more energetic than the $m_U/2\upi=6$ VSHF that precedes the instability.

%
%When making comparisons among simulations of the NL, QL, and S3T systems, the existence of multiple equilibria must be considered. We present in \textsection \ref{sec:NLphenom} and \textsection \ref{sec:modelcomparison} example simulations of the NL and QL systems that form $m_U/2\upi=6$ VSHF structures reflecting the fastest-growing eigenmode of the S3T system. We note, however, that the NL system, when initialized from rest, frequently forms stable VSHFs with $m_U/2\upi=7$ or $m_U/2\upi=8$, and that in fact these higher wavenumber structures occur somewhat more frequently than the $m_U/2\upi=6$ structure that we primarily discuss. This tendency of the NL system to form a VSHF with a somewhat different scale than that which is predicted by the S3T system when identical parameter values are used has been explained by \citet[]{Constantinou:2014fh}, in the context of barotropic beta-plane turbulence, to result from the modification of the background spectrum of the NL system by the perturbation-perturbation nonlinear interactions that are retained in the NL system but not in the QL or S3T systems. 
%
\section{Reflection of the S3T Bifurcation in the NL and QL Systems} \label{sec:bifurcation_comparison}

In \textsection \ref{sec:modelcomparison} we compared the behaviour of the NL, QL, and S3T systems with all parameter values fixed. Comparing the three systems in this way allows for a detailed comparison of the structures of the mean state and of the turbulent spectra to be made. 
However, our analysis of the S3T system has revealed phenomena, including the bifurcation associated with the initial formation of the VSHF, that can be analyzed only by allowing variation of the control parameters. 
We now compare the behaviour of the three systems as a function of the excitation strength, $\varepsilon$, in terms of the fraction of the total kinetic energy of the flow that is associated with the VSHF. We define this fraction as $\text{zmf}=\overline{K}/(\overline{K}+K')$, for zonal mean flow (zmf) index, borrowing this definition from studies of barotropic beta-plane turbulence \citep{Srinivasan:2012im,Constantinou:2014fh}. 
We note that in the context of barotropic turbulence an alternate approach based on regime diagrams has also been used to characterize the transition of turbulence to states dominated by zonal jets \citep{Galperin:2010ix}. %, based on regime diagrams, is  emergence of turbulent jets to turbulent states dominated by turbulent jets %which measures the fraction of the total kinetic energy that is contained in the mean flow.

\begin{figure}
%\centerline{\includegraphics[scale=.35]{Fig21_v1p0.png}} 
 %\centerline{\includegraphics{Fig21_v2p0.pdf}} 
  %\centerline{\includegraphics{Fig21_v2p1.pdf}} % Images in 100% size
    %\centerline{\includegraphics{Fig21_rev_v1p0.pdf}} % Images in 100% size
    \centerline{\includegraphics[scale=.35]{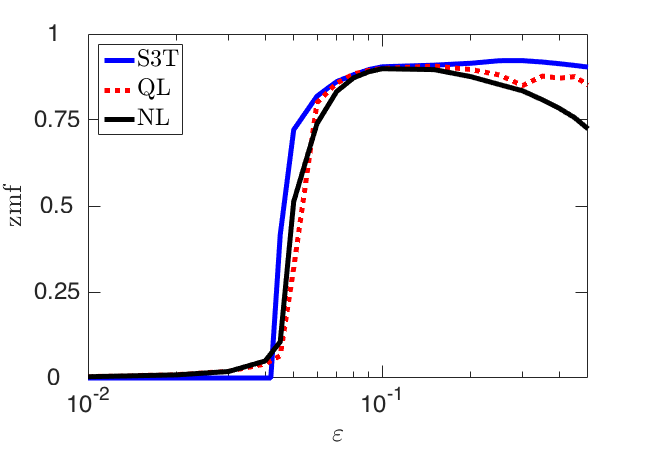}} % Images in 100% size

   \caption{Equilibrium zmf indices in the NL, QL, and S3T systems as functions of $\varepsilon$, with other parameters as in figure \ref{fig:NL_Phenom_FlowSnapshots}. The zmf index measures the fraction of the total kinetic energy of the flow that is associated with the VSHF. This figure shows that the bifurcation through which VSHF forms in the deterministic S3T system is reflected in the behaviour of the NL and QL systems, which show an abrupt increase in the fraction of the total kinetic energy contained in the VSHF near the S3T bifurcation point.}
\label{fig:combined_bifurcation_plot}
\end{figure}

Figure \ref{fig:combined_bifurcation_plot} shows the equilibrium zmf value as a function of $\varepsilon$ for the NL, QL, and S3T systems. As all three systems possess multiple equilibria, there is some ambiguity as to the meaning of the equilibrium energies. For the S3T system we show the maximum zmf obtained when the system is initialized with a perturbative VSHF at each unstable wavenumber $m_U$. For the NL and QL systems we initialize from rest and show the time average of zmf over the final 10 time units of a $t\in[0,450]$ integration. As many long integrations are required for this comparison, the simulations shown in this section are spun up at low resolution and the resulting turbulence is interpolated to the standard resolution of $512^2$ grid points to initialize a simulation of the final 10 time units. 

As discussed in \textsection \ref{sec:scaleselection}, the S3T system passes through a bifurcation near $\varepsilon \approx0.04$ ($Fr=.24$). This bifurcation is reflected in the zmf indices of the QL and NL systems. For $\varepsilon\lesssim 0.04$ the VSHF accounts for only a few percent of the total kinetic energy of the flow. As $\varepsilon$ increases beyond the S3T bifurcation point, the zmf index increases rapidly and the VSHF becomes energetically dominant. As was found in \textsection \ref{sec:modelcomparison}, the S3T VSHF is the most energetic and the QL VSHF tends to be more energetic than the NL VSHF. 
The eventual decrease of the zmf indices in the QL and NL systems as $\varepsilon$ is increased may be due to the tendency of those systems to maintain VSHF structures with $m_U/2\upi=6$, even when this is not the most energetic VSHF structure. 
The S3T curve does not show this decrease as we choose the most energetic VSHF equilibrium to define the S3T equilibrium zmf. 
This maximally energetic VSHF equilibrium often has a lower vertical wavenumber than that of the fastest growing eigenmode, as discussed in \textsection \ref{sec:multiple_equilibria}. 
Developing a complete understanding of the behavior of, and correspondence between, the QL, NL, and S3T systems in the limit of strong excitation is beyond the scope of the present study but is an important avenue for future investigation. %of future work. % he subject of future work. 
Note also in figure \ref{fig:combined_bifurcation_plot} the characteristic increase in fluctuating VSHF amplitude in the NL and QL cases as the bifurcation point is approached. 
This results from excitation of the reflection in QL and NL of the stable modes of the S3T system, which have no analytical expression in the QL and NL systems themselves.
These modes are excited by the noise inherent in the QL and NL systems while no such excitation is seen in the noise-free S3T system \citep{Constantinou:2014fh}.

\section{Conclusions}\label{sec:conclusions}

In this work we studied the formation and maintenance of the VSHF and associated density layers in stratified turbulence by applying SSD to the stochastically excited 2D Boussinesq system. 
Although highly simplified, the 2D Boussinesq system has previously been shown to reflect the properties of VSHF emergence in 3D \citep{Smith:2001uo,Smith:2002wg}. 
Our analysis focused on the strongly stratified regime in which the VSHF is known to develop and that is also the regime relevant to geophysical turbulent jets such as the EDJs. 
Using the S3T implementation of SSD, we showed that VSHFs form spontaneously in this system through the mechanism of cooperative interaction between the turbulence and the mean state. 
While wave-mean flow interaction has previously been hypothesized to be the mechanism responsible for the formation and maintenance of the EDJs \citep{Muench:1999dy,Ascani:2015dd}, the analytical structure required for constructing a comprehensive theory connecting turbulence to the formation and equilibration of these coherent structures was lacking.
The 2D Boussinesq system is a minimal dynamical model that captures horizontal structure formation in stratified turbulence, analogous to the role played by the barotropic beta-plane system in planetary-scale jet formation. 
Unlike the beta-plane system, the 2D Boussinesq equations do not have a conserved potential vorticity and so this stratified turbulence provides a test of the role played by conservation laws in the formation of jets. In agreement with previous studies, we find that VSHF formation occurs robustly in spite of the absence of vorticity as a conserved quantity.  

An aspect of horizontal mean structure formation in stratified turbulence highlighted in this work is the formation of horizontal mean density layers. 
When the VSHF emerges in turbulence, it typically dominates the velocity field at equilibrium and is clearly visible in the instantaneous flow. The associated changes in the stratification, however, are relatively weak and are obscured by turbulent fluctuations. Horizontal averaging reveals the structure of the modified stratification, which agrees well with the predictions of the S3T system. Stratified turbulence thus provides an example in in which the mean state is characterized by examples of both `manifest' and `latent' \citep{Berloff:2009ct} structures simultaneously.

The primary contribution of this work is to explain the dynamics of VSHF formation and equilibration in stratified turbulence using SSD. 
We developed and applied the S3T equations for this system and showed that the behaviour observed in nonlinear simulations mirrors that of the S3T system. 
S3T provides a deterministic and autonomous dynamical system that describes the formation, temporal evolution, and equilibration of the statistical state of the turbulence at second order. 
In S3T the third cumulant, which is associated with perturbation-perturbation nonlinearity, is set to zero and the ergodic assumption equating horizontal and ensemble averages is made. 
The S3T system provides tools, concepts, and insights for understanding turbulent structure formation. For example, test function analysis was used in \textsection \ref{sec:testfunction} to calculate the statistical mean turbulent perturbation fluxes in the presence of an imposed horizontal mean structure. This tool yields the insight that the VSHF forms by modifying the fluxes in such a way as to reinforce the VSHF structure, and explains the specific horizontal mean structure maintained in turbulent equilibrium as being the structure for which the fluxes balance dissipation while not distorting the structure itself. Analysis of the S3T system also allows for the identification of phenomena that are difficult to capture or anticipate in the presence of turbulent fluctuations. For example, the linear stability analysis carried out in \textsection \ref{sec:scaleselection} shows that VSHF formation occurs via a linear instability of the SSD of the underlying homogeneous turbulent state. The growth rate of this instability crosses zero as the strength of the stochastic excitation is increased beyond a critical threshold, resulting in a supercritical bifurcation. 
This bifurcation behaviour is reflected in the dynamics of both the associated quasilinear (QL) and fully nonlinear (NL) systems. 
As an additional example, the S3T system predicts the existence of multiple simultaneously stable turbulent equilibria with different horizontal mean structures. This property of stratified turbulence has not previously been emphasized and might be unexpected from the perspective of nonlinear cascade constrained by conservation laws. From the perspective of S3T as an autonomous and nonlinear dynamical system the existence of multiple equilibria is not surprising.

The authors acknowledge valuable comments from Navid Constantinou and thank John Taylor for providing the DIABLO code. The authors also acknowledge Boris Galperin and several anonymous reviewers whose suggestions helped to improve the manuscript. J.\ G.\ F.\ was partially supported by a doctoral fellowship from the Natural Sciences and Engineering Research Council of Canada. B.\ F.\ F.\ was partially supported by the U.S. National Science Foundation (NSF) under Grant Nos. NSF AGS-1246929 and NSF AGS-1640989.

\appendix
\section{A Reduced Model of 2D Boussinesq Turbulence Illustrating S3T}\label{sec:appendixA}

In this Appendix we construct a severely truncated low-order model (LOM) of the stochastically excited 2D Boussinesq system. A similar approach has been applied to the stochastically excited barotropic beta-plane system \citep{Majda:1999dm}. We first formulate the model, which is expressed using coupled ordinary differential equations, and then proceed to derive the S3T equations for this system. This demonstration serves to illustrate the analytical techniques used in this paper in the context of a simple set of equations. Moreover, we find that this severely truncated model accurately captures certain aspects of the full 2D system.

To obtain the LOM we choose the stochastic excitation, $\sqrt{\varepsilon}S$, to excite a single standing wave mode so that $S\propto \sin(kx)\sin(mz)$ and analyze the interaction between the excited mode and a VSHF with vertical wavenumber $m_U$. We neglect the horizontal mean buoyancy, $B$, as we focus on the linear instability responsible for VSHF formation in which $B$ plays no role (see \textsection \ref{sec:scaleselection}), and we set $\nu=0$. We write the perturbation streamfunction, $\psi'$, the perturbation buoyancy, $b'$, and the VSHF, $U$, in the form of low-order Fourier truncations as
\begin{eqnarray}
\psi'(x,z,t) &=& \psi_e \sin(kx)\sin(mz)+\psi_{+}\cos(kx)\cos\left((m+m_U)z\right), \label{eq:LOM1} \\
b'(x,z,t) &=& b_e \cos(kx)\sin(mz)+b_{+}\sin(kx)\cos\left((m+m_U)z\right), \label{eq:LOM2} \\
U(z,t) &=& U\sin(m_U z). \label{eq:LOM3}
\end{eqnarray}
We choose to retain these terms because the interaction between $U$ and the excited wave, ($\psi_e,b_e$), produces sum and difference wavenumber components including the sheared wavenumber component, $(\psi_{+}$, $b_{+})$. The difference wavenumber component, $(\psi_{-}$, $b_{-})$, is also produced. For simplicity of the present development we write equations with only the $+$ terms included, but the results we show in this Appendix are calculated using a version of the LOM that includes both the $+$ and $-$ components.

To obtain the equations of motion for the coefficients we substitute the expansion (\ref{eq:LOM1})-(\ref{eq:LOM3}) into the QL equations (\ref{eq:QL1})-(\ref{eq:QL4}) and project each term onto the structure functions. For example, the contribution to the $\psi_e$ equation from the mean flow interaction terms in the vorticity equation is given by
\begin{multline}
%\left(-\frac{k^2+m^2}{4}\right)^{-1}\int_0^1 \text{d} x\int_0^1\text{d}z \left(\sin(kx)\sin(mz)\right)\left(-U\partial_x \Delta \psi'+(\partial_x \psi')U_{zz}\right) \\
%= -\frac{1}{2}k(k^2+m(m+2m_j))U\psi_{+}.\end{multline}
\left(-\frac{k_e^2}{4}\right)^{-1}\int_0^1 \text{d} x\int_0^1\text{d}z \left(\sin(kx)\sin(mz)\right)\left(-U\partial_x \Delta \psi'+(\partial_x \psi')U_{zz}\right) \\
= -\frac{1}{2}\frac{k}{k_e^2}(k_+^2-m_U^2)U\psi_{+},\end{multline}
in which $k_e^2=k^2+m^2$ and $k_+^2=k^2+(m+m_U)^2$. The LOM is most compactly written in vector-matrix form. Defining the state vectors of the excited and sheared wave components as $\boldsymbol{\phi}_e=(\psi_e,b_e)^T$ and $\boldsymbol{\phi}_{+}=(\psi_{+},b_{+})^T$ we obtain
\begin{eqnarray}
\left( \begin{array}{c} \dot{\boldsymbol{\phi}}_e \\ \dot{\boldsymbol{\phi}}_{+} \end{array} \right) &=& \left(\begin{array} {cc} \mathsfbi{W}_e & 0 \\ 0 & \mathsfbi{W}_{+} \end{array} \right) \left( \begin{array}{c} \boldsymbol{\phi}_e \\ \boldsymbol{\phi}_{+} \end{array} \right) + U \left(\begin{array} {cc} 0 & \mathsfbi{L}_{e,+} \\ \mathsfbi{L}_{+,e} & 0 \end{array} \right) \left( \begin{array}{c} \boldsymbol{\phi}_e \\ \boldsymbol{\phi}_{+} \end{array} \right)+\sqrt{\varepsilon} \boldsymbol{\xi}, 
\label{eq:LOMQL1} \\
%\dot{U} &=& -r_m U + \frac{1}{4}km_j(2m+m_j)\psi_e \psi_{+}. \label{eq:LOMQL2}
\dot{U} &=& \frac{1}{4}k(k_+^2-k_e^2)\psi_e \psi_{+} - r_m U, \label{eq:LOMQL2}
\end{eqnarray}
%where $\varepsilon$ is the energy injection rate and $\boldsymbol{\xi}=\left(\eta(8/k_e^2)^{1/2},0,0,0\right)^T$ where $\eta$ is Gaussian white noise with unit variance. 
where $\varepsilon$ is the energy injection rate and $\boldsymbol{\xi}=\left(2\sqrt{2}\eta/k_e,0,0,0\right)^T$ where $\eta$ is Gaussian white noise with unit variance. 
The operators $\mathsfbi{W}_e$ and $\mathsfbi{W}_{+}$ encode the gravity wave  dynamics of the excited and sheared components and are given by
%\begin{eqnarray}
%\mathsfbi{W}_e &=& \left( \begin{array}{cc}-1 & k/k_e^2 \\ -kN_0^2 & -1 \end{array} \right), \\
%\mathsfbi{W}_{+} &=& \left( \begin{array}{cc}-1 & -k/k_{+}^2 \\ k N_0^2 & -1 \end{array} \right).
%\end{eqnarray}
\begin{align}
\mathsfbi{W}_e = \left( \begin{array}{cc}-1 & k/k_e^2 \\ -kN_0^2 & -1 \end{array} \right), &&
\mathsfbi{W}_{+} = \left( \begin{array}{cc}-1 & -k/k_{+}^2 \\ k N_0^2 & -1 \end{array} \right).
\end{align}
The operators $\mathsfbi{L}_{e,+}$ and $\mathsfbi{L}_{+,e}$ encode the interactions between the VSHF and the perturbations and are given by
%\begin{eqnarray}
%%\mathsfbi{L}_{e,+} &=& \left( \begin{array}{cc} -\frac{k}{2k_e^2} (k^2+m(m+2m_j)) & 0 \\ 0 & \frac{k}{2} \end{array} \right), \\
%%\mathsfbi{L}_{+,e} &=& \left( \begin{array}{cc} \frac{k}{2k_{+}^2}(k^2+m^2-m_j^2) & 0 \\ 0 & -\frac{k}{2} \end{array} \right).
%\mathsfbi{L}_{e,+} &=& \left( \begin{array}{cc} -\frac{k}{2k_e^2} (k^2_+-m_j^2) & 0 \\ 0 & \frac{k}{2} \end{array} \right), \\
%\mathsfbi{L}_{+,e} &=& \left( \begin{array}{cc} -\frac{k}{2k_{+}^2}(m_j^2-k_e^2) & 0 \\ 0 & -\frac{k}{2} \end{array} \right).
%\end{eqnarray}
\begin{align}
\mathsfbi{L}_{e,+} = \left( \begin{array}{cc} -\frac{k}{2k_e^2} (k^2_+-m_U^2) & 0 \\ 0 & \frac{k}{2} \end{array} \right), &&
\mathsfbi{L}_{+,e} = \left( \begin{array}{cc} -\frac{k}{2k_{+}^2}(m_U^2-k_e^2) & 0 \\ 0 & -\frac{k}{2} \end{array} \right).
\end{align}

\begin{figure}
  %\centerline{\includegraphics[scale=.32]{Fig22_v1p0.png}} 
    \centerline{\includegraphics{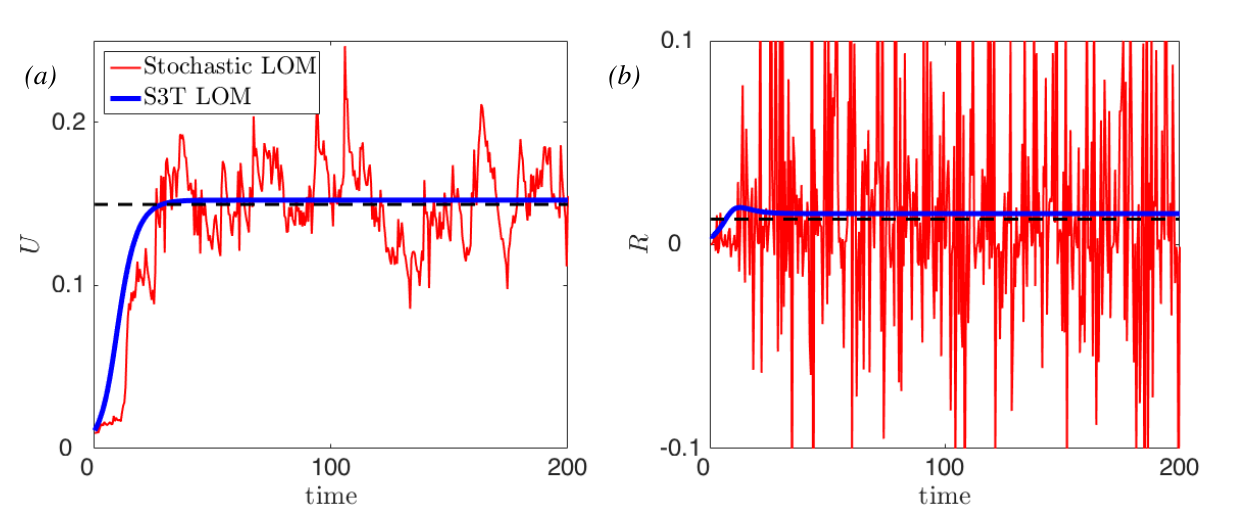}} 
  \caption{Evolution of the LOM system state in the original stochastic system (thin curves) and the corresponding S3T system (thick curves). (a) $U$, the VSHF. (b) $R$, the perturbation momentum flux divergence. Time average values over $t\in[100,200]$ in the stochastic system are indicated by dashed lines. The values of the control parameters are $\varepsilon=.01$, $N_0^2=100$, $r_m=0.1$, and $(k,m,m_U)=2\upi(6,3,7)$, corresponding to $Fr=.42$. This figure shows that the trajectory of the stochastic LOM is made complicated by large fluctuations but that the average behaviour of the system is captured by the deterministic S3T dynamics.}
\label{SSD_vs_StochasticLOM_Figure}
\end{figure}

Equation (\ref{eq:LOMQL1}) is the LOM analog of the QL perturbation equation (\ref{eq:QLmatrixform}). As in the QL system the VSHF, $U$, forms spontaneously in the LOM under certain parameter conditions due to feedbacks between $U$ and the perturbation statistics. Figure \ref{SSD_vs_StochasticLOM_Figure} shows the time evolution of $U$ and the perturbation momentum flux divergence (which we denote $R$, for Reynolds stress) in the LOM. The VSHF develops by $t=25$ and exhibits red noise fluctuations. The momentum flux divergence fluctuates rapidly, sometimes strongly opposing $U$. The time average values of $U$ and $R$ are indicated by the black dashed lines. These results demonstrate the complexity of the LOM `turbulence'. The statistical equilibrium state is characterized by the presence of large fluctuations that obscure the processes that generate and maintain the VSHF. 
We note that this example also demonstrates that VSHFs can form in stochastically excited flows in which the vertical wavenumber of the VSHF (in the present case, $m_U/(2\upi)=7$) is not contained in the excitation spectrum (which, in the present case, contains only $m/(2\upi)=3$). 

We now illustrate the S3T closure technique in the simplified context of the LOM. Defining the complete perturbation state vector as $\boldsymbol{\phi}=(\boldsymbol{\phi}_e^T,\boldsymbol{\phi}_{+}^T)^T$, the instantaneous covariance matrix of the perturbations, prior to ensemble averaging, is $\mathsfbi{C}_{stoch}=\boldsymbol{\phi}\boldsymbol{\phi}^T$. By It\^{o}'s lemma,
\begin{equation}
\dot{\mathsfbi{C}}_{stoch}=\mathsfbi{A}(U)\mathsfbi{C}_{stoch}+\mathsfbi{C}_{stoch} \mathsfbi{A}(U)^T+\varepsilon \mathsfbi{Q}+\boldsymbol{\phi}\boldsymbol{\xi}^T+\boldsymbol{\xi}\boldsymbol{\phi}^T. \label{eq:LOMCstoch}
\end{equation}
Here $\mathsfbi{Q}$ is the ensemble mean excitation covariance, which is the $4\times4$ matrix with $\mathsfbi{Q}_{11}=8/k_e^2$ and all other entries zero, and the operator $\mathsfbi{A}(U)$ is defined as 
\begin{equation}
\mathsfbi{A}(U) = \left(\begin{array} {cc} \mathsfbi{W}_e & 0 \\ 0 & \mathsfbi{W}_{+} \end{array} \right) + U \left(\begin{array} {cc} 0 & \mathsfbi{L}_{e,+} \\ \mathsfbi{L}_{+,e} & 0 \end{array} \right).
\end{equation}
We obtain the S3T dynamics by taking the ensemble averages of (\ref{eq:LOMQL2})  and (\ref{eq:LOMCstoch}) under the ergodic assumption that horizontal and ensemble averages are equivalent so that $U=\langle U \rangle$. The stochastic terms in (\ref{eq:LOMCstoch}) vanish upon averaging and the S3T equations of motion are
\begin{eqnarray}
\dot{\mathsfbi{C}} &=& \mathsfbi{A}(U)\mathsfbi{C}+\mathsfbi{C} \mathsfbi{A}(U)^T+\varepsilon \mathsfbi{Q}, \label{eq:LOMS3T1}\\
\dot{U} &=& R-r_m U , \label{eq:LOMS3T2}
\end{eqnarray}
%where $\mathsfbi{C}=\langle \mathsfbi{C}_{stoch}\rangle$, $R=(1/4)km_j(2m+m_j) \mathsfbi{C}_{13}$, and $\mathsfbi{C}_{13}=\langle \psi_e \psi_{+}\rangle$. 
where $\mathsfbi{C}=\langle \mathsfbi{C}_{stoch}\rangle$, $R=(1/4)k(k_+^2-k_e^2)\mathsfbi{C}_{13}$, and $\mathsfbi{C}_{13}=\langle \psi_e \psi_{+}\rangle$. The thick curves in figures \ref{SSD_vs_StochasticLOM_Figure} (a) and (b) show the time evolution of the S3T state. The S3T dynamics captures the time evolution of the VSHF as well as the time average of the rapidly-fluctuating perturbation momentum flux divergence. 

Although working in the S3T formalism introduces some abstraction, the S3T equations provide understanding by enabling direct interpretation and analysis of the second-order statistical relationships that are explicit in S3T. A statistical quantity of central interest is $\langle \psi_e \psi_{+}\rangle$, which is proportional to $R$ and so directly drives the VSHF. The S3T dynamics of $\langle \psi_e \psi_{+}\rangle$ are
%\begin{multline}
%\frac{d}{dt} \langle \psi_e \psi_{+} \rangle = -2 \langle \psi_e \psi_{+} \rangle + \frac{k}{k_e^2}\langle b_e \psi_{+} \rangle - \frac{k}{k_{+}^2}\langle \psi_e b_{+} \rangle \\
%+ \frac{k}{2k_{+}^2}(k^2+m^2-m_j^2)U\langle \psi_e^2 \rangle - \frac{k}{2k_e^2}(k^2+m(m+2m_j))U\langle \psi_{+}^2\rangle, \label{eq:LOMS3TReqn}
%\end{multline}
\begin{multline}
\frac{d}{dt} \langle \psi_e \psi_{+} \rangle = -2 \langle \psi_e \psi_{+} \rangle + \frac{k}{k_e^2}\langle b_e \psi_{+} \rangle - \frac{k}{k_{+}^2}\langle \psi_e b_{+} \rangle \\
- \frac{k}{2k_{+}^2}(m_U^2-k_e^2)U\langle \psi_e^2 \rangle - \frac{k}{2k_e^2}(k_+^2-m_U^2)U\langle \psi_{+}^2\rangle, \label{eq:LOMS3TReqn}
\end{multline}
%\begin{equation}
%\frac{d}{dt} \langle \psi_e \psi_{+} \rangle = -2 \langle \psi_e \psi_{+} \rangle + \frac{k}{k_e^2}\langle b_e \psi_{+} \rangle - \frac{k}{k_{+}^2}\langle \psi_e b_{+} \rangle - \frac{k}{2k_{+}^2}(m_j^2-k_e^2)U\langle \psi_e^2 \rangle - \frac{k}{2k_e^2}(k_+^2-m_j^2)U\langle \psi_{+}^2\rangle, \label{eq:LOMS3TReqn}
%\end{equation}
which is the $(1,3)$ component of (\ref{eq:LOMS3T1}). The direct feedback between $U$ and $R$ is expressed in the fourth term on the right-hand side (RHS) of (\ref{eq:LOMS3TReqn}). For our parameter choices $(m_U^2-k_e^2)>0$ so this feedback suppresses VSHF formation. The flux divergence $R$ is instead produced by covariances involving the buoyancy field, expressed in the second and third terms on the RHS of (\ref{eq:LOMS3TReqn}). These covariances are in turn produced through direct interactions with $U$ which are expressed in other components of (\ref{eq:LOMS3T1}).

\begin{figure}
 % \centerline{\includegraphics[scale=.31]{Fig23_v1p0.png}} 
%   \centerline{\includegraphics[scale=.28]{Fig23_v1p1_PPT.pdf}} 
   \centerline{\includegraphics{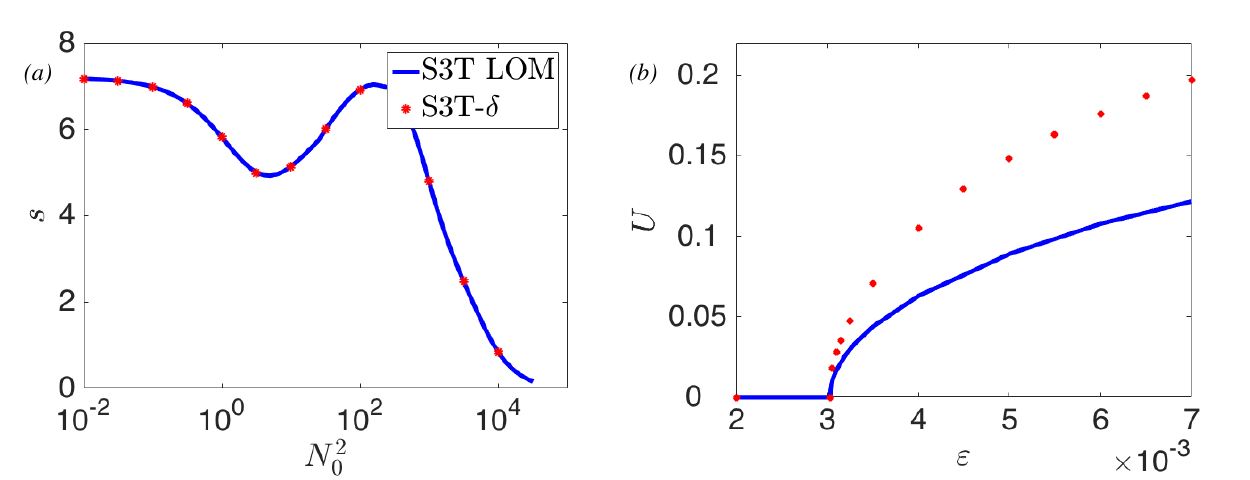}} 
  \caption{Comparison between the S3T LOM (solid curves), in which a standing wave is stochastically excited, and the full S3T dynamics when a closely related homogeneous excitation is chosen, referred to as S3T-$\delta$ (dots). (a) Growth rate of the linear instability responsible for VSHF formation, maximized over all VSHF wavenumbers, as a function of $N_0^2$ for $\varepsilon=0.5$. (b) Equilibrium amplitude of the $m_U/2\upi=7$ VSHF as a function of $\varepsilon$ for $N_0^2=100$. The control parameters are $(k,m)=2\upi(6,3)$ and $r_m=0.1$. This figure shows that the S3T LOM accurately captures the instability responsible for VSHF formation, indicating that the highly simplified LOM has correct physics at the linear level and can be used to understand the process of VSHF formation. The VSHF equilibration process, however, is not captured correctly even for weakly supercritical excitation strengths.}
\label{Comparison_Figure_Full_new}
\end{figure}

VSHF formation in the LOM can be understood through linear stability analysis of the S3T system in analogy with \textsection \ref{sec:scaleselection}. The fixed point of the S3T equations corresponding to turbulence without a VSHF is obtained by solving (\ref{eq:LOMS3T1}) with $U=0$, which gives
%\begin{eqnarray}
%\mathsfbi{C}^{\star}_{11}&=&\frac{2\varepsilon}{k_e^2}\left(2- \frac{k^2 N_0^2}{k_e^2+k^2 N_0^2}\right), \\
%%\mathsfbi{C}^{\star}_{12}&=&-{2\varepsilon k N_0^2}{k_e^2+k^2N_0^2}, \\
%%\mathsfbi{C}^{\star}_{22}&=&\frac{2\varepsilon k^2 N_0^2}{k_e^2+k^2N_0^2},
%\mathsfbi{C}^{\star}_{12}&=&-\frac{2\varepsilon k N_0^2}{k_e^2+k^2N_0^2}, \\
%\mathsfbi{C}^{\star}_{22}&=&\frac{2\varepsilon k^2 N_0^4}{k_e^2+k^2N_0^2},
%\end{eqnarray}
\begin{align}
\mathsfbi{C}^{\star}_{11}=\frac{2\varepsilon}{k_e^2}\left(2- \frac{k^2 N_0^2}{k_e^2+k^2 N_0^2}\right), &&
\mathsfbi{C}^{\star}_{12}=-\frac{2\varepsilon k N_0^2}{k_e^2+k^2N_0^2}, &&
\mathsfbi{C}^{\star}_{22}=\frac{2\varepsilon k^2 N_0^4}{k_e^2+k^2N_0^2},
\end{align}
with $\mathsfbi{C}^{\star}_{21}=\mathsfbi{C}^{\star}_{12}$ and the other elements of $\mathsfbi{C}^{\star}$ being zero. Linearizing equations (\ref{eq:LOMS3T1})-(\ref{eq:LOMS3T2}) about the fixed point $(\mathsfbi{C},U)=(\mathsfbi{C}^{\star},0)$ we find that VSHF perturbations, $\delta U$, evolve together with covariance matrix perturbations of the form
\begin{equation}
\delta \mathsfbi{C}=\left( \begin{array}{cc} 0 & \mathsfbi{\delta C}^{e,+} \\  (\mathsfbi{\delta C}^{e,+})^T & 0 \end{array}\right),
\end{equation}
independently of perturbations to the other elements of $\mathsfbi{C}$, according to the linearized equations
\begin{eqnarray}
%\delta \dot{\mathsfbi{C}}^{e,+} &=& \mathsfbi{W}_e \mathsfbi{\delta C}^{e,+}+\mathsfbi{\delta C}^{e,+}(\mathsfbi{W}_{+})^T+\delta U \mathsfbi{C}^{\star}(\mathsfbi{L}_{+,e})^T, \\
%\dot{\delta U} &=& -r_m \delta U +(1/4)km_j(2m+m_j)\mathsfbi{\delta C}^{e,+}_{11}.
\delta \dot{\mathsfbi{C}}^{e,+} &=& \mathsfbi{W}_e \mathsfbi{\delta C}^{e,+}+\mathsfbi{\delta C}^{e,+}(\mathsfbi{W}_{+})^T+\delta U (\mathsfbi{C}^{e,e})^{\star}(\mathsfbi{L}_{+,e})^T, \\
\dot{\delta U} &=& -r_m \delta U +(1/4)k(k_+^2-k_e^2)\mathsfbi{\delta C}^{e,+}_{11},
\end{eqnarray}
in which $(\mathsfbi{C}^{e,e})^{\star}$ denotes the upper-left nonzero $2\times2$ submatrix of $\mathsfbi{C}^{\star}$. This system of five linear ODEs can be rearranged as a $5\times 5$ matrix-vector system and the eigenvalues and eigenvectors can be calculated as usual. For sufficiently strong excitation, the dominant eigenvalue has a positive real part and the $U=0$ fixed point is unstable to eigenmodes associated with VSHF formation. Figure \ref{Comparison_Figure_Full_new} (a) (solid curve) shows the instability growth rate as a function of $N_0^2$, maximized over all mean flow wavenumbers $m_U$.

At the level of the linear instability responsible for VSHF formation, the S3T dynamics of the LOM captures the dynamics of the full S3T system (\ref{eq:S3TLIN1})-(\ref{eq:S3TLIN3}) when the excitation of the full system is appropriately chosen. Figure \ref{Comparison_Figure_Full_new} (a) (dots) shows the instability growth rate in the full S3T dynamics when $\sqrt{\varepsilon}S$ excites only the four Fourier components with $(k,m)=2\upi(\pm6,\pm3)$, which is a homogeneous but anisotropic excitation. We refer to this configuration of the S3T system as $\text{S3T-}\delta$, as the excitation spectrum consists of delta functions at the excited wavenumbers. Although this excitation is not identical to the  LOM excitation, which is not homogeneous, the correspondence between the growth rates indicates that the LOM accurately captures the dynamics of the full S3T system in this case. We note that, in contrast to the results of \textsection \ref{sec:scaleselection}, the VSHF growth rate remains positive as $N_0^2\to0$ for this choice of anisotropic excitation.

Although the LOM correctly captures the linear instability that produces the VSHF, it fails to capture the finite-amplitude equilibration of the VSHF. Figure \ref{Comparison_Figure_Full_new} (b) shows the fixed point value of $U$ as a function of $\varepsilon$ for the S3T dynamics of the LOM alongside $\text{max}(U)$ for the $\text{S3T-}\delta$ system. Although the VSHF in the LOM S3T dynamics forms through a bifurcation at the same value of $\varepsilon$ as in the full S3T dynamics, the LOM does not capture the equilibrium amplitude of the VSHF even very near the bifurcation point. This failure of the LOM occurs because the dynamics of weakly supercritical VSHF equilibration, as explained in \textsection \ref{sec:equilibration}, are related to the negative curvature of the flux divergence as a function of the VSHF strength (see figure \ref{fig:WeakEquil_SimpleBalance_Plot}). This curvature is due to the production via wave-mean flow interaction of perturbations with vertical wavenumbers that are not included in the LOM.

\section{The Covariance Matrix of Homogeneous Turbulence}\label{sec:appendixB}

In this Appendix we show details of the analytical solution of the time-independent Lyapunov equation (\ref{eq:S3TFP1}). We define the $N\times N$ submatrices of $\mathsfbi{C}^{\star}_n$ as
\begin{equation}
\mathsfbi{C}_n^{\star}=\left( \begin{array}{cc} \mathsfbi{C}_{\psi \psi,n} & \mathsfbi{C}_{\psi b,n} \\ \mathsfbi{C}_{\psi b,n}^{\dagger} & \mathsfbi{C}_{b b,n} \end{array}\right).
\end{equation}
Defining $\mathsfbi{R}_n=\mathsfbi{I}-\nu \mathsfbi{\Delta}_n$ and $\mathsfbi{M}_n =\mathsfbi{\Delta}_n^{-1}$ we expand (\ref{eq:S3TFP1}) into the three independent matrix equations
\begin{align}
-\mathsfbi{R}_n \mathsfbi{C}_{\psi \psi,n} +\text{i}k_n\mathsfbi{M}_n \mathsfbi{C}_{\psi b,n}^{\dagger}-\mathsfbi{C}_{\psi \psi,n} \mathsfbi{R}_n -\text{i}k_n\mathsfbi{C}_{\psi b,n} \mathsfbi{M}_n &= -\varepsilon \mathsfbi{Q}_{\psi \psi,n}, \label{eq:FPexp1} \\
-\mathsfbi{R}_n \mathsfbi{C}_{\psi b,n} +\text{i}k_n\mathsfbi{M}_n \mathsfbi{C}_{b b,n}+\text{i}k_nN_0^2 \mathsfbi{C}_{\psi \psi,n}-\mathsfbi{C}_{\psi b,n}\mathsfbi{R}_n &= 0  \label{eq:FPexp2}, \\
-\text{i}k_nN_0^2 \mathsfbi{C}_{\psi b,n}-\mathsfbi{R}_n \mathsfbi{C}_{b b,n}+\text{i}k_nN_0^2\mathsfbi{C}_{\psi b,n}^{\dagger}-\mathsfbi{C}_{b b,n}\mathsfbi{R}_n &= 0.  \label{eq:FPexp3}
\end{align}
%We note that for our boundary conditions and excitation the matrices $\mathsfbi{R}_n$, $\mathsfbi{M}_n$, and $\mathsfbi{Q}_{\psi \psi,n}$ are real, symmetric, and circulant, and that circulant matrices form a commutative algebra.
We note the following: \emph{i)} For our boundary conditions and excitation the matrices $\mathsfbi{R}_n$, $\mathsfbi{M}_n$, and $\mathsfbi{Q}_{\psi \psi,n}$ are real, symmetric, and circulant. \emph{ii)} Inverses, products, and sums of circulant matrices are circulant. \emph{iii)} Circulant matrices commute with one another \citep{Davis:2012tw}.
The form of (\ref{eq:FPexp1})-(\ref{eq:FPexp3}) suggests that we seek solutions with $\mathsfbi{C}_{\psi b,n}=\text{i}\mathsfbi{\tilde{C}}_{\psi b,n}$ and real $\mathsfbi{C}_{\psi \psi,n},\mathsfbi{\tilde{C}}_{\psi b,n},\mathsfbi{C}_{b b,n}$. We further seek solutions in which $\mathsfbi{C}_{\psi \psi,n}$, $\mathsfbi{\tilde{C}}_{\psi b,n}$, and $\mathsfbi{C}_{b b,n}$ are also circulant and symmetric, corresponding to homogeneous turbulence. Using these properties we rewrite (\ref{eq:FPexp1})-(\ref{eq:FPexp3}) as
\begin{align}
-2\mathsfbi{R}_n \mathsfbi{C}_{\psi \psi,n}+2k_n\mathsfbi{M}_n \mathsfbi{\tilde{C}}_{\psi b,n}&=-\varepsilon \mathsfbi{Q}_{\psi \psi,n}, \label{eq:FPfinalsimp1} \\
-2\mathsfbi{R}_n \mathsfbi{\tilde{C}}_{\psi b,n}+k_n N_0^2\mathsfbi{C}_{\psi \psi,n}+k_n\mathsfbi{M}_n \mathsfbi{C}_{b b,n} &=0 ,\label{eq:FPfinalsimp2} \\
-2\mathsfbi{R}_n \mathsfbi{C}_{b b,n} + 2k_n N_0^2 \mathsfbi{\tilde{C}}_{\psi b,n}&=0. \label{eq:FPfinalsimp3}
\end{align}
Equations (\ref{eq:FPfinalsimp1})-(\ref{eq:FPfinalsimp3}) can be solved to give
\begin{align}
\mathsfbi{C}_{b b,n}&=k_n N_0^2 \mathsfbi{R}_n^{-1} \mathsfbi{\tilde{C}}_{\psi b,n},  \label{eq:FPsoln1} \\
\mathsfbi{\tilde{C}}_{\psi b,n}&=-k_nN_0^2\left[-2\mathsfbi{R}_n+k_n^2N_0^2 \mathsfbi{M}_n\mathsfbi{R}_n^{-1}\right]^{-1}\mathsfbi{C}_{\psi \psi,n} , \label{eq:FPsoln2} \\
\mathsfbi{C}_{\psi \psi,n}&=\left\{2\mathsfbi{R}_n+2k_n^2 N_0^2 \mathsfbi{M}_n \left[-2\mathsfbi{R}_n+k_n^2N_0^2 \mathsfbi{M}_n\mathsfbi{R}_n^{-1}\right]^{-1}\right\}^{-1} \varepsilon \mathsfbi{Q}_{\psi \psi,n}. \label{eq:FPsoln3}
\end{align}
Equations (\ref{eq:FPsoln1})-(\ref{eq:FPsoln3}) can be inverted to obtain $\mathsfbi{C}^{\star}_n$ explicitly in terms of $\mathsfbi{Q}_{\psi \psi,n}$ and constitute our final solution for the homogeneous turbulent fixed point.

\bibliographystyle{jfm}
%\bibliography{mybib.bib}
%\bibliography{BibFileFromPapers.bib}
%\bibliography{Working_Bibfile_resub.bib}
\bibliography{Working_Bibfile_resub}

\end{document}